%% file: ThesisFinal.tex
\documentclass[12pt]{report}
\usepackage{epsfig}
\usepackage{epstopdf}
\usepackage{suthesis-2e}
\usepackage{epsfig}
\usepackage{amsthm}
\usepackage{amsfonts}
\usepackage{amssymb}
\usepackage{amsmath}
\usepackage{graphicx}
\usepackage{graphics}
\usepackage{lscape}
\usepackage{pdflscape}

\begin{document}
\bibliographystyle{unsrt}
\title{Modeling of Ultra-Short Soliton Propagation in Deterministic and Stochastic Nonlinear Cubic Media}
\author{Levent Kurt}
\dept{Physics}
\submitdate{April 2011}
\tablespagefalse

\principaladvisor{Professor Sultan Catto \\ Chair of Examining Committee}
\firstreader{Professor Tobias Sch{\"a}fer - Co-Advisor}


\beforepreface
\prefacesection{Abstract}
\input{Abstract}

\prefacesection{Acknowledgements}
\input{Acknowledgements}

\prefacesection{Dedication}

\input{Dedication}
\afterpreface

\chapter{Introduction}
\input{Chap1_Introinc}

\chapter{Pulse Propagation in Nonlinear Cubic Media}
\input{Chap2_Maxwellinc}

\chapter{Nonlinear Schr\"odinger Equation}
\input{Chap3_NLSEinc}
\chapter{Short Pulse Equation}
\input{Chap4_SPEinc}

\chapter{Solitons}

\input{Chap5_Solitonsinc}
\chapter{Stochastic Short Pulse Equation}
\input{Chap6_StochasticSPEinc}
\chapter{Numerical Methods}
\input{Chap7_NumericalMethodsinc}
\chapter{Conclusion}

\input{Chap8_Conclusion}

\bibliography{master}

\end{document}

%% file: Abstract.tex
We study the short pulse dynamics in the deterministic and stochastic environment in this thesis. The integrable short pulse equation is a modeling equation for ultra-short pulse propagation in the infrared range in the optical fibers. We investigate the numerical proof for the exact solitary solution of the short pulse equation. Moreover, we demonstate that the short pulse solitons approximate the solution of the Maxwell equation numerically. Our numerical experiments prove the particle-like behavior of the short pulse solitons. Furthermore, we derive a short pulse equation in the higher order.

A stochastic counterpart of the short pulse equation is also derived through the use of the multiple scale expansion method for more realistic situations where stochastic perturbations in the dispersion are present. We numerically show that the short pulse solitary waves persist even in the presence of the randomness. The numerical schemes developed demonstrate that the statistics of the coarse-graining noise of the short pulse equation over the slow scale, and the microscopic noise of the nonlinear wave equation over the fast scale, agree to fairly good accuracy.

%% file: Acknowledgements.tex
First and foremost I wish to express my deep gratitude to my mentor and advisor, Sultan Catto. His guidance, encouragement, and enthusiasm for science and art always inspired me. I always found his anecdotes of the lives of prominent scientists to be intriguing. I became aware of the great contributions made to science and physics in general by the prominent physicist Feza G\"ursey, the former advisor of Prof. Catto himself, only as a result of my interactions with Prof. Sultan Catto. I took it upon myself to research the work and life of Dr.G\"ursey and came to realize what a truly inspirational figure he must have been for Prof Sultan Catto, many others during his time, and now myself. Without Prof. Sultan Catto's guidence and thoughtfulness, I don't believe I would be able to in the position I am in today.  

I would like to thank my advisor Tobias Sch\"afer. It has been an honor to be his first Ph.D. student. With his enthusiasm and his great efforts to explain ideas and concepts clearly and simply, he helped tremendously in working out numerics enjoyable for me. Whenever we met, the unending smile on his face always seemed to liven my mood and impart a resurgence in my energies. He provided invaluable encouragement and help throughout my thesis work. I would have been lost without him.

I would like to thank all my committee members and acknowledge all of those who contributed greatly to my physics education at the Graduate Center of the City University of New York. I wish to especially express my warm and sincere thanks to Prof. Ramzi Khuri, Prof. Gregory Aizin and Prof. Ming-Kung Liou.
  
I would like to acknowledge my classmates, colleagues and friends as well. I had the privilege of living in the most wonderful and amazing city in the world, New York City. My time in New York during the period of my Ph.D. study was made enjoyable due to the company of my friends, particularly, Amish Khalfan, Murali Devi, Duran Kurt and \c{S}amil Emre \"{O}\u{g}\"{u}n. I wish to express my sincere thanks to Amish and Murali for their valuable discussions about physics. The loss of my dear friend, \c{S}amil Emre, deeply hurt me. I dedicate this thesis to the memory of \c{S}amil Emre \"{O}\u{g}\"{u}n. May peace be upon him. 

I would like to thank my close friends back in Turkey. They have been a source of  real friendships and strong moral values in contributing to my intellectual growth and personal identity.   

I would like to thank my brother, Onur Kurt. His tireless support helped me work through this trying period of my life.  

Finally, I would like to extend my deepest gratitude and warmest thanks to my parents for all their love and encouragement throughout the years. My parents have always been a tremendous source of love and support.

%% file: Dedication.tex
This thesis is dedicated to my friend, \c{S}amil Emre \"{O}\u{g}\"{u}n. I miss him deeply.

%% file: Chap1_Introinc.tex
\section{Introduction}

Physical phenomena whose states change over time are modeled by different mathematical techniques. Often, one is interested in the behavior of the physical system in time, or finds it necessary to study of the dynamical behavior of the physical system. Modeling equations are generally given in partial-differential equation (PDE) forms \cite{farlow:1993}. The wave equation, the heat equation and the Schr\"odinger equation are only a few famous partial differential equations that we encounter in physics and the applied sciences. If the modeling equation for a given dynamical system is integrable, such a system is known as an integrable system. These types of models are not only used in the natural sciences and engineering, but also in social sciences such as financial and economic forecasting and environmental modeling \cite{salsa:2008}. Not all physical phenomena can be modeled in a deterministic manner. Deterministic dynamical systems need not be linear. Modeling equations may also appear in nonlinear differential forms \cite{debnath:2005} whose solution may not be found analytically. Numerical techniques are then utilized to study such dynamical systems \cite{mathworks:online,hunt-lipsman-rosenberg:2001}. On the other hand, some other phenomena appear to be highly stochastic. Stochasticity may arise from an internal or external mechanism in the system and may have some visible impact in the evolution of the system state. Under the influence of randomness, the system is not deterministic anymore and it requires a probabilistic and statistical approach to 
study the evolution of the system \cite{lemons:2002}. Governing equations for such systems can be written in stochastic differential form \cite{gardiner:1985} or in corresponding Fokker-Planck equation form \cite{risken:1989}.

Interaction of light with matter is an important physical phenomenon that has played a central role in the recent history of technological advances. Light is an electromagnetic wave. A wave is simply a quantity that varies with time and position. It is a periodic disturbance with many oscillations, and propagates through a medium such as empty space or an optical fiber \cite{scholarpedia:online}. A wave is generally modeled mathematically by using partial differential equations (PDEs) with a wave value $u(x,t)$, an independent variable time $t$ and one or more independent spatial variables $x$.  Waves can fall under two categories: linear waves and nonlinear waves. Linear waves are those modeled by linear equations. The classical wave equation in three dimensions is very well-known example: 
\begin{equation}
\frac{1}{c^2}\frac{\partial^2 u}{\partial t^2}= \mathbf{\nabla}u \,,
\end{equation}
whose solution is a  linear wave. Linear PDEs are easy to solve as opposed to nonlinear PDEs. Since the superposition principle applies to linear waves, linear combinations of the simple solutions can be used to form more complex solutions for complicated linear problems. However, the principle of superposition does not apply to the second category of waves, namely, nonlinear waves. Nonlinear waves are described by nonlinear equations. The nonlinear Schr\"odinger equation (NLSE), the  Korteweg-de Vries equation (KdV) and the sine-Gordon equation (sG) are some of the very well-known and well-studied examples of integrable nonlinear equations whose solutions are solitary waves, also called solitons. The inverse scattering theorem is used to solve integrable nonlinear PDEs \cite{ablowitz-segur:1981,debnath:2005}. Although the nonlinear KdV and sG equations have exact solitary wave solutions, there are some other types of nonlinear wave equations whose solutions display singularities or those decaying in time and space (smooth dispersive solutions). 

The discovery of optical fibers is a milestone in modern day communication technology and has been used in many optical systems since then. Compared with the electrical transmission system based on copper wires, fiber optic communication utilizes huge amount of data to transmit signals. That serves as the main reason why optical fibers based communication systems have played a major role in the advent of the information age. The set of Maxwell's equations is an elegant description of light and matter interaction, and can be reduced to a single nonlinear wave equation for studying optical phenomena. However, the exact solution of the nonlinear wave equation is unknown as of yet. Nevertheless, the wave equation can be approximated to simpler forms expressed as nonlinear partial differential equations to describe light propagation in optical fibers. 

The asymptotic multiple scale expansion of the nonlinear wave equation yields the cubic nonlinear Schr\"odinger equation \cite{newell-moloney:1992}. The assumption that is made in the derivation of the NLSE is that the pulse width is large in comparison to the carrier wavelength \cite{schaefer-wayne:2004}. The NLSE is the governing equation for pulse propagation in optical fibers and possesses exact solitary wave solutions \cite{mollenauer-gordon:2006}. As the name suggests, the NLSE is just the nonlinear version of the famous quantum mechanical Schr{\"o}dinger equation. It can also be applied to some other physical phenomena such as hydrodynamics and quantum condensates \cite{sulem-sulem:1999}. 

The short solitary waves, or short pulses, are used to study the nonlinear effects in optical fibers due to their unique solitonic properties. The width of those short pulses generally ranges from 10 nano-seconds to 10 femto-seconds \cite{agrawal:2007}. As the pulse shortens even further, the NLSE fails to be a good modeling equation \cite{schaefer-wayne:2004,rothenberg:1992,sulem-sulem:1999}. A nonlinear partial differential equation derived by T.Schafer and C.E. Wayne \cite{schaefer-wayne:2004} may be used to model the propagation of ultra-short pulses \cite{karasawa-nakamura-etal:2001,hasegawa-kodama:1995} in nonlinear cubic media (optical fibers) and is given, up to a scale transformation, as     
\begin{equation}
u_{xt}=u+\frac{1}{6}u^3_{xx} \,.
\end{equation}
This equation is called the short pulse equation (SPE). The SPE is integrable \cite{sakovich-sakovich:2005}, and possesses exact one-soliton \cite{sakovich-sakovich:2006} and multi-soliton \cite{matsuno:2007} solutions. There has been an intensive research on the SPE over the past few years due to its integrability and its exact soliton solutions. The Hamiltonian structures \cite{brunelli:2005,brunelli:2006} and conserved quantities \cite{erbas:2008} of the SPE are found. The periodic solutions are also obtained \cite{parkes:2008}. The vector short pulse equation (VSPE) is introduced \cite{pietrzyk-kanatt-bandelow:2008} and the integrability of the VSPE is studied as well \cite{sakovich:2008}. The regularized short pulse equation (RSPE) is derived \cite{costanzino-manukian-jones:2008}. The existence of multi-pulses \cite{manukian-costanzino-etal:2009}, traveling waves, and solitary wave solutions \cite{costanzino-manukian-jones:2008,costanzino:2006} of the RSPE are studied. 

The study of the short pulse equation in both deterministic and stochastic media is the main focus for the rest of the chapters. We present Maxwell's equations and the nonlinear wave equation in lieu of the optical phenomena discussed in chapter two. The discussion of the multiple scale expansion method and the NLSE is presented in chapter three.  The derivation of the SPE followed by a numerical analysis is provided in chapter four. We also derive a higher order SPE in this chapter. In chapter five, we investigate SPE solitons and their particle-like properties. The following chapter deals with a relatively new area concerning stochasticity. The derivation of the stochastic SPE, and the discussion on coarse-graining noise are given in chapter six as well. Since the numerical analysis is an important portion in our work, we reserve an entire chapter for numerical methods employed in our numerical experiments. The closing chapter provides prospect for research pursuits regarding short pulse dynamics.

%% file: Chap2_Maxwellinc.tex
The theory of electromagnetic wave propagation in dispersive nonlinear media has played a major role in the advent of the 20th century's information technology. In this chapter, we will derive a basic equation that governs propagation
of optical pulses in nonlinear cubic media such as single-mode fibers in one dimension. This equation is the starting point of our discussion in this thesis. 

\section{Maxwell's Equations}

The theory of wave motion is an important mathematical model in many 
areas of physics and engineering. A large number of real world applications can be explained using the solutions of the wave equation. We look at the wave theory in the perspective of light and obtain a one dimensional model associated with the light propagation in nonlinear cubic media. Mathematically, the basis of wave theory is the wave equation which is a second-order partial differential equation.  One can derive a wave equation either in a linear form or nonlinear form depending upon the polarization of the material \cite{boyd:1992,schaefer-wayne:2004}. The linear wave equation is no longer sufficient to describe the propagation of light in a medium when the intensity of light becomes large enough. In such situations, light waves interact with one another and with the optical medium leading to the emergence of nonlinear effects. These nonlinear phenomena require an extension of the linear theory, and a nonlinear response of optical materials to the optical excitations must be taken into account. In the one-dimensional case, the displacement of the wave is assumed to be a scalar function and the pulse is a scalar wave. One must refer to Maxwell's equations to grasp a deep understanding of wave motion in a variety of situations. Maxwell's equations are the governing equations for the propagation of optical pulses in nonlinear media and are given as a set of equations in differential form in three dimensions as \cite{jackson:1999}
\begin{equation}\label{MaxEqn3}
\begin{aligned}
\mathbf{\nabla} \times \textbf{E} & =  - \frac{\partial \textbf{B}}{\partial t} 
\qquad &\textbf{$\nabla$} \cdot \textbf{D} & = & \rho 
\\
\textbf{$\nabla$} \times \textbf{H} &= \textbf{J} + \frac{\partial\textbf{D}}{\partial t} \qquad &\textbf{$\nabla$} \cdot\textbf{B}&=&0 
\end{aligned}
\end{equation}
where $\rho$ is charge density, $\textbf{J}$ is current density. The magnetizing field or magnetic field $\textbf{H}$ and electric field $\textbf{E}$ are related with the corresponding magnetic flux density or magnetic induction $\textbf{B}$ and the electric displacement field $\textbf{D}$ through
\begin{eqnarray}\label{magflux}
\mathbf{B} &=& \mu_0\mathbf{H}+\mathbf{M} \\
\mathbf{D} &=& \epsilon_0 \mathbf{E} + \mathbf{P}
\end{eqnarray} 
where $\mathbf{P}$ is electric polarization, $\mathbf{M}$ is magnetic polarization, $\epsilon_0$ is the electric constant and $\mu_0$ is the magnetic constant. Let's briefly mention the physical meanings of these quantities before deriving the wave equation from Maxwell's equations. The electric constant $\epsilon_0$ is also called the permittivity of free space and describes the interaction of free space with an electric field. Therefore, permittivity is a physical quantity and relates the electric charge to the mechanical quantities such as force. Similarly, the magnetic constant $\mu_0$ is also a physical constant which is sometimes called the permeability of free space. It is the degree of magnetization that vacuum can obtain in response to an applied magnetic field. As for the electric polarization, it is simply the average electric dipole moment per unit volume and can be formulated as $\mathbf{P}=\sum_{i=1}^{N}\mathbf{p}_i$, where $p_i$ is the dipole moment of the $i$th molecule and $N$ is the number of molecules per unit volume. Notice that the average of the sum of the dipole moments of all molecules comes from the fact that $N$ is the statistically averaged large number of molecules. In general, the polarization vector $\mathbf{P}$ depends upon the local value of the electric field strength $E$ nonlinearly. The total polarization induced by electric dipoles in a material as a response to an applied electric field satisties the general form \cite{boyd:1992} and can be written as a power series in the electric field strength 
\begin{equation}\label{Polarization_General}
\mathbf P = \epsilon_0 ( \chi^{(1)} . \mathbf E + \chi^{(2)} : \mathbf E^2 + \chi^{(3)} \vdots \mathbf E^3 + ... )
\end{equation}
where $\chi^{(j)}$ is the jth order susceptibility. The nonlinear susceptibility is a quantity that is used to determine the nonlinear polarization of a medium and is a very useful quantity when describing nonlinear optical phenomena. In a similar fashion, $\mathbf{M}$ is the magnetic polarization and is simply an average magnetic dipole moment per unit volume. There is no further explaination required for this concept because it is zero for nonmagnetic materials. By manipulating Maxwell's equations, we can obtain a number of the properties of light such as the relationship between the $\mathbf{E}$ and $\mathbf{B}$ fields. Maxwell's equations require light to be a transverse wave, i.e., the vector displacements $\mathbf{E}$ and $\mathbf{B}$ are perpendicular to the direction of propagation $k$. It will be seen in the next section that we choose $\mathbf{E}$ and $\mathbf{B}$ fields based on this condition. In light of the physical meanings of these concepts, Maxwell's equations can now be rearranged to display explicitly the time and coordinate dependence of the wave amplitude.

\section{The Nonlinear Wave Equation}

To derive a wave equation that describes pulse propagation in nonlinear cubic media such as optical fibers in one dimension, we choose the electric field $\mathbf{E}$ in the $z$ direction and the $\mathbf{H}$ field in the $y$ direction such that  
\begin{equation}\label{1d elec field}
\mathbf{E} = u(x,t){\mathbf{\hat{k}}}, \qquad \mathbf{H} = H(x,t){\mathbf{\hat{j}}}
\end{equation}
where $u(x,t)$ and $H(x,t)$ are the magnitudes of electric field and magnetic fields respectively. Remember the choice of perpendicular electric and magnetic fields to each other and to the direction of propagation is the requirement of Maxwell's equations. It should be noted that $\rho=0$ and $\textbf{J}=0$ in the absence of free charges, and $\mathbf{M}=0$ for nonmagnetic media. Because we consider a nonmagnetic nonlinear cubic medium with no free charge in our model, we set $\rho=0$ and $\textbf{J}=0$, and $\mathbf{M}=0$ in (\ref{MaxEqn3}) and (\ref{magflux}) respectively. Once we substitute (\ref{1d elec field}) in (\ref{MaxEqn3}), we reduce the set of three dimensional equations (\ref{MaxEqn3}) to a set in one dimension
\begin{equation}\label{u_x}
\begin{aligned}
\frac{\partial u}{\partial x} &= \mu_0\frac{\partial H}{\partial t} \\
\frac{\partial H}{\partial x} &= \frac{\partial (\epsilon_0 u + p)}{\partial t} 
\end{aligned}
\end{equation}
where $p=p(x,t)$ is the magnitude of polarization along the $z$-direction. To obtain an equation for the description of pulse dynamics in terms of the magnitude of the applied electric field, we take the derivative of the first equation in (\ref{u_x}) with respect to $x$ and the derivative of the second equation in (\ref{u_x}) with respect to $t$. Subsequently, elimination of the $H$ terms and a basic manipulation in the set lead to the wave equation
\begin{equation}\label{wave_eqn1d}
\frac{\partial^2u}{\partial x^2} = \frac{1}{c^2} \frac{\partial^2 u}{\partial t^2} + \mu_0 \frac{\partial^2p}{\partial t^2} 
\end{equation}  
where $c$ is the speed of light in vacuum and the relation $c = 1 / \sqrt{\mu_0\epsilon_0}$ is used. Equation (\ref{wave_eqn1d}) is the standard wave equation that describes the propagation of linearly polarized light in one dimensional medium with $u=u(x,t)$ being the magnitude of the applied electric field and $p$ being the polarization of the medium in response to the electric field. Equation (\ref{wave_eqn1d}) is not a good representation of wave motion because it contains both magnitudes of the electric field and polarization. However, we already know from the previous section that polarization can be expanded in terms of the electric field strength. Such a relation between polarization $\mathbf{P}$ and electric field $\mathbf{E}$ can be used to obtain an equation in terms only of the magnitude of the electric field. The total polarization induced by electric dipoles in a material as a response to an applied electric field satisfies the general form (\ref{Polarization_General}). The main contribution to polarization comes from the linear term. For media that have inversion symmetry, the quadratic term in the expansion of the polarization vanishes. Nonlinear cubic media (for example, silica fibers) do exhibit such symmetry and no quadratic susceptibility contributes to nonlinear effects. Hence, the third order susceptibility is the origin of the lowest-order nonlinear effects in optical fibers. Nonlinearity is very important for the application of optical data processing, and the physics of nonlinear effects can be extracted by studying the response of the applied optical field to the medium. Since the contribution of higher order nonlinearities to the total polarization is negligibly small, we truncate the series at the third order of susceptibility, and therefore consider only the nonlinear effects up to third order. Let the polarization be split into two terms such that
\begin{equation}\label{polarization}
P(x,t) = P_L(x,t) + P_{NL}(x,t)
\end{equation}
where the linear part and the nonlinear part of the polarization are given as
\begin{eqnarray}\label{linear_polarization}
P_L(x,t)    &=& \epsilon_0 \left( \int_{-\infty}^t \chi^{(1)}(t-\tau)u(x,\tau)\,d\tau \right) \\
P_{NL}(x,t) &=& \epsilon_0 \left( \chi^{(3)}u(x,t)^3 \right).
\end{eqnarray}
$\chi^{(3)}$ is the third order susceptibility and assumed to be a constant. For the linear part, since we assume that the medium response is local, retardation in the medium's response to the applied electric field must be taken into account. Note that the relation (\ref{linear_polarization}) is valid in the electric-dipole approximation, which can simply be described as the first order approximation.
One might expect only the second order nonlinearity to be comparable to the linear response, however, due to inversion symmetry this contribution vanishes in optical fibers. We must also note that it is experimentally known that the nonlinear effects are weak in silica fibers \cite{hasegawa-kodama:1995}. For the reason that nonlinear effects (third order susceptibility) are small in optical fibers, the nonlinear polarization $P_{NL}$ in (\ref{polarization}) will be treated as a small perturbation to the total polarization in the derivation of the wave equation. This leads to a major simplification, and it is a good treatment for our present disscussion. Therefore, we manipulate equation (\ref{wave_eqn1d}) with $P_{NL}=0$ first and the include nonlinear term later. Equation (\ref{wave_eqn1d}) becomes linear when $P_{NL}=0$. If a Fourier transform is applied to the linear equation
\begin{equation}\label{ln_wave_eqn}
\partial_x^2 \hat{u}(x,w) + \frac{1}{c^2}w^2\hat{u}(x,w) = \mu_0 (-w^2) \hat{P_L}(x,w) \, ,
\end{equation}
then a simple calculation of the Fourier transform of the linear polarization (\ref{linear_polarization}) gives 
\begin{eqnarray}
\hat{P}_L(x,w) &=& \int_{-\infty}^{+\infty}P_L(x,t)e^{-iwt}\,dt \nonumber \\
				     &=&  \epsilon_0 \int_{-\infty}^{+\infty} \int_{-\infty}^{+\infty} \chi^{(1)}(t-\tau) u(x,\tau) e^{-iwt}\,dt d\tau \nonumber \\
&=&\epsilon_0 \int_{-\infty}^{+\infty} u(x,\tau)\,d\tau \int_{-\infty}^{+\infty} \chi^{(1)}(t-\tau) e^{-iwt}\,dt \nonumber \\
&=&\epsilon_0 \int_{-\infty}^{+\infty} u(x,\tau)\,d\tau \int_{-\infty}^{+\infty} \chi^{(1)}(t) e^{-iw(t+\tau)}\,dt \nonumber \\
&=&\epsilon_0 \int_{-\infty}^{+\infty} u(x,\tau) e^{-iw\tau}\,d\tau \int_{-\infty}^{+\infty} \chi^{(1)}(t) e^{-iwt}\,dt \nonumber \\
&=&\epsilon_0\hat{\chi}^{(1)}(w)\hat{u}(x,w)
\end{eqnarray}
With this simple form of the linear polarization in hand, we can rewrite equation (\ref{ln_wave_eqn}) as 
\begin{equation}\label{ln_wave_eqn_2}
\partial_x^2 \hat{u}(x,w) + \frac{1}{c^2}w^2\hat{u}(x,w) = \frac{1}{c^2}(-w^2)\left[\hat{\chi}^{(1)}(w)\right]\hat{u}(x,w) 
\end{equation}
where  the relation $c = 1 / \sqrt{\mu_0\epsilon_0}$ is used to obtain the coefficient on the right hand side. Now the question is how to deal with the linear susceptibility $\hat{\chi}^{(1)}$ in this equation. In general, the linear and nonlinear susceptibilities depend on the frequencies of the applied fields. Since there is an assumption of an instantaneous response in the nonlinear part of the polarization, 
we therefore take the nonlinear susceptibility to be a small constant. However, the response of the linear part of the optical field is local, and the linear susceptibility is a function of the atomic transition frequency. The context in which we are describing propagation of a pulse in a nonlinear cubic medium is a semi-classical theory. This indicates that molecules and atoms of the medium in which the pulse propagates are explained by quantum mechanics, and light itself is governed by the laws of classical electrodynamics. Therefore, the quantum mechanical density matrix method applies to the derivation of the linear susceptibility in equation (\ref{ln_wave_eqn_2}). Because the derivation of the linear susceptibility is tedious and requires a good deal of quantum mechanics, we ask the reader to refer to the literature \cite{boyd:1992} for the detailed derivation. Instead, we mention briefly how to approximate the linear susceptibility. Notice that polarization $\mathbf{P}$ is parallel to the electric field $\mathbf{E}$ in the medium. It is not hard to see as a consequence of this that the linear susceptibility can be expressed as a scalar ($\hat P =\epsilon_0 \hat \chi \hat E$). Loosely speaking, if the material is modeled as a free atom interacting with an electro-magnetic field, the linearized susceptibility of the medium in Fourier space can be approximated as 
\begin{equation}\label{linearSuscep}
\hat \chi^{(1)}(w)\approx\sum_n f_{na} \frac{c_0}{(w^2_{na}-w^2)-2i\gamma_{na}\omega}  
\end{equation}
where $f_{na}$ is the oscillator strength, $c_0$ is some constant, $w_{na}$ is the resonant frequency of the medium and $\gamma_{na}$ is a small damping
coefficient added to keep the susceptibility finite at the resonant frequency.
Although equation (\ref{linearSuscep}) is a simplified version of the linear susceptibility, the exact form of $\chi^{(1)}$ is not important for our results. What is important is that we make an approximation of $\chi^{(1)}$ such that $\chi^{(1)}$ can be expressed as a polynomial in $\lambda$. We study the propagation of light in the infrared range with wavelengths of 1600-3000 $nm$, where $nm$ stands for nanometer and the conversion relation is $1 nm = 10^{-9} m$. In this regime for silica fibers, a further $\chi^{(1)}$ approximation can be done such that the linear susceptibility  can be approximated to a form of
\begin{equation}\label{linearSuscepApprox}
\hat{\chi}^{(1)}(w)\approx\hat{\chi}^{(1)}(\lambda)=\hat{\chi}_0^{(1)} - \hat{\chi}_2^{(1)} \lambda^2
\end{equation}
where $\hat{\chi}^{(1)}_0 = 1.1104\ \mu m^{-2}$  and $\hat{\chi}^{(1)}_2 = 0.011063\ \mu m^{-2}$.  Numerical studies based on the experimental data show that equation (\ref{linearSuscepApprox}) is a very good approximation of the equation (\ref{linearSuscep}) for the pulses with wavelengths ranging from 1600 $nm$ to 3000 $nm$. Using the relation between frequency and wavelength $\lambda =2\pi c / \omega $ and substituting equation (\ref{linearSuscepApprox}) in the equation (\ref{ln_wave_eqn_2}) leads to
\begin{equation}\label{wave_eqn_1d_final_fourier}
\partial_x^2 \hat{u} + \frac{1+\hat{\chi}_0^{(1)}}{c^2}w^2\hat{u} -  (2 \pi)^2  \hat{\chi}_2^{(1)} \hat{u}=0 \,.
\end{equation}
This is the wave equation in the Fourier domain with the linear component of polarization only. As a final step, we apply the inverse Fourier transform to the equation (\ref{wave_eqn_1d_final_fourier}) first and add the small perturbation (nonlinear part of the polarization) to the equation later. The wave equation in its final form ultimately becomes
\begin{equation}\label{wave_eqn_1d_final}
\partial_x^2 u = \frac{1}{c_1^2}\partial_t^2 u + \frac{1}{c_2^2}u + \frac{1}{c^2} \chi^{(3)}  \partial_t^2 u^3 
\end{equation}
where $c_1= c / \sqrt{1+\hat{\chi}_0^{(1)}}=2.065\times 10^8 m/s$, $c_2=1 / 2 \pi \sqrt{\hat{\chi}_2^{(1)}}=1.59\mu m$, $c = 1 / \sqrt{\mu_0\epsilon_0}$ is the speed of light and $\chi^{(3)}$  is the nonlinear susceptibility. Equation (\ref{wave_eqn_1d_final}) is the Maxwell equation in one dimension for pulse propagation in the infrared regime 1600-3000 $nm$.


\section{Rescaling the Nonlinear Wave Equation}
In chapter four, we derive an approximate equation replacing the nonlinear wave equation for pulse propagation in nonlinear cubic media in the infrared regime with the set of equations in (\ref{u_x}). The new governing equation arises from a scaled version of the Maxwell equation. Therefore, we show in this section how to scale the Maxwell equation (\ref{wave_eqn_1d_final}) in a physical way.

Let us make a coordinate transformation of the form
$(x,t) \rightarrow (\xi,\tau)$ with
\begin{equation}
x=\hat{x}\xi, \qquad t=\hat{t}\tau
\end{equation}
where $\hat{x}$ and $\hat{t}$ are properly chosen constants. The double time and space derivatives in the wave equation (\ref{wave_eqn_1d_final}) must be modified according to the new coordinates such that
\begin{eqnarray}\label{XiTau2ndDer}
\frac{\partial^2}{\partial x^2} = \frac{1}{\hat{x}^2}\frac{\partial^2}{\partial \xi^2}\\
\frac{\partial^2}{\partial t^2} = \frac{1}{\hat{t}^2}\frac{\partial^2}{\partial \tau^2}
\end{eqnarray}
Inserting the derivatives (\ref{XiTau2ndDer}) in the wave equation (\ref{wave_eqn_1d_final}) yields an equation in the new coordinates 
\begin{equation}\label{wave_eqn_NewScales}
u_{\xi\xi}=\frac{\hat{x}^2}{c_1^2\hat{t}^2}u_{\tau\tau}+\frac{\hat{x}^2}{c_2^2}u+\frac{\hat{x}^2\chi^{(3)}}{\hat{t}^2}(u^3)_{\tau\tau}
\end{equation}   
From a mathematical point of view, it does not really matter how to choose the constants $\hat{x}$ and $\hat{t}$. However, the physics of the problem imposes conditions on these constants, and they cannot be chosen arbitrarily from the physics point of view. Noting that the nonlinear wave equation (\ref{wave_eqn_1d_final}) describes the short pulses propagating in a nonlinear cubic medium, these constants must be determined based on short pulse parameters. In addition, there is also another condition we must take into account. As we will see in chapter four, the coefficient of the $u_{\tau \tau}$ term has to be set to one in order to derive a short pulse equation over a slow time scale that is introduced through a multiple scale expansion. Therefore, we first make the choice of
\begin{equation}\label{UnitCoeff_spe}
\frac{\hat{x}^2}{c_1^2\hat{t}^2}=1
\end{equation}
as a condition for the derivation of the short pulse equation from the Maxwell equation. To incorporate the physics of the short pulse equation into the rescaling, let us briefly discuss the physical description of a short pulse. A short pulse with wavelength $\lambda=1.55$ $\mu m$ has an angular frequency $w=1.94\times 10^{14}$ $rad/s$ ( $\lambda \nu = c$ ). The relationship between angular frequency and the period of a wave can be found using $w=2\pi \nu=2\pi/T$. The period of a short pulse with an angular frequency $1.94\times 10^{14}$ $rad/s$ equals to $T=2\pi/w=32.4$ femtoseconds ( $fs$ ). This means we are dealing with a femtosecond time scale in the short pulse dynamics. Hence, we choose the time coefficient $\hat{t}$ to be comparable to the femtosecond time unit such that $\hat{t}=1$ $fs$. One can, on the other hand, choose another $\hat{t}$ value in the femtosecond unit differing from the one we have made here. The choice of $\hat{t}=1$ $fs$ and the condition of (\ref{UnitCoeff_spe}) impose the value of $\hat{x}$. A basic calculation gives the value of $\hat{x}$ as $\hat{x}=2.065\times10^{-7}\,m \approx 206$ $nm$. Before computing all the coeffients of the rescaled wave equation (\ref{wave_eqn_NewScales}), one has to know the physical value of the nonlinear susceptibility. In silica fiber, non-linear susceptibility $\chi^{(3)}$ is given as $1.28\times 10^{-19}$ $m^2/W$ for a wave with a wavelength $\lambda = 1.55$ $\mu m$ \cite{liu:2005,czichos-saito-smith:2006}.
Using the values $\hat{t}=1$ $fs$, $\hat{x}=206$ $nm$ and $\chi^{(3)}=1.28\times 10^{-19}$ $m^2/W$, one can calculate the unitless coefficents
\begin{equation}
\begin{aligned}
a&=\frac{\hat{x}^2}{c_2^2}=0.01678573 \\
b&=\frac{\hat{x}^2\chi^{(3)}}{\hat{t}^2}=0.005431808\,.
\end{aligned}
\end{equation}
Finally, the nonlinear wave equation (\ref{wave_eqn_NewScales}) is re-written in its final form after renaming independent variables $\xi$ \& $\tau$ as $x$ \& $t$, respectively, as
\begin{equation}\label{maxwell_1d}
u_{xx}=u_{tt}+au+b(u^3)_{tt}
\end{equation}
where $a=0.01678573$ and $b=0.005431808$. Notice that the coefficient $b$ is a measure of nonlinearity in this new scale. This is the form of the Maxwell equation we will use in the derivation of the deterministic and the stochastic short pulse equation. Before proceeding into the next chapter, it should be noted that we shall always refer to this equation as the nonlinear wave equation or Maxwell equation in the rest of the work.

%% file: Chap3_NLSEinc.tex
We discuss the multiple-scale expansion method and the derivation of the nonlinear Schr\"odinger equation (NLSE) in this chapter. The sections contained herein will also highlight NLSE solitons and the limitations of the NLSE.  


\section{Multi-Scale Expansion}
 
The method of multiple scales is a widely used technique to solve a wide variety of problems in physics, engineering and applied sciences \cite{kevorkian-cole:1996,hinch:1991}. When regular perturbation approaches fail, multiple scale expansion is used to approximate periodic solutions to differential equations. In addition to the NLSE and the SPE, the wide spectrum \cite{nayfeh:1973} of the application of multiple scaling includes problems such as weakly linear and nonlinear vibrations governed by differential equations, the Earth-Moon-spaceship problem in the context of orbital mechanics, the role of different time scales in flight mechanics, the propagation of waves on a spherical shell in the context of solid mechanics, the investigation of the evolution of multi-phase modes for the Klein-Gordon equation, the interaction of random waves in dispersive media, propagation of nonlinear waves in a cold plasma and approximation for the Thomas-Fermi model in atomic physics. For the reason of such a diverse application of the multiple scale expansion technique, we shall discuss the method in further detail so as to provide gainful insight into the subject. Let us commence with the classical example to illustrate multiple scales. We follow the standard example of a damped harmonic oscillator. This can be found in any conventional book regarding pertubative techniques such as "Introduction to Perturbation Methods," by M.H. Holmes \cite{holmes:1995}.  Consider
\begin{equation}\label{eqn_dampedOscillator}
y^{''} + \epsilon y^{'} + y = 0, \quad \mathrm{for} \; t>0 \,,
\end{equation}
where the initial conditions are given as
\begin{equation}
y(0)=0 \quad \mathrm{and} \quad y^{'}(0)=1.
\end{equation}
This is the equation of a damped oscillator whose analytical solution is well known. The exact solution of (\ref{eqn_dampedOscillator}) is given as
\begin{equation}\label{Damped_sho_ExactSoln}
y(t)=\frac{1}{\sqrt{1-\epsilon^2/4}}\, e^{-\epsilon t/2}\, \sin (t \sqrt{1-\epsilon^2/4})
\end{equation}
Notice that $\epsilon$ is a small parameter.  We will get the solution of the above equation by both the regular expansion technique and multiple-scale expansion, and then compare them with the exact solution. Such a comparison illuminates power and beauty of the multiple scaling. Let us assume that the function $y(t)$ can be expanded as
\begin{equation}\label{y_expansion}
y(t)=y_0(t)+\epsilon y_1(t) + \epsilon^2y_2(t)+...
\end{equation} 
If equation (\ref{y_expansion}) is inserted into the equation (\ref{eqn_dampedOscillator}) and the terms of different orders in $\epsilon$ are rearranged, we will find
\begin{equation}
y^{''}_0(t)+y_0(t)+\epsilon[y^{''}_1(t)+y_0^{'}(t)+y_1(t)]+\epsilon^2[y_2^{''}(t)+y_1^{'}(t)+y_2(t)]+...=0
\end{equation} 
To satisfy the equality, the terms of each order of $\epsilon$ including the zeroth order must be set to zero. In the zeroth order, we will arrive at the following equation
\begin{equation}\label{eqn_sho}
y^{''}_0(t)+y_0(t)=0
\end{equation}
with the initial conditions $y_0(0)=0$ and $y_0^{'}(0)=0$. This is the equation of very well-known case, namely, classical undamped simple harmonic oscillator with a unit frequency. Equation (\ref{eqn_sho}) is analytically solvable and the exact solution is found for the given initial conditions to be
\begin{equation}
y_0(t)=\sin(t) \,.
\end{equation}  
In a similar fashion, setting the term of the order $\epsilon$ to zero gives 
\begin{equation}\label{eqn_damped_order1}
y^{''}_1(t)+y_1(t)=-y_0^{'}(t)
\end{equation}
with the initial conditions $y_1(0)=0$ and $y_1^{'}(0)=0$. This represents the equation of the damped simple harmonic oscillator, and its form is same as that of the equation (\ref{eqn_dampedOscillator}). The damped harmonic oscillator can be solved by looking for trial oscillatory solutions of the form $\exp(t)$ or $A\cos(t)+B\sin(t)$
because these functions reproduce themselves when differentiated. Using the trial form of a $A\cos(t)+B\sin(t)$ solution along with the initial conditions generates the exact solution of the equation (\ref{eqn_damped_order1}) as
\begin{equation}\label{eqn_damped_order1_Soln}
y_1(t)=-\frac{1}{2}\,t\sin(t) \,.
\end{equation}  
Noting that $\epsilon$ is a small constant, we truncate the expansion (\ref{y_expansion}) at the first order and write the approximate solution for the equation (\ref{eqn_dampedOscillator}) as
\begin{equation}\label{y_ApproxSoln_RegExpansion}
y(t) \approx y_0(t) + \epsilon \, y_1(t) = \sin(t) - \frac{1}{2}\epsilon\, t \sin(t).
\end{equation}
In comparison with the analytical solution (\ref{Damped_sho_ExactSoln}), we observe a significant difference between the two. The regular expansion approximate solution (\ref{y_ApproxSoln_RegExpansion}) grows in an unbounded manner in time because of the second term. The factor $\epsilon t$ in the second term on the right-hand side of equation (\ref{y_ApproxSoln_RegExpansion}) can take values such that it can make the whole term as large as the first term or even larger. For this reason, this term is called the secular term. On the other hand, the analytical solution decays exponentially with time. The condition required for having a valid approximate solution via the regular expansion method is $t \ll 1/\epsilon$.

Let us now solve the same problem using multiple scales. The exact solution (\ref{Damped_sho_ExactSoln}) has an oscillatory part that occurs in a time scale on the order of $\epsilon^0$, i.e., $O(1)$. However, the exponentially decaying part has a dependence of $\epsilon\, t$ so that it takes place on a time scale of the order of $O(1/\epsilon)$. This means the envelope of the exact solution, which is governed by the exponential function, changes much more slowly than the oscillatory part of the solution, which is governed by the sine term. One can then articulate the need for two different scales in the expansion instead of one. This is precisely what we are embarking upon now. Let us introduce two time scales such that
\begin{eqnarray}
t_1 &=& t \label{New_t_scales} \\
t_2 &=& \epsilon t
\end{eqnarray}
Notice that since we truncate the regular expansion (\ref{y_expansion}) at the second term, we keep the multiple scaling at two time scales only. In the general case, the exponential term depends on $\epsilon t$, $\epsilon^2 t$, $\epsilon^3 t$,... as it can easily be seen by expanding the exponent in the series form. Therefore, the general case requires multiple scales in the form of $t_n=\epsilon^nt$. Note also that the new variables $t_1$ and $t_2$ will be treated as independent variables. As a consequence of introducing new variables (\ref{New_t_scales}), derivatives in equation (\ref{eqn_dampedOscillator}) must be changed according to the chain rule so that the first derivative now becomes
\begin{eqnarray}\label{New_t_1stDer}
\frac{\partial }{\partial t} &=& \frac{\partial t_1}{\partial t} \frac{\partial }{\partial t_1} + \frac{\partial t_2}{\partial t} \frac{\partial }{\partial t_2} \nonumber \\
 &=& \frac{\partial }{\partial t_1} + \epsilon \, \frac{\partial }{\partial t_2} 
\end{eqnarray}
and the second derivative becomes
\begin{equation}\label{New_t_2ndDer}
\frac{\partial^2 }{\partial t^2} = \frac{\partial^2}{\partial t_1^2} + 2  \epsilon \, \frac{\partial}{\partial t_1 \partial t_2} + \epsilon^2 \, \frac{\partial^2}{\partial t_2^2} 
\end{equation}  
where $\partial t_1/\partial t=1$ and $\partial t_2/\partial t=\epsilon$ are used. Once we insert (\ref{New_t_1stDer}) and (\ref{New_t_2ndDer}) into the equation (\ref{eqn_dampedOscillator}), we obtain
\begin{equation}\label{eqn_damped_NewScale}
\left( 1+\frac{\partial^2}{\partial t_1^2}+\epsilon  \left[\frac{\partial}{\partial t_1} + 2\frac{\partial^2}{\partial t_1 \partial t_2}\right]+ \epsilon^2 \left[ \frac{\partial^2}{\partial t_2^2} + \frac{\partial }{\partial t_2}\right] \right) y(t)=0
\end{equation}
where the initial conditions are $y(t)=0$ and $(\partial/\partial t_1 + \epsilon \, \partial / \partial t_2) y(t)=1$ for $t_1=t_2=0$ ( see the derivative relation (\ref{New_t_1stDer}) ). Let's now replace the expansion (\ref{y_expansion}) with the form
\begin{equation}\label{New_y_expansion}
y(t)=y_0(t_1,t_2)+\epsilon y_1(t_1,t_2) + \epsilon^2y_2(t_1,t_2)+...
\end{equation}  
Inserting the equation (\ref{New_y_expansion}) into the equation (\ref{eqn_damped_NewScale}) and rearranging the terms lead to an equation at the zeroth order of $\epsilon$
\begin{equation}\label{eqn_sho_NewScales}
\frac{\partial^2 y_0(t_1,t_2)}{\partial t_1^2}+y_0(t_1,t_2)=0
\end{equation}
with initial conditions $y_0=0$ and $ \partial y_0 /\partial t_1 =1$ at $t_1=t_2=0$. Notice that we have a partial differential equation now, whereas we previously had an ordinary differential equation in the regular expansion method at the zeroth order. Partial differential equation formalism will actually prevent the secular term appearing in the solution as it will be seen shortly. The equation (\ref{eqn_sho_NewScales}) is the equation of a simple harmonic oscillator as before. The general solution for this equation can be expressed as
\begin{equation}\label{eqn_sho_NewScales_Soln}
y_0(t_1,t_2)=a_0(t_2)\sin(t_1)+b_0(t_2)\cos(t_1)
\end{equation}
where $a_0(0)=1$ and $b_0(0)=0$. The coefficients $a_0(t_2)$ and $b_0(t_2)$ are functions of $t_2$ and will be determined after we handle the next order. At the next order, i.e., $O(\epsilon)$, we have the equation
\begin{equation}\label{eqn_damped_order1_NewScale}
 \left[\frac{\partial^2}{\partial t_1^2} + 1\right] y_1(t_1,t_2)= -\left[ 2\frac{\partial^2}{\partial t_1 \partial t_2} + \frac{\partial}{\partial t_1} \right] y_0(t_1,t_2) 
\end{equation}
where the initial conditions are $y_1=0$ and $\partial y_1 /\partial t_1 = - \partial y_0 / \partial t_2$ at $t_1=t_2=0$. If we substitute equation (\ref{eqn_sho_NewScales_Soln}) in the equation (\ref{eqn_damped_order1_NewScale}), we find
\begin{equation}\label{eqn_damped_order1_NewScales}
\frac{\partial^2 y_1}{\partial t_1^2} + y_1 = ( 2b_0^{'}(t_2)+b_0)\sin(t_1) - (2a_0^{'}(t_2)+a_0(t_2))\cos(t_1)
\end{equation}
The solution of this equation is given as
\begin{eqnarray}\label{eqn_damped_order1_NewScales_Soln}
y_1 &=& a_1(t_2)\sin(t_1)+b_1(t_2)\cos(t_1) \nonumber \\
 &-& \frac{1}{2}(2b_0^{'}(t_2)+b_0)t_1\cos(t_1) - \frac{1}{2} (2a_0^{'}(t_2)+a_0(t_2))t_1\sin(t_1)
\end{eqnarray}
where $a_1(0)=0$ and $b_1(0)=0$. Remember that we had a secular term in the first order solution of the regular expansion. Therefore, there is no reason for there not to be one or more secular terms in the solution (\ref{eqn_damped_order1_NewScales_Soln}) as well. One can observe the direct dependence of the solution (\ref{eqn_damped_order1_NewScales_Soln}) on $t_1$ and argue that the terms with $t_1\cos(t_1)$ and $t_1\sin(t_1)$ in (\ref{eqn_damped_order1_NewScales_Soln}) are secular terms in an analogous way to the treatment of the term $t\sin(t)$ as a secular term in the solution (\ref{y_ApproxSoln_RegExpansion}). Removing these terms will prevent the growth of the first order solution (\ref{eqn_damped_order1_NewScales_Soln}). The secular terms can be disregarded by choosing
\begin{equation}\label{secular_condition}
\begin{aligned}
2b_0^{'}(t_2)+b_0(t_2)=0 \\
2a_0^{'}(t_2)+a_0(t_2)=0
\end{aligned}
\end{equation}
It is not difficult to find the solution of these two equations. Using the initial conditions $a_0(0)=1$ and $b_0(0)=0$, we obtain the solutions as
\begin{equation}\label{secular_coeff}
\begin{aligned}
a_0(t_2)&=e^{-t_2/2} \\
b_0(t_2)&=0
\end{aligned}
\end{equation}
This is the solvability condition of the first order. This means one has to choose the coefficients as in (\ref{secular_coeff}) in order to remove the secular terms appearing in the solution (\ref{eqn_damped_order1_NewScales_Soln}). Notice that the secular terms can be removed without solving the equation (\ref{eqn_damped_order1_NewScales}) by just setting the coefficients of the trigonometric functions on the right-hand side to zero. If we truncate the series (\ref{New_y_expansion}) at the first order and tie everthing together, the approximate solution of the problem (\ref{eqn_dampedOscillator}) obtained by the multiple scale expansion in the order of $\epsilon$  will be
\begin{eqnarray}\label{y_ApproxSoln_MultiScales}
y(t) &\approx& y_0(t_1,t_2)+\epsilon y_1(t_1,t_2) \nonumber \\
    &=& e^{-\epsilon t_2/2}\sin(t_1)+\epsilon y_1(t_1,t_2) \nonumber \\
& \approx & e^{-\epsilon t_2/2}\sin(t_1).
\end{eqnarray}
When this solution and the exact solution are compared, one can see that the solution (\ref{y_ApproxSoln_MultiScales}) is a much better solution than the one expressed in (\ref{y_ApproxSoln_RegExpansion}). Note that for solutions up to the order of $\epsilon^2$, one has to introduce one more time scale, namely, $t_3=\epsilon^2t$. The introduction of one more time scale would give an approximate solution that is valid over a larger time scale (that is, up to $O(1/\epsilon^2)$), but not necessarily more precise over time interval up to $O(1/\epsilon)$ than the approximate solution with two time scales. We must also note that two time scales $t_1$ and $t_2$ in the solution (\ref{y_ApproxSoln_MultiScales}) are called  fast scale and  slow scale respectively. 

In general, the time scales depend on the nature of the problem. More complex time scaling may be required for some nonlinear problems. Although the exact form of different scales may not be clear immediately, multiple scale expansion is a powerful method to remove the secular terms in the expansion of the approximate solutions of many nonlinear problems. 


\section{Derivation of the NLSE}

The nonlinear Schr{\"o}dinger equation is a model that describes physical nonlinear systems and may be applied to many nonlinear phenomena such as nonlinear optics, nonlinear acoustics, quantum condensates, hydrodynamics and heat pulses in solids \cite{sulem-sulem:1999}. There are many different forms of the nonlinear Schr\"odinger equation available in the literature. The nonlinear Schr\"odinger equation has the general form 
\begin{equation}
i\frac{\partial u}{\partial z}+ \nabla^2u=f(|u|^2)u\,,
\end{equation}
where $z$ is the propagation direction, $i=\sqrt{-1}$, $f$ is the general function that represents the problem, $\nabla^2$ is the Laplacian operator that can be in one, two or three transverse dimensions. Here, we study the NLSE in the context of propagation of pulses in a cubic nonlinear medium in one transverse direction. It is a very well-known fact that the propagation of optical pulses in optical fibers is governed by the NLSE \cite{hasegawa-kodama:1995}. In the case of a cubic nonlinear medium such as optical fibers, the function $f$ is given as $f(|u|^2)=\gamma|u|^2$, where $\gamma$ is a constant coefficient. 

One can derive the NLSE by using multiple scales \cite{newell-moloney:1992,mollenauer-gordon:2006}. If the multiple scale expansion of the form 
\begin{equation}
\begin{aligned}
u(z,t)&=\epsilon g(z,t) + \epsilon g^{*}(z,t)  +... \\
g(z,t)&=A_0(z_1,z_2,...,t_1)e^{i(\tilde{\beta}z_0-wt_0)} \\
z_n&=\epsilon^n z, \quad t_0=t, \quad t_1=\epsilon t 
\end{aligned}
\end{equation}
is inserted in the Maxwell equation (\ref{maxwell_1d}), we obtain the nonlinear Sch\"odinger equation up to a scale transformation in $1+1$ dimension as
\begin{equation}\label{NLSE}
-i \frac{\partial u}{ \partial z } = \frac{1}{2}\frac{\partial^2u}{\partial t^2} \pm {\mid u \mid}^2 u,  
\end{equation}
where $u(z,t)$ is the magnitude of the applied field (electric field) and is a complex quantity, $z$ is the propagation direction, and $t$ is time. The choice of sign depends on the cubic (Kerr) coefficient of the nonlinear material. The one dimensional NLSE describes the wave propagation in fluids and plasmas as well and emerges as a mean field equation for many-body bosonic systems in quantum field theory \cite{sulem-sulem:1999}.

The first term in the right hand side of the equation (\ref{NLSE}) describes the effects of dispersion. When acting by itself, it does nothing to change the frequency spectrum of the pulse and serves only to broaden or narrow the pulse in time. However, when the nonlinear term (the pulse intensity envelope times $u$ itself) acts by itself, it does nothing to change the pulse shape in time and serves only to broaden or narrow the pulse in the frequency domain.


\section{The NLSE Solitons}

The nonlinear Schr{\"o}dinger equation is a simple partial differential equation with complete integrability \cite{zakharov-shabat:1971}. Therefore, NLSE can be solved exactly using the inverse scattering method \cite{kivshar-agrawal:2003}. Exact solutions of the NLSE are solitary wave type. This means the effects of dispersion and nonlinearity cancel one another and this balance prevents the pulse from broadening or blowing up as it propagates \cite{kivshar-agrawal:2003}. Solitary wave solutions are also called solitons. The choice of the sign in the equation (\ref{NLSE}) gives rise to the different type of solitary waves. The soliton solution corresponding to the choice of $+$ sign is called a bright spatial soliton, whereas the soliton solution corresponding to the choice of $-$ sign is called a dark soliton. These types of solutions are sometimes called spatial solitons.

Let's consider (\ref{NLSE}) with the plus sign for the nonlinear term;
\begin{equation}\label{NLSE+}
i \frac{\partial u}{ \partial z } + \frac{1}{2}\frac{\partial^2u}{\partial t^2} + { |u|}^2 u =0.  
\end{equation}
This equation is exactly integrable \cite{zakharov-shabat:1971} and the analytical solution in the general form \cite{mollenauer-gordon:2006} is given as 
\begin{equation}\label{NLSE_BrightSoliton}
u(z,t)=a\,\mathrm{sech}[a(t-vz)]\exp[ivt+i(a^2-v^2)z/2]
\end{equation}
where $a$ is the soliton amplitude and $v$ is the velocity of the soliton. This is the bright soliton solution characterized by the two parameters $a$ and $v$. The general solution reduces to a special solution known as the fundamental solution in the limit $v=0$. The fundamental solution in this limit would be
\begin{equation}\label{NLSE_BrightSoliton_Special}
u(z,t)=a\,\mathrm{sech}(at)\exp(ia^2z/2).
\end{equation}
This is a particular pulse of an amplitude $a$ whose mean frequency is the central frequency. The hyperbolic secant function determines the shape of the envelope of the solitary wave, and the exponential function is the phase term. The phase term in the solution (\ref{NLSE_BrightSoliton_Special}) has no dependence on $t$, and therefore, the soliton is completely nondispersive. This means that its shape does not change with $z$ either in the temporal domain or in the frequency domain \cite{mollenauer-gordon:2006}. The dispersive term, which affects the pulse only in the time domain, and the nonlinear term, which affects the pulse only in the frequency domain, cancel each other in a way that stable propagation of solitons occur just leaving only a phase shift of the whole pulse behind.
The fundamental soliton represents the fundamental mode of the optical waveguide created by the propagating pulse. If the input pulse has the correct shape in a way that it satisfies the relation (\ref{NLSE_BrightSoliton_Special}), all of its energy will be contained in this mode, and the pulse will propagate without change in its shape. 

If the sign of the nonlinearity is taken to be minus in (\ref{NLSE}), the nonlinear Schr\"odinger equation becomes
\begin{equation}\label{NLSE-}
i \frac{\partial u}{ \partial z } + \frac{1}{2}\frac{\partial^2u}{\partial t^2} - {\mid u \mid}^2 u =0.
\end{equation}
This equation is also integrable \cite{mollenauer-gordon:2006} and gives rise to the solution in the general form
\begin{equation}\label{NLSE_DarkSoliton}
u(z,t)=u_0[B\mathrm{tanh}(u_0B(t-Au_0z))+iA]\mathrm{exp}(-iu_0^2z) \, ,
\end{equation}
where the parameters $A$ and $B$ satisfy $A^2+B^2=1$ and $u_0$ is the background amplitude. This solution is called a dark soliton. To write the solution (\ref{NLSE_DarkSoliton}) in terms of one parameter instead of two, the relations $A=\sin\phi$ and $B=\cos\phi$ can be used. Here the angle $\phi$ is the half the angle of the total phase shift of $2\phi$. When the dark soliton is characterized by $\phi$ only, we can write the solution as
\begin{equation}
|u|^2=u_0^2[1-\cos^2\phi \, \mathrm{sech}^2 (u_0\cos \phi (t-u_0 \sin\phi \, z))].
\end{equation}
The term $u_0\sin\phi$ represents the velocity of the soliton in the $z$ direction. The special dark soliton solution can be obtained by setting $\phi=0$ such that
\begin{equation}\label{NLSE_DarkSoliton_Special}
u(z,t)=u_0 \, \mathrm{tanh}(u_0t)\exp(-iu_0^2z).
\end{equation}
This special dark soliton does not move against the background. Therefore, it is a stationary dark soliton and given the name, black soliton. The reason why it is called a black soliton is because the soliton (\ref{NLSE_DarkSoliton_Special}) has a $\pi$ phase shift at $x=0$ and the intensity drops to zero. In the general case (\ref{NLSE_DarkSoliton}), the intensity does not drop to zero at the center and the soliton is refered to as a gray soliton in such situtations. The phase of a dark soliton changes across its width unlike the phase of a special bright soliton which remains constant during propagation. 

Finally, we should note that in many situations the physical constants such as the speed of light and pulse width are incorporated in the definition of special soliton units, which measure distance, time and power only \cite{mollenauer-gordon:2006}. The special soliton units make nonlinear equations look simple. As a consequence of this simple manifestation, adaptation of numerical schemes and understanding solitonic features become easier. There are also other types of NLSE solitons such as dispersion managed solitons and temporal solitons. The reader can find many works related to the other types of NLSE solitons and their interaction \cite{boyd:1992,mollenauer-gordon:2006}.


\section{Limitations of the NLSE} 

In this section, we discuss conditions under which the nonlinear Schr\"odinger equation fails to be an approximation to Maxwell's equations. The derivation of the NLSE is based on the multiple scaling technique as shown before. In the derivation of the NLSE, we make the assumption that the nonlinearity of the polarization has a small contribution to the total polarization and is treated as a perturbation in the expansion of the total polarization. There are cases in which the coupling of the frequencies generated by the medium may increase the effects of the nonlinearity on the propagation of the pulse, and phase mismatch may occur as a result \cite{kivshar-agrawal:2003}. In such cases, the nonlinear Schr\"odinger equation fails. The details of the breakdown of this assumption is beyond the scope of this work. 

The NLSE is a scalar equation. To derive the scalar NLSE, we also make the assumption that the electric field maintains its polarization along the fiber length. This is a fairly good approximation and is valid in many practical situations \cite{newell-moloney:1992}. 

The third assumption that is made in the derivation of the NLSE is the slowly varying envelope approximation (SVEA) in the propagation direction. This approximation is sometimes called paraxial approximation. The paraxial approximation is valid if the width of the pulse is much longer than the light (pulse) wavelength \cite{newell-moloney:1992}. The slowly varying approximation separates the rapidly varying part of the electric field from the slowly varying envelope. Such a seperation is not possible in deriving the NLSE if the envelope is assumed not to vary with propagation direction $z$ on a scale much longer than the wavelength. As the pulse width decreases, this approximation begins to breakdown, and the NLSE is not the governing equation of the ultra-short pulses anymore \cite{rothenberg:1992,schaefer-wayne:2004}. In the present discussion, the third assumption is the cause of the breakdown of the nonlinear Schr\"odinger equation in describing the of ultra-short pulses. In the next chapter, we will derive a governing equation to replace the NLSE for ultra-short pulse propagation in a nonlinear cubic media.

%% file: Chap4_SPEinc.tex
This chapter is one of the core chapters of this thesis. We will first show the derivation of the short pulse equation (SPE). The analytical solution of the short pulse equation will be discussed next. We will complete this chapter with the discussion of numerical analyses and higher order terms of the short pulse equation.


\section{Derivation of the SPE}

In the derivation of the short pulse equation, we begin with a multiple scale expansion \textit{ansatz} of the form \cite{schaefer-wayne:2004}
\begin{equation}\label{u_expansion}
u(x,t)=\epsilon A_0(\phi,x_1,x_2,...)+\epsilon^2 A_1(\phi,x_1,x_2,...)+...
\end{equation}
with
\begin{equation}\label{xt_expansion}
\phi=\frac{t-x}{\epsilon}, \qquad x_n=\epsilon^n x.
\end{equation}
The idea is to observe the effects of dispersion and nonlinearity on the pulse over a  different time scale and see if there exists an equation that is easier in the mathematical sense and that can model the pulse propagation in the new scales.
In this section we only consider the expansion up to the order of $\epsilon$, which is a small constant expansion parameter. This means we only introduce two space scales $x_0=x$ and $x_1=\epsilon x$. Note that we treat these new variables independently. The equation (\ref{maxwell_1d}) includes double time and space derivatives of the $u$ function with respect to the original scales $x$ and $t$. These derivatives must be changed with respect to the new variables in the application of the multiple scale expansion to the Maxwell equation (\ref{maxwell_1d}). The transformation of the derivatives follow the chain rule once again, and the space and time derivatives are written respectively as
\begin{equation}\label{spe_NewScale_1stDer}
\begin{aligned}
\frac{\partial }{\partial x} &= \frac{\partial \phi}{\partial x} \frac{\partial }{\partial \phi} + \frac{\partial x_1}{\partial x} \frac{\partial }{\partial x_1}= -\frac{1}{\epsilon} \frac{\partial }{\partial \phi} + \epsilon \frac{\partial }{\partial x_1} \\
\frac{\partial }{\partial t} &= \frac{\partial \phi}{\partial t} \frac{\partial }{\partial \phi} + \frac{\partial x_1}{\partial t} \frac{\partial }{\partial x_1} = \frac{1}{\epsilon} \frac{\partial }{\partial \phi}
\end{aligned}
\end{equation}
where $\partial \phi / \partial x = - 1 / \epsilon$, $ \partial x_1 / \partial x =\epsilon $, $ \partial \phi / \partial t=1 / \epsilon$ and  $ \partial x_1 / \partial t =0$, and the second derivatives becomes
\begin{equation}\label{spe_NewScale_2ndDer}
\begin{aligned}
\frac{\partial^2 }{\partial x^2} &= \frac{1}{\epsilon^2} \frac{\partial^2}{\partial \phi^2} - 2 \frac{\partial}{\partial \phi} \frac{\partial}{\partial x_1} + \epsilon^2 \frac{\partial^2}{\partial x_1^2} \\
\frac{\partial^2 }{\partial t^2} &= \frac{1}{\epsilon^2} \frac{\partial^2}{\partial \phi^2} 
\end{aligned} 
\end{equation}
If the relations (\ref{u_expansion}) and (\ref{spe_NewScale_2ndDer}) are substituted into (\ref{maxwell_1d}), we obtain
\begin{equation}
\begin{aligned}  	
\frac{1}{\epsilon^2} \frac{\partial^2}{\partial \phi^2} \left( \epsilon A_0+\epsilon^2 A_1+...\right) - 2 \frac{\partial}{\partial \phi} \frac{\partial}{\partial x_1} \left( \epsilon A_0+\epsilon^2 A_1+...\right) \\
+ \epsilon^2 \frac{\partial^2}{\partial x_1^2}\left( \epsilon A_0+\epsilon^2 A_1+...\right) = \frac{1}{\epsilon^2} \frac{\partial^2}{\partial \phi^2} \left( \epsilon A_0+\epsilon^2 A_1+...\right) \\
+ a \left( \epsilon A_0+\epsilon^2 A_1+...\right) + \frac{b}{\epsilon^2}\frac{\partial^2}{\partial \phi^2} \left( \epsilon A_0+\epsilon^2 A_1+ \dots \right)^3 \,.
\end{aligned}
\end{equation}
Once we rearrange all the terms, we see that the terms up to $O(1/\epsilon)$ and $O(1)$ cancel out. As mentioned before, we only keep the terms up to $O(\epsilon)$. Collecting the terms of $O(\epsilon)$ on each side results in
\begin{equation}
\epsilon \left( \frac{\partial^2}{\partial \phi^2} A_2 - 2 \frac{\partial}{\partial \phi} \frac{\partial}{\partial x_1} A_0 \right) = \epsilon \left(\frac{\partial^2}{\partial \phi^2} A_2 + a A_0 + b \frac{\partial^2}{\partial \phi^2} A_0^3 \right) \,.
\end{equation}
Finally, if we cancel out like terms, we obtain the following equation
\begin{equation}\label{SPE_original}
-2 \partial_{\phi} \partial_{x_1} A_0 = a A_0 + b \partial^2_{\phi} A_0^3 \,,
\end{equation}
where the coefficients $a$ and $b$ remain the same with $a \approx 0.0168$ and $b \approx 0.00543$. This equation is called the short pulse equation (SPE) and derived by T. Sch\"afer and C. Wayne \cite{schaefer-wayne:2004}. The SPE rests at the heart of the ultra-short pulse dynamics. It is the governing equation for ultra-short pulses in silica fibers in the infrared regime.


\section{Transformation of the SPE}

We will now show how the original form of SPE (\ref{SPE_original}) can be transformed to the form of
\begin{equation}\label{SPE_sak}
U_{XT}=U+\frac{1}{6}(U^3)_{XX}
\end{equation} 
We start with a general discussion of transformations of differential equations \cite{sakovich-sakovich:2007,rabelo-tenenblat:1990} and bring the transformation of SPE into this discussion along the way. 
The exact solution of the SPE has been derived simply for this form \cite{sakovich-sakovich:2006}. The exact solitary wave solution cannot be accepted as the exact solution of the original SPE (\ref{SPE_original}) without making a proper transformation between the two forms. The transformation of SPE is of great importance especially when we check numerically whether the exact solution satisfies Maxwell equation from which we derive the SPE through multi-scale expansion. 

The system of equations
\begin{equation}\label{pde_rabelo}
u_{xt}=\left( (\alpha g(u)+\beta)u_x \right) _x \pm g^\prime(u)
\end{equation}
\begin{equation}\label{ode_rabelo}
g^{\prime\prime} (u) +\mu g(u) = \theta
\end{equation}
are given by M.L. Rabelo \cite{sakovich-sakovich:2007} to describe pseudo-spherical surfaces. Here $\alpha$, $\beta$, $\mu$ and $\theta$ are arbitrary constants, and $g(u)$ is a solution of the linear ordinary differential equation of (\ref{ode_rabelo}) and $g^\prime(u)$ stands for the derivative of $g$ with respect to $u$. The partial differential equation of the form (\ref{pde_rabelo}) can be transformed to the form (\ref{SPE_sak}) for $U(X,T)$ through the following transformations 
\begin{equation}\label{trans_rabelo}
u=c_1U(X,T)+c_2, \qquad X=c_3x+c_4t, \qquad T=c_5t
\end{equation} 
where $c_1$,$c_2$,...,$c_5$ are properly chosen constants with $c_1c_3c_5\neq 0$ and $\alpha \neq 0$. 

First, we will determine whether we can write the original form of the SPE (\ref{SPE_original}) in the form given by (\ref{pde_rabelo}). If $\mu=0$ in equation (\ref{ode_rabelo}), then we have
\begin{equation}
g(u)=\frac{1}{2}\theta u^2 +\gamma u + \delta
\end{equation} 
with arbitrary constants $\gamma$ and $\delta$. Once we substitute $g(u)$ and its first and second derivatives with respect to $u$ in equation (\ref{pde_rabelo}) for which we also choose the positive sign in the last term, we obtain
\begin{equation}
u_{xt}=\frac{\partial}{\partial x}\left( u_x(\frac{1}{2}\alpha \theta u^2 + \alpha \gamma u + \alpha \delta +\beta) \right) + \theta u +\gamma.
\end{equation}
Let's further differentiate each term in the parenthesis. After collecting like terms, we find
\begin{equation}\label{pde_2}
u_{xt}=u_{xx}\left(\frac{1}{2}\alpha \theta u^2 + \alpha \gamma u + \alpha \delta +\beta \right) + (u_x)^2\left( \alpha \theta u + \alpha \gamma \right) + \theta u +\gamma
\end{equation}
Choosing arbitrary constants $\gamma=0$ and $\alpha\delta+\beta = 0$ reduces equation (\ref{pde_2}) to
\begin{equation}\label{pde_3}
u_{xt}= \theta u + \alpha \theta u (u_x)^2 +\frac{1}{2}\alpha \theta u^2 u_{xx}.
\end{equation}
The short pulse equation and equation (\ref{pde_3}) are actually equivalent. A simple manipulation of the SPE is required to show this equivalency. In carrying this out, the last term of (\ref{SPE_original}) can be expanded in the following manner
\begin{equation}\label{last_term}
\partial^2_\phi(A_0^3)=\partial_\phi[3(A_0)^2(A_0)_\phi]=6(A_0){(A_0)_\phi}^2+3(A_0)^2(A_0)_{\phi\phi}\,,
\end{equation}
and the short pulse equation (\ref{SPE_original}) can be re-written as
\begin{equation}\label{SPE_original_2}
(A_0)_{\phi x_1}= -\frac{a}{2}A_0+(-3b)A_0(A_\phi)^2 + \frac{(-3b)}{2}A_0^2(A_0)_{\phi\phi} \,.
\end{equation}
By comparing equations (\ref{SPE_original_2}) and (\ref{pde_3}), one can easily see that these equations are equivalent if we have
\begin{equation}\label{theta_alpha}
u=A_0, \quad x=\phi, \quad t=x_1, \quad \theta=-\frac{a}{2}, \quad \alpha=-\frac{3b}{\theta}=\frac{6b}{a}.
\end{equation}
This concludes the transformation of equation (\ref{SPE_original}) to the form in (\ref{pde_rabelo}). We will now cast equation (\ref{SPE_original}) into the form given by (\ref{SPE_sak}). To accomplish this, we choose the coefficents in (\ref{trans_rabelo}) to be 
\begin{equation}
c_1=\frac{1}{\sqrt{\alpha}}, \quad c_2=0, \quad c_3=1, \quad c_4=0, \quad c_5=\theta 
\end{equation}
and rewrite equation (\ref{trans_rabelo}) as
\begin{equation}\label{trans_rabelo_2}
u=\frac{1}{\sqrt{\alpha}}U(X,T), \qquad X=x, \qquad T=\theta t \,.
\end{equation} 
Combining equations (\ref{theta_alpha}) and (\ref{trans_rabelo_2}) leads to the transformation of the kind
\begin{equation}\label{spe_trans}
A_0=\frac{1}{\sqrt{\alpha}}U(X,T)=\sqrt{\frac{a}{6b}}U(X,T), \quad \phi=X, \quad x_1=\frac{1}{\theta}T=-\frac{2}{a}T \,.
\end{equation}
This is the transformation rule needed to shift from one form of the short pulse equation to the other form. It is rather straightforward to check whether or not we have obtained the right transformation rule. Once we substitute equation (\ref{spe_trans}) into equation (\ref{SPE_original_2}) and carry out the derivatives, we obtain the simplified equation 
\begin{equation}
\begin{aligned}\label{pde_4}
U_{XT}&=U + U(U_X)^2+\frac{1}{2}U^2U_{XX} \\
      &=U + \frac{1}{6}\left(6U(U_X)^2+3U^2U_{XX}\right) \,.
\end{aligned}
\end{equation}
The second term on the right-hand side is the same expansion we observed in equation (\ref{last_term}). As a final step, we write equation (\ref{pde_4}) in its simplest form
\begin{equation}
U_{XT}=U+\frac{1}{6}U^3_{XX} \,.
\end{equation}
To summarize, the original form of the SPE may be transformed to a new form through a transformation of the kind such that
\begin{equation*}
-2\partial_\phi \partial_{x_1} A_0 = a A_0 + b \partial^2_\phi(A_0^3) 
\end{equation*}
\begin{equation}\label{spe_Transformation} 
U_{XT}=U+\frac{1}{6}U^3_{XX}
\end{equation}
\begin{equation*}
A_0(\phi,x_1)=\sqrt{\frac{a}{6b}}U(X,T), \qquad \phi=X, \qquad x_1=-\frac{2}{a}T.
\end{equation*}


\section{Analytical Solution of the SPE}

The short pulse equation (\ref{SPE_sak}) is an integrable equation \cite{sakovich-sakovich:2005}. Therefore, its analytical solution can be derived by several different methods. We do not show in this section how the derivation of the exact solution can be performed. However, we give a brief discussion and interpretation of the analytical result. The proof of integrability shows that the SPE possesses a Lax pair of the Wadati-Konno-Ichikawa type \cite{sakovich-sakovich:2005}. Such a property allows for the method of inverse scattering to be used in solving nonlinear differential equations including the SPE. However, the exact solution of the SPE has not yet been found using the inverse scattering theorem.  The solution has been found using a transformation between the short pulse equation (\ref{SPE_sak}) and the sine-Gordon equation \cite{sakovich-sakovich:2006}. The sine-Gordon equation is a very well known and very well studied equation whose analytical solution is a soliton \cite{dodd-eilbeck-gibbon-morris:1982,campbell:1987}. The sine-Gordon equation  
\begin{equation}\label{sineGordonEqn}
z_{yt}(y,t)=\sin(z)
\end{equation}
is equivalent to the SPE through a chain of transformations \cite{sakovich-sakovich:2006} of the form
\begin{equation}\label{sineGordon_spe_transformation}
\begin{aligned}
v(x,t)&=(u_x^2+1)^{-1/2} \\
x=w(y,t), & \quad v(x,t)=w_y(y,t) \\
z(y,t)&=\mathrm{arccos}(w_y)\,,
\end{aligned}
\end{equation}
where the subscripts denote the derivatives with respect to $x$ and $y$. The transformation (\ref{sineGordon_spe_transformation}) is used to study the short pulse equation. The exact single-valued, non-singular solitary wave solution of the SPE in (\ref{SPE_sak}) has been derived from a bound state kink and a anti-kink solution of the sine-Gordon equation (SGE) by means of a transformation of the kind in (\ref{sineGordon_spe_transformation}), and is given as \cite{sakovich-sakovich:2006}
\begin{equation}\label{spe_soln}  
\begin{aligned}
U&=4mn\frac{m\sin\psi\sinh\phi+n\cos\psi\cosh\phi}{m^2\sin^2\psi+n^2\cosh^2\phi} \\ 
X&=Y+2mn\frac{m\sin2\psi - n\sinh2\phi}{m^2\sin^2\psi+n^2\cosh^2\phi}
\end{aligned}
\end{equation} 
with
\begin{equation}\label{spe_parameters}
\phi=m\left(Y+T\right),\quad \psi=n\left(Y-T\right), \quad n=\sqrt{\left(1-m^2\right)}, \quad 0<m<1 \,. 
\end{equation}
The exact solution (\ref{spe_soln}) represents a nonsingular pulse if and only if $m<m_{cr}$, where
\begin{equation}\label{mCritical}
m_{cr}=\sin\frac{\pi}{8} \approx 0.383\,.
\end{equation}
$m$ is a parameter that determines how short the pulse is. When $m$ reaches its critical value, the pulse becomes as short as approximately three cycles of its central frequency. At the critical value of $m$, the SPE solution (\ref{spe_soln}) becomes singular even though it remains single-valued.

If $m$ is small, for instance $m=0.05$, then the solitary wave solution of the SPE can be approximated to a simpler form, which can be easily done. As $m\rightarrow0$, $n=\sqrt{1-m^2}\approx 1$, and when the Taylor series expansion is applied to the hyperbolic function sinh$\phi$ in (\ref{spe_soln}), all sine and sinh terms go to zero. As a result, the solution (\ref{spe_soln}) can be simplified to the form   
\begin{equation}\label{spe_approx_soln} 
\begin{aligned}
U& \approx 4m\frac{\cos(Y-T)}{\cosh(m(Y+T))}=4m\cos(Y-T)\mathrm{sech}(m(Y+T)) \\ 
X& \approx Y \,.
\end{aligned}
\end{equation}
Notice that the approximate SPE solution (\ref{spe_approx_soln}) is similar to the bright soliton of the NLSE (\ref{NLSE_BrightSoliton}) or (\ref{NLSE_BrightSoliton_Special}). In either case, whether it is the true solution (\ref{spe_soln}) or an approximate solution (\ref{spe_approx_soln}), the hyperbolic function nonetheless determines the envelope of the wave packet while the trigonometric function is responsible for oscillations. It should be noted that we will use the single solitary wave solution (\ref{spe_soln}) in the numerical analysis of short pulse dynamics. 

Multi-soliton solutions of the SPE have also been constructed \cite{matsuno:2007} through a systematic procedure. This procedure starts with a hodographic transformation rule between the SPE and the sine-Gordon equation. The hodograph transformation is a method used to transform nonlinear partial differential equations into linear partial differential equations. Using the breather solutions (solitons) of the sine-Gordon equation, the solutions of the SPE can be written as a system of linear partial diffential equations governing the inverse mapping to the original coordinates. The analytical integration can be done for this linear system of partial differential equations, which yields the analytical multi-loop and multi-breather solutions. The one-breather solution (solitary wave solution) is given by 
\begin{equation}\label{spe_soln_Matsuno}
\begin{aligned}
u(y,t)=2i\left(\ln\frac{f^{'}}{f}\right)_t \\
x(y,t)=y-2\left(\ln f^{'}f\right)_t+d 
\end{aligned}
\end{equation}
with
\begin{equation}\label{spe_soln_Matsuno_With}
\begin{aligned}
f=1+ie^{\xi_1}+ie^{\xi_1^{*}}+\left(\frac{b}{a}\right)^2 e^{\xi_1+\xi_1^{*}}\\
f^{'}=f^{*} \\
\xi_{1} = \theta +i \chi \\
\theta=a \left( y+\frac{1}{a^2+b^2} t \right) +\lambda \\
\chi=b \left( y-\frac{1}{a^2+b^2} t \right) +\mu \,,
\end{aligned}
\end{equation}
where $a$ and $b$ are positive constants, $\mu$ and $\lambda$ are real constants, and $d$ is the integration constant. Note that the choices of different values of $d$ make a difference only in the location of the pulse. Equations (\ref{spe_soln_Matsuno}) and (\ref{spe_soln_Matsuno_With}) represent the compact form of the one-soliton solution of the short pulse equation (\ref{SPE_sak}) obtained by a different technique than the one used to arrive at the solution of the form (\ref{spe_soln}). In order to show that the solutions (\ref{spe_soln}) and (\ref{spe_soln_Matsuno}) are equivalent, we will explicitly write the solution (\ref{spe_soln_Matsuno}) in the parametric form as  \cite{matsuno:2007}
\begin{equation}\label{spe_soln_Matsuno_Parametric}
\begin{aligned}
u(y,t)&=\frac{4ab}{a^2+b^2}\frac{b\sin(\chi)\cosh\left(\theta+\ln\frac{b}{a}\right) - a\cos(\chi)\sinh\left(\theta+\ln\frac{b}{a}\right)}{b^2\cosh^2\left(\theta+\ln\frac{b}{a}\right)+a^2\cosh^2(\chi)}\\
x(y,t)&=y-\frac{2ab}{a^2+b^2}\frac{a\sin(2\chi)+b\sinh\left(2\theta+2\ln\frac{b}{a}\right)}{{b^2\cosh^2\left(\theta+\ln\frac{b}{a}\right)+a^2\cosh^2(\chi)}} -\frac{4a}{a^2+b^2}+d \,.
\end{aligned}
\end{equation}
This solution is multi-valued without any conditions imposed. Since we are searching for a single-valued nonsingular solution, the condition giving rise to a one soliton single-valued solution is
\begin{equation}\label{Matsuno_1solitonCondtion}
-\sqrt{2}+1<\frac{a}{b}\frac{\cos(\chi)}{\cosh\left(\theta+\ln\frac{b}{a}\right)}<\sqrt{2}-1 \,.
\end{equation}
Recall that we have imposed $a$ and $b$ positive constants beforehand. However, the condition (\ref{Matsuno_1solitonCondtion}) imposes an additional constraint on $a$ and $b$ such that $0<a/b<\sqrt{2}-1$. Notice that the phase $\theta$ governs the envelope of the soliton whereas the phase $\chi$ is responsible for the oscillations. The comparison of the solutions (\ref{spe_soln}) and (\ref{spe_soln_Matsuno_Parametric}) shows that the two are equivalent. The advantage of using the solution (\ref{spe_soln_Matsuno_Parametric}) is that we can vary the speed of pulse since the speed is defined to be $c=1/(a^2+b^2)$. This flexibility is very useful when studying the solitonic features numerically. 

In a similar fashion, multi-soliton solutions of the SPE can be constructed \cite{matsuno:2007}. We present the multi-soliton solution here because we intend to discuss the collisons of the SPE solitons in chapter five. The parametric multi-soliton solution can be expressed in the compact form as 
\begin{equation}\label{spe_soln_Matsuno_MultiSoln}
\begin{aligned}
u(y,t)=2i\left(\ln\frac{f^{'}}{f}\right)_t \\
x(y,t)=y-2\left(\ln f^{'}f\right)_t+d 
\end{aligned}
\end{equation}
with
\begin{equation}\label{spe_soln_Matsuno_MultiSoln_With}
\begin{aligned}
f= \sum_{\mu=0,1} \mathrm{exp} [ \sum_{j=1}^N \mu_j \left( \chi_j+\frac{\pi}{2}i \right)+ \sum_{1 \leq j \leq N} \mu_j \mu_k \gamma_{jk} ] \\
f^{'}= \sum_{\mu=0,1} \mathrm{exp} [ \sum_{j=1}^N \mu_j \left( \chi_j-\frac{\pi}{2}i \right)+ \sum_{1 \leq j \leq N} \mu_j \mu_k \gamma_{jk} ] \\
\xi_j=p_jy+\frac{1}{p_j}t+\xi_{j0}, \qquad (j=1,2,...,M)\\
e^{\gamma_{jk}}=(\frac{p_j-p_k}{p_j+p_k})^2, \qquad (j,k=1,2,...,M; j \neq k) \\
p_{2j-1}=p_{2j}^{*} \equiv a_j+ib_j, \quad a_j>0, \quad b_j>0, \quad (j=1,2,...,M) \\
\xi_{2j-1,0}=\xi_{2j,0}^{*}\equiv \lambda_j+i\mu_j,   \quad (j=1,2,...,M) \\
\theta_j=a_j(y+c_jt)+\lambda_j,  \quad (j=1,2,...,M) \\
\chi_j=b_j(y-c_jt)+\mu_j,  \quad (j=1,2,...,M) \\
c_j=\frac{1}{a_j^2+b_j^2},  \quad (j=1,2,...,M) \,,
\end{aligned}
\end{equation}
where $p_j$ and $\xi_{j0}$ are arbitrary parameters satisfying the conditions $p_j\neq\pm p_k$ for $j\neq k$, $i$ is the imaginary number such that $i=\sqrt{-1}$, $N$ is an arbitrary positive integer, and $M=N/2$ is the number determining the multi-soliton solutions (one soliton, two solitons, etc). $\sum_{\mu=0,1}$ means summation over all possible combinations of $\mu_1=0,1$, $\mu_2=0,1$,...,$\mu_N=0,1$. Notice that when $M=1$ (or $N=2$), multi-soliton solutions given by equations (\ref{spe_soln_Matsuno_MultiSoln}) and (\ref{spe_soln_Matsuno_MultiSoln_With}) reduce to the one-soliton solution in (\ref{spe_soln_Matsuno}) and (\ref{spe_soln_Matsuno_With}). To derive the two-soliton solution, one must choose $N=4$ and $M=2$. The condition for a single-valued multi-breather solution is
\begin{equation}\label{MultiSolitonCondition}
0<\sum_{j=1}^M\frac{a_j}{b_j}<\sqrt{2}-1
\end{equation}  

It is worthwhile to note that single and multi-loop solutions of the SPE have been derived as well \cite{sakovich-sakovich:2006,matsuno:2007}. Although these loop solutions show solitonic features, loop solutions are not single-valued. We are only interested in the nonsingular, single-valued solitary wave solutions in the context of, and in application to, nonlinear optics.

\section{Numerical Analysis of the SPE}

The short pulse equation is an exactly solvable equation whose analytical solution is a soliton as mentioned in the previous section. We have numerically shown that these solitary waves persist in the short pulse equation if they are used as initial conditions.  
\begin{figure}[htp]
\begin{center}
\scalebox{0.7}{\includegraphics{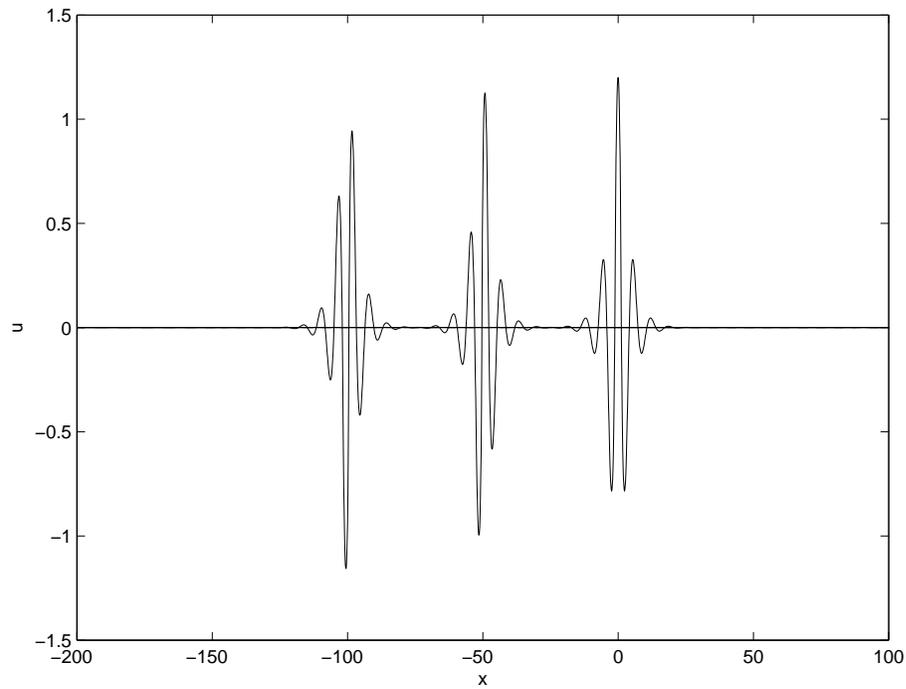}}
\caption{The numerical solution of the SPE at t=50 and t=100 units of propagation distance.}
\label{SPEpropagation}
\end{center}
\end{figure} 
Figure \ref{SPEpropagation} shows that if an ultra-short solitary wave (\ref{spe_soln}) is chosen as the initial condition, this initial wave propagates stably in the SPE (\ref{SPE_sak}). In this experiment, we pick the soliton parameter $m=0.3$ so that we have the very short solitary wave. For the other values of $m$ such as $m=0.2$ and $m=0.05$, which accordingly generate solitary waves with different sizes, we observe stable propagation as well. We see the initial solitary wave at $t=0$ in figure \ref{SPEpropagation}. Note that $x$ is the temporal variable in our case. The propagation of the initial pulse is shown at $t=50$ and $t=100$. 

Let us also compare these results with the results obtained from the exact solution to see whether there is any deviation from the exact shape and size of the soliton as it moves along the line. By doing so, we compute the exact solutions at $t=50$ and $t=100$ using the analytical solution (\ref{spe_soln}), and compare them with the corresponding numerical results.  
\begin{figure}[htp]
\begin{center}
\scalebox{0.5}{\includegraphics{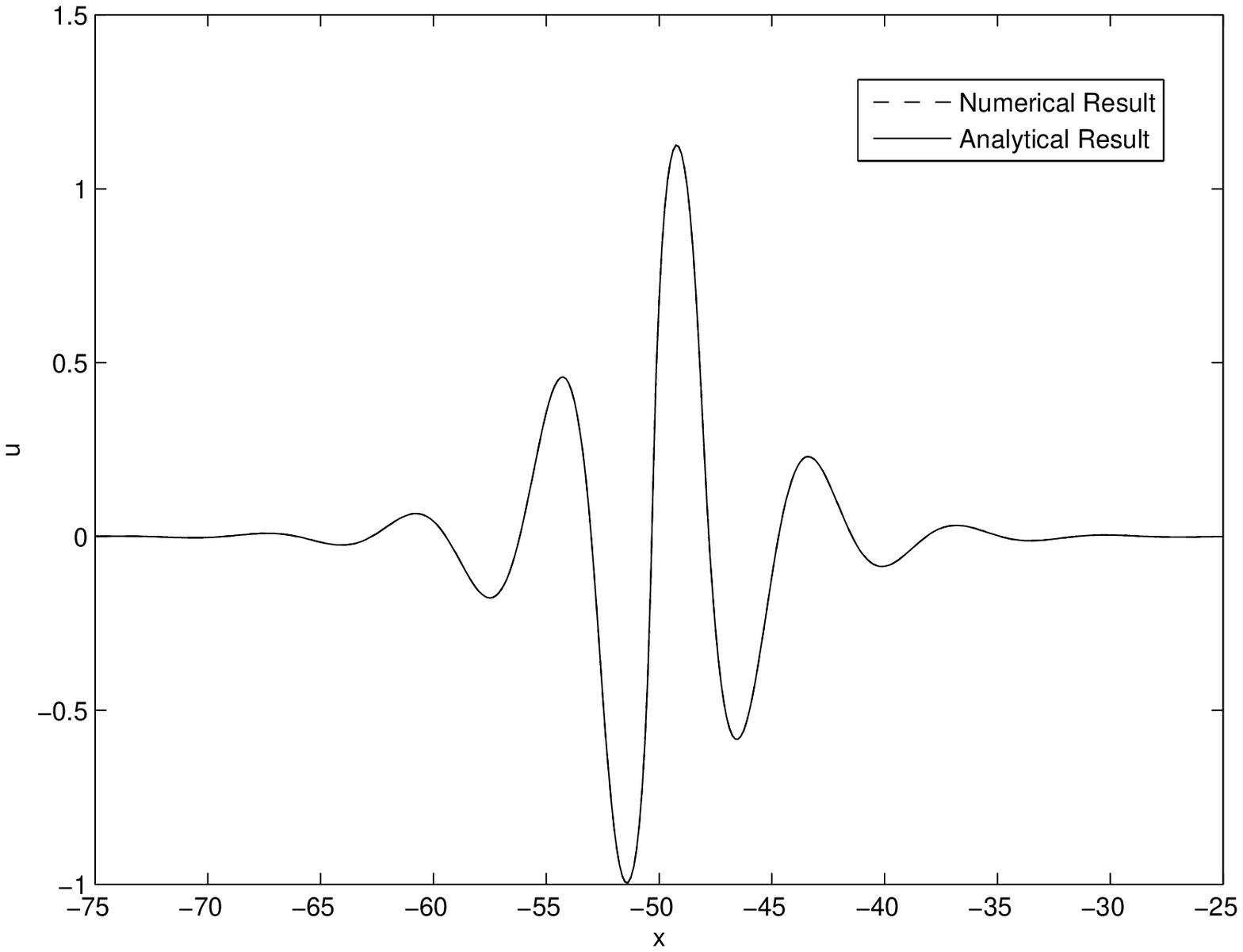}}
\scalebox{0.5}{\includegraphics{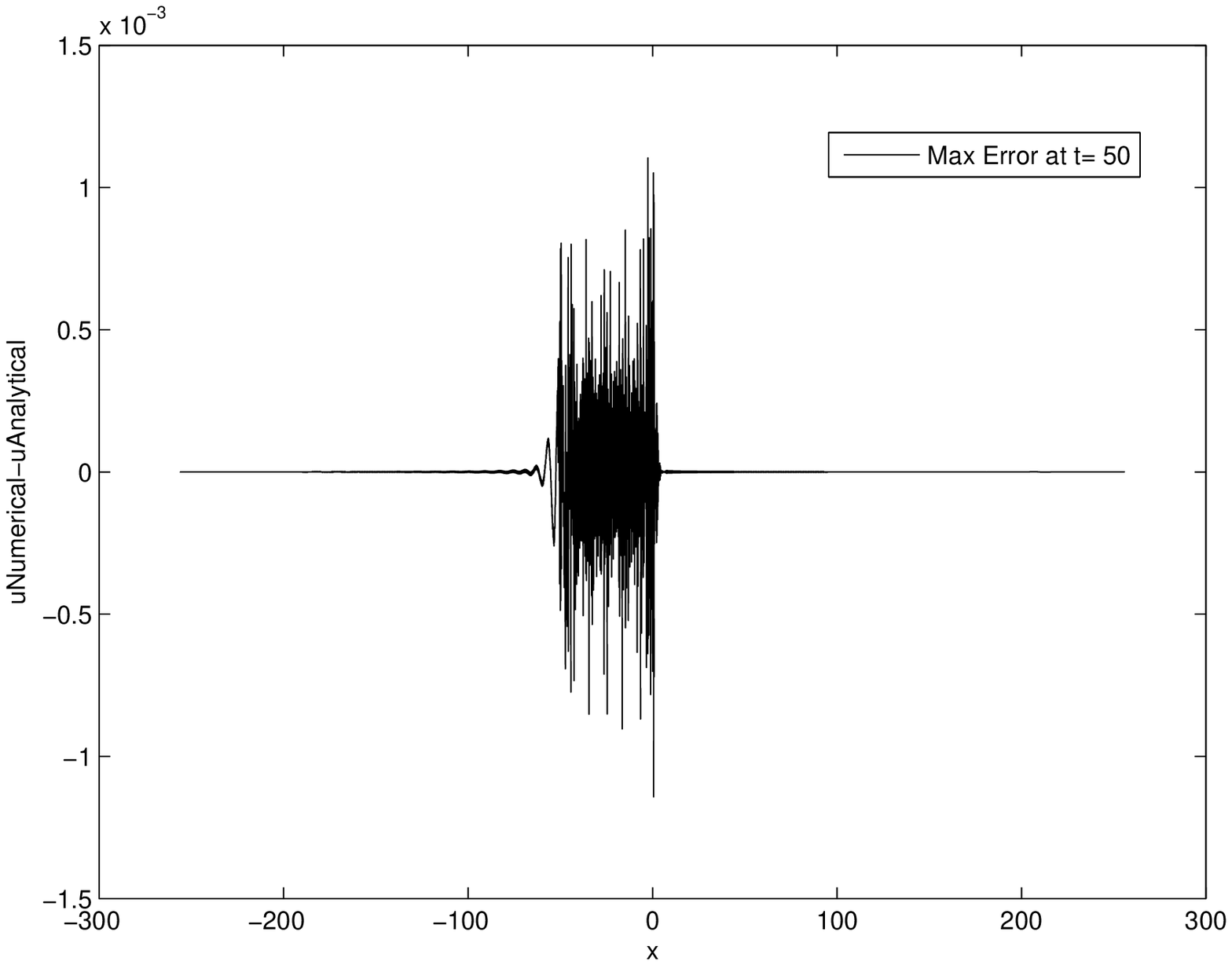}}
\caption{Comparison of the exact solitary wave solution to the numerical result at $t=50$ distance units (upper graph). Maximum error between the exact solution and the numerical solution at the same distance (bottom graph)}
\label{SPEpropagation50Comparison}
\end{center}
\end{figure} 

\begin{figure}[htp]
\begin{center}
\scalebox{0.5}{\includegraphics{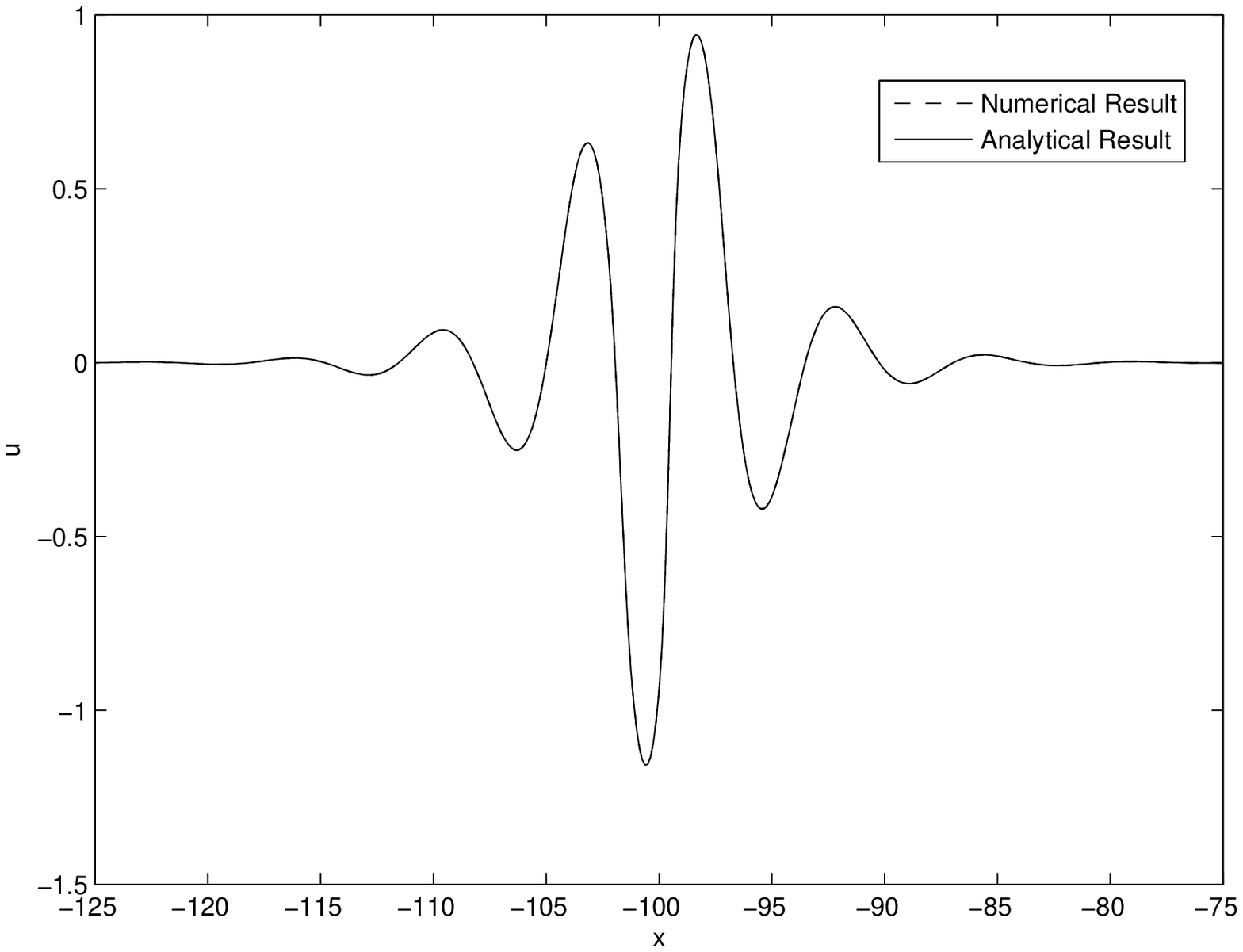}}
\scalebox{0.5}{\includegraphics{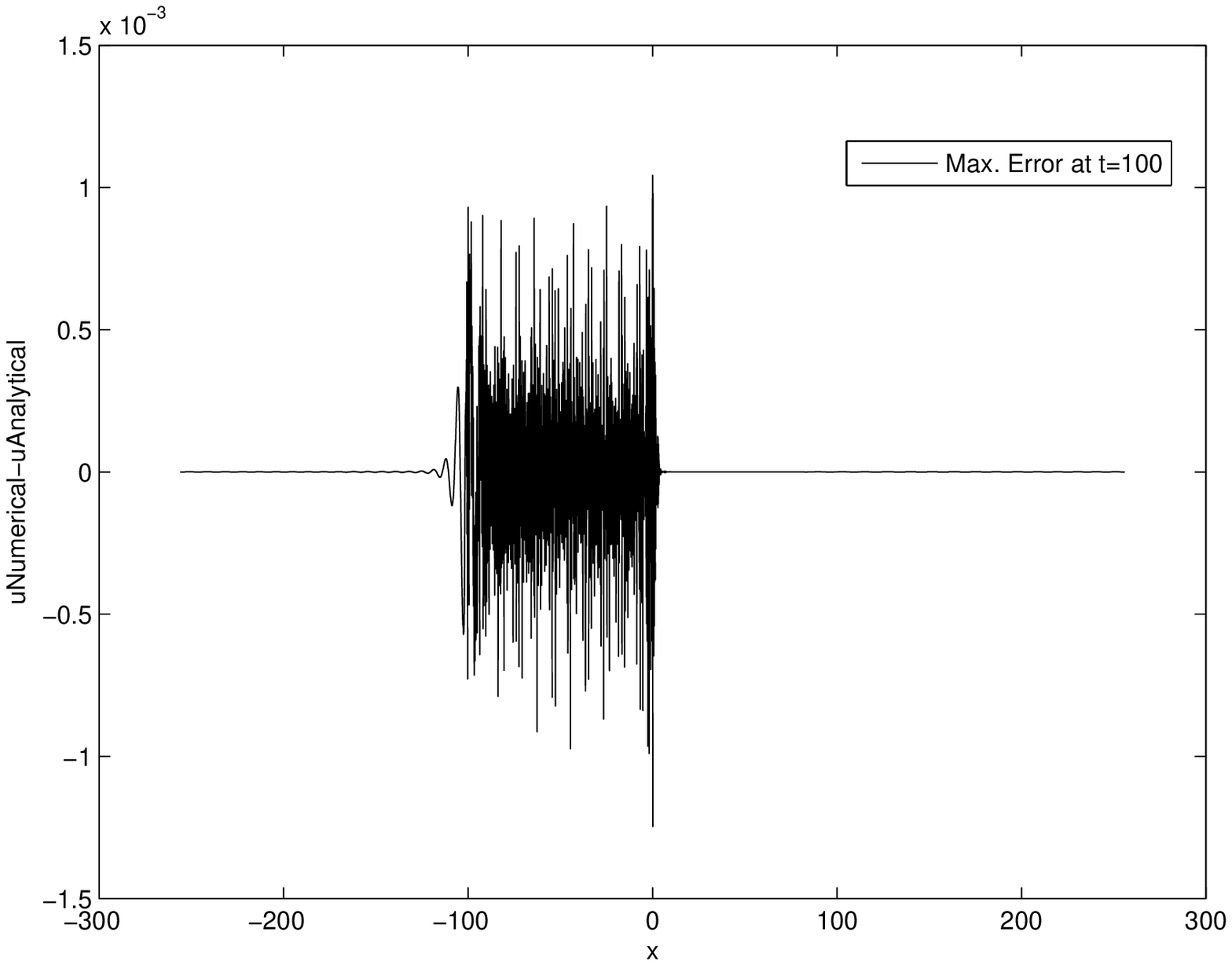}}
\caption{Comparison of the exact solitary wave solution to the numerical result at $t=100$ distance units (upper graph). Maximum error between the exact solution and the numerical solution at the same distance (bottom graph)}
\label{SPEpropagation100Comparison}
\end{center}
\end{figure} 
Figures \ref{SPEpropagation50Comparison} and \ref{SPEpropagation100Comparison} exhibit the comparisons of these two solitons at $t=50$ and $t=100$ respectively. Since there are no observable differences between the exact and the numerical results at both $t=50$ and $t=100$, we also plot the maximum error for each case. The error in both cases is almost zero, and therefore we can just relate this negligible differences to the experimental error. To reiterate, an initial solitary wave obtained from the exact result for any values of the soliton parameter $m$ propagates stably in the short pulse equation, and this serves as numerical proof for the analytical solution of the SPE.    
 
An interesting question to pose is to ask what happens if the exact solitary wave solution (\ref{spe_soln}) is used as an initial condition in the original equation (\ref{maxwell_1d}). It has already been shown  \cite{ chung-jones-etal:2005} that the exact solitary wave of the short pulse equation persists in the linear wave equation ($b=0$ in (\ref{maxwell_1d})). We have also shown numerically that these solitons propagate stably in the nonlinear wave equation \cite{kurt-schaefer:2010}.
\begin{figure}[htp]
\begin{center}
\scalebox{0.7}{\includegraphics{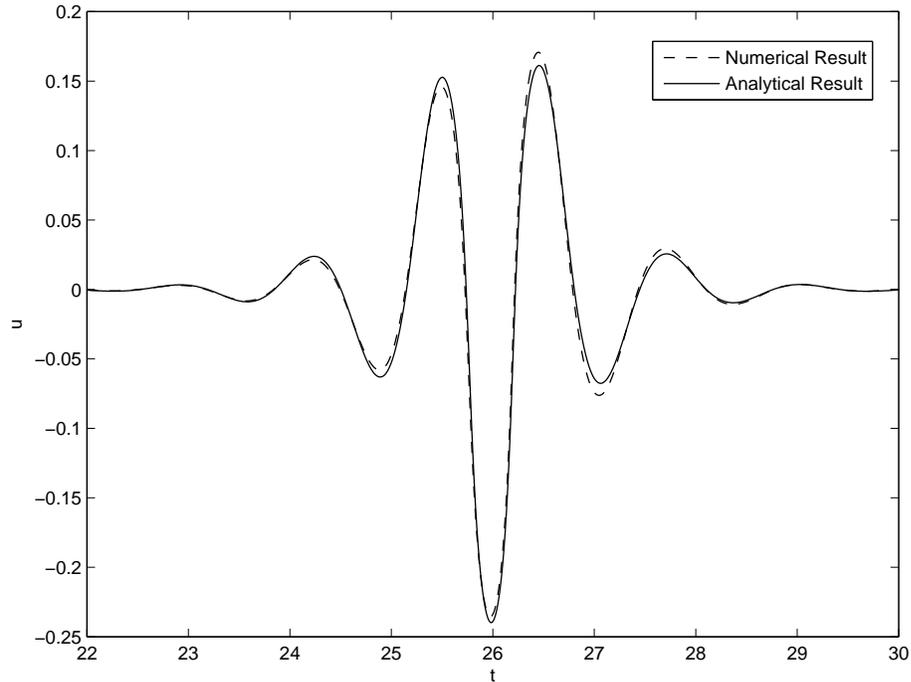}}
\caption{Evolution of the SPE soliton in the Maxwell equation }
\label{MaxwellSPEcomparison}
\end{center}
\end{figure} 
Figure \ref{MaxwellSPEcomparison} displays the exact initial SPE soliton propagation at $x=25$ and its comparison to the analytical result at $x=25$. Note that the evolution variable is switched to $t$ in the nonlinear equation. The solid line shows the analytical solution at $x=25$, whereas the dashed line shows the numerical solution at the same distance. As one can see from figure \ref{MaxwellSPEcomparison}, the initial SPE soliton persists in the nonlinear Maxwell equation. 

It may noteworthy to say a few more words about the details of the numerical experiment. We modify the initial SPE soliton according to the multiple scale expansion (\ref{u_expansion}) and (\ref{xt_expansion}) such that the magnitude of the initial pulse becomes $u = \epsilon \,u(x/\epsilon,0)$ at $t=0$. Accordingly, the analytical result has to be modified as well. Therefore, we have the analytical solution at $t=25$ distance as $\mathrm{u_{ana}} = \epsilon \, \mathrm{u}\left(\right[(x-25)/\epsilon\left],-\epsilon \, 25\right)$. We choose the soliton parameter $m=0.3$ and the expansion parameter $\epsilon=0.2$. Note also that SPE is the leading order $O(\epsilon)$ approximation and choosing the propagation distance $O(1/\epsilon^2) = 25$ units in the numerical experiment is more than enough to observe any abnormalities in the propagation.

Before closing this section, let us also discuss the error accumulation when the initial SPE soliton propagates in the nonlinear wave equation. Since the SPE solitons are not the exact solutions of the nonlinear wave equation, we expect, in general, a growing difference between the numerical and the analytical results with the propagation distance. For instance, the maximum error at $t=25$ is $0.0186$ unit. A more elucidating picture for this discussion is to plot the $L^1$, $L^2$ and $L^{\infty}$ norms of $f=u_{Numerical}-u_{Exact}$ versus propagation distance, and the norms of $f$ are defined respectively as
\begin{equation}
\begin{aligned}
\|f\|_{L^1}&=\left(\int |f(t)|\,dt\right) \\
\|f\|_{L^2}&=\left(\int |f(t)|^2\,dt\right)^{1/2}  \\
\|f\|_{L^{\infty}}&=\max(|f(t)|)\,.
\end{aligned}
\end{equation}
\begin{figure}[htp]
\begin{center}
\scalebox{1}{\includegraphics{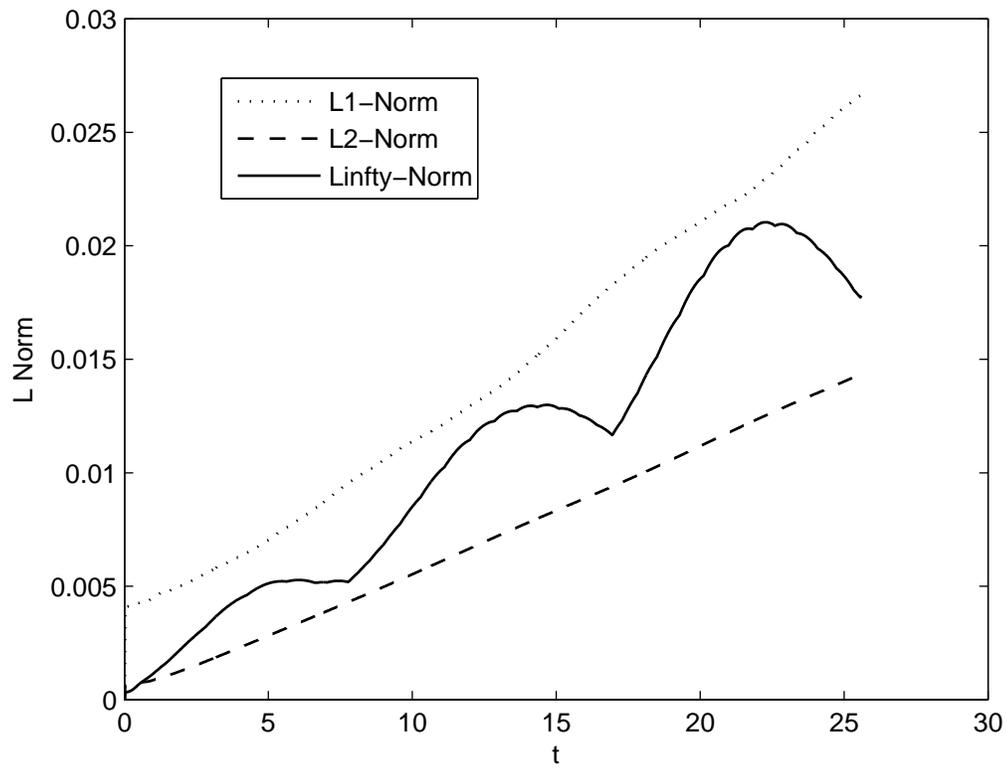}}
\caption{ The growth of the deviations for the evolution of the SPE soliton in the nonlinear wave equation as defined by $L^1$, $L^2$ and $L^{\infty}$ norms.}
\label{LNorms}
\end{center}
\end{figure} 
In figure \ref{LNorms}, the dotted line shows $L^1$ norm, the dashed line shows $L^2$ norm and the solid line shows the $L^{\infty}$ norm. The solution of the leading order equation of the multiple scale expansion fails to approximate the solution of the nonlinear wave equation at the larger propagation distances as can be seen by figure \ref{LNorms} \cite{kurt-schaefer:2010}.

\section{The Higher Order SPE}

If a multiple scale expansion of the form (\ref{u_expansion}) with the scale transformation (\ref{xt_expansion}) is applied to the Maxwell equation (\ref{maxwell_1d}), the evolution of ultra-short pulses in optical fibers is expressed by the function $A_0(\phi,x_1)$ over the scales $(\phi,x_1)$ instead of the function $u(x,t)$ over the scales $(t,x)$. If one wants to improve the accuracy of the expansion, one can take into account the dependence of the $A_0$ function on $x_3$ as introduced by the multiple scale expansion (\ref{u_expansion}) and (\ref{xt_expansion}). In the derivation of the short pulse equation, we only consider the expansion up to the order of $\epsilon$, and the evolution equation is expressed over two variables, $\phi$ and $x_1$. In this section, we consider the higher order expansion terms so that the $A_0$ function has another dependence, i.e., $A_0(\phi,x_1) \rightarrow A_0(\phi,x_1,x_3)$. To insure that, we introduce more space variables such that $x_2=\epsilon^2x$ and $x_3=\epsilon^3x$ according to the scale transformation (\ref{xt_expansion}). It is imperative to note that we follow the same procedure here as we followed in section $4.1$. Note also that the presence of $x_0$ in the expansion leaves the result unchanged. Therefore, we only keep the new space variables $x_1$, $x_2$ and $x_3$ here. If the procedure in section $4.1$ is repeated, the terms of $O(1/\epsilon)$ and $O(\epsilon)$ canceled out, and the terms of $O(\epsilon)$ generate the short pulse equation (\ref{SPE_original}). By choosing $A_1=0$ and $\partial A_0/\partial x_2 =0$ (i.e, $A_0$ is independent of $x_2$), the terms of $O(\epsilon^2)$ will canceled out as well. Finally, the terms of $O(\epsilon^3)$ become 
\begin{equation}\label{equation_Order3}
\frac{\partial^2 A_0}{\partial^2 x_1} - 2 \frac{\partial^2 A_0}{\partial \phi \partial x_3}=0 \,.
\end{equation}   
This equation underlines the dependence of $A_0$ on $x_3$. Now it remains to be seen how exactly one incorporates the $x_3$ dependence in the short pulse equation. To write a single evolution equation for $A_0$ using the terms of $O(\epsilon)$ and $O(\epsilon^3)$, we will combine the two equations given by (\ref{SPE_original}) and (\ref{equation_Order3}). By doing so, let us first integrate the SPE (\ref{SPE_original}) with respect to $\phi$,
\begin{equation}\label{phiIntegration_spe}
(A_{0})_{x_1}=-\frac{a}{2} \int_{-\infty}^\phi A_0 d\phi - \frac{b}{2} (A_0)^3_{\phi}. 
\end{equation} 
Notice that we have the double $x_1$ derivative in equation (\ref{equation_Order3}). We now take the derivative of (\ref{phiIntegration_spe}) with respect to $x_1$
\begin{equation}\label{A0_DoubleX1Der}
\begin{aligned}
(A_{0})_{x_1 x_1}=&-\frac{a}{2} \int_{-\infty}^\phi (A_{0})_{x_1} d\phi - \frac{b}{2} 
\bigg(3(A_0)^2(A_{0})_{x_1}\bigg)_{\phi} \\
=& -\frac{a}{2}  \int_{-\infty}^\phi \bigg( -\frac{a}{2} \int_{-\infty}^\phi A_0 d\phi - 
\frac{b}{2} (A_0)^3_{\phi}   \bigg) d\phi - \\
&\frac{b}{2} \bigg(3(A_0)^2 \big( - \frac{a}{2} \int_{-\infty}^\phi A_0 d\phi - \frac{b}{2} (A_0)^3_{\phi}  \big)   \bigg)_{\phi} \\
=& \frac{a^2}{4} \int_{-\infty}^\phi \int_{-\infty}^\phi A_{0} d^2\phi + \frac{ab}{4} (A_0)^3 +\\ &\frac{3b}{2} \bigg((A_0)^2 \big( \frac{a}{2} \int_{-\infty}^\phi A_0 d\phi + \frac{b}{2} (A_0)^3_{\phi}  \big)   \bigg)_{\phi}
\end{aligned}
\end{equation}
If we substitue the expression for $(A_{0})_{x_1 x_1}$ in (\ref{A0_DoubleX1Der}) into equation (\ref{equation_Order3}) and integrate it with respect to $\phi$, we obtain
\begin{equation}\label{A0_X3}
\begin{aligned}
2(A_0)_{x_3}= \frac{a^2}{4} \int_{-\infty}^\phi \int_{-\infty}^\phi \int_{-\infty}^\phi A_{0} d^3\phi + \frac{ab}{4} \int_{-\infty}^\phi (A_0)^3 d\phi + \\
\frac{3ab}{4} (A_0)^2 \int_{-\infty}^\phi A_0 d\phi + \frac{3b^2}{4} (A_0)^2 (A_0)^3_{\phi}
\end{aligned}
\end{equation}
Introducing a new variable such that $\varkappa=x_1$ and $\epsilon^2\varkappa =x_3$, we find
\begin{equation}\label{NewA0}
(A_0)_{\varkappa}=(A_0)_{x_1}+\epsilon^2(A_0)_{x_3}.
\end{equation}
Finally, we obtain $(A_0)_{x_1}$ and $(A_0)_{x_3}$ from the relations (\ref{phiIntegration_spe}) and (\ref{A0_X3}) respectively and substitute them into
equation (\ref{NewA0}) so that
\begin{equation}\label{HOspe}
\begin{aligned}
(A_0)_{\varkappa}=-\frac{\chi_0}{2} \int_{-\infty}^\phi A_0 d\phi - \frac{\chi_3}{2} (A_0^3)_{\phi} +
\frac{\epsilon^2}{2} \Bigg( \frac{\chi_0^2}{4} \int_{-\infty}^\phi \int_{-\infty}^\phi \int_{-\infty}^\phi A_0 d^3\phi + \\
\frac{\chi_0 \chi_3}{4} \int_{-\infty}^\phi A_0^3 d\phi + \frac{3 \chi_0 \chi_3}{4} A_0^2 \int_{-\infty}^\phi A_0 d\phi + \frac{3 \chi_3^2}{4} A_0^2 (A_0^3)_{\phi} \Bigg)
\end{aligned}
\end{equation}
This represents the higher order short pulse equation. Notice that $A_0({\varkappa},\phi)$ is the magnitude of the electric field following the introduction of the new variable $\varkappa$. For higher orders, we expect to improve our numerical results. The numerical validation of the higher order SPE remains an open problem, and one in which we hope to tackle in the future.

%% file: Chap5_Solitonsinc.tex
Solitons manifest themselves in many branches of modern science such as nonlinear optics, plasma physics, hydrodynamics and biology \cite{campbell:1987,lomdahl:1984,scott-chu-mcLaughlin:1973}. They have been used extensively in optical communication \cite{hasegawa-kodama:1995,agrawal:2007,mollenauer-gordon:2006} since the discovery of the nonlinear Schr\"odinger equation. Although we have touched on various soliton solutions in the preceding chapters, we provide a historical synopsis of solitons in the present chapter, and also delve into a brief discussion of the most famous solitons such as KdV solitons. We aim to show and interpret our numerical results and in-so-doing validate the solitonic properties of the SPE solitons.


\section{What is a Soliton?}

A soliton is a wave packet or a pulse that maintains its shape as it propagates with a constant speed in a medium because of a delicate balance between dispersive effects and nonlinearity \cite{dodd-eilbeck-gibbon-morris:1982}. Although the term soliton was introduced in the 1960s, physical solitary waves were first observed in water waves by J.S.Russell in 1834. The observed nondispersive water waves were just the analog of the latter optical solitons. The description of Scott Russell's solitary water waves was published in his paper in 1844 \cite{mollenauer-gordon:2006}, which includes the following quote:\\
 
\textit{
I was observing the motion of a boat which was rapidly drawn along a
narrow channel by a pair of horses, when the boat suddenly stopped--not
so the mass of water in the channel which it had put in motion; it accumulated
round the prow of the vessel in a state of violent agitation, then
suddenly leaving it behind, rolled forward with great velocity, assuming
the form of a large solitary elevation, a rounded, smooth and well-defined
heap of water, which continued its course along the channel apparently
without change of form or diminution of speed. I followed it on horseback,
and overtook it still rolling on at a rate of some eight or nine miles
an hour, preserving its original figure some thirty feet long and a foot to a 
foot and a half in height. Its height gradually diminished, and after a chase
of one or two miles I lost it in the windings of the channel. Such, in the
month of August 1834, was my first chance interview with that singular
and beautiful phenomenon which I have called the Wave of Translation}.\\

The name soliton was given later to such wave translation observed in 1834. In response to the observation of the wave of translation by John Scott Russell, Boussinesq's equation and KdV equation were derived by Joseph Boussinesq, and Diederik Korteweg and Gustav de Vries in 1872 and 1895 respectively to describe the wave of translation mathematically \cite{kumar-tiwari}. Both equations can be solved exactly and the simplest solutions turn out to be solitary waves. The third-order KdV serves as a model for waves on shallow water surfaces, and can be given in standard form as
\begin{equation}
u_t+6uu_z+(u)_{zzz}=0 \,,
\end{equation}
where $z$ is the propagation direction and $t$ is the time variable. The exact solution of the KdV equation is a soliton and can be written as
\begin{equation}\label{KdV_soliton}
u(z,t)=a\,{\mathrm{sech}}^2(\sqrt{a/2}(z-2at)) \,,
\end{equation}  
where $a$ is a constant number representing the amplitude of the initial soliton.
\begin{figure}[htp]
\begin{center}
\scalebox{1}{\includegraphics{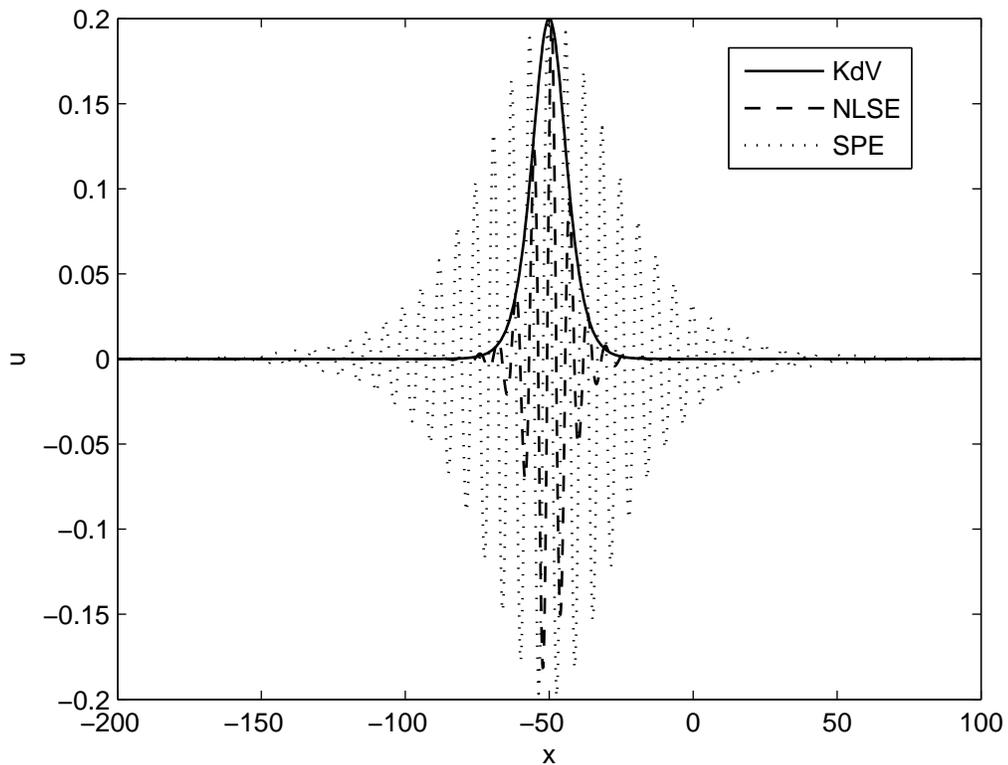}}
\caption{Comparison of the SPE, NLSE and KdV solitons}
\label{SolitonsComparison}
\end{center}
\end{figure} 

It may be enlightening to compare all the solitons we have thus far mentioned, i.e., the SPE, NLSE and KdV solitons. In figure \ref{SolitonsComparison}, we show the SPE, NLSE and KdV solitons at the propagation distance $t=50$ (in accordance with the SPE transformation, $t$ is expressed in units of distance). We directly observe the similarity among the KdV soliton (\ref{KdV_soliton}), the NLSE fundamental soliton (\ref{NLSE_BrightSoliton}) and the SPE approximate solution (\ref{spe_approx_soln}) in that they all exhibit a similar envelope shape albeit the KdV soliton has no oscillatory contribution. It is worthy to note that upon varying the parameter $m$ up to a critical value ($m_{cr}\approx 0.383$), we can alter the width of the envelope of the SPE solitons.  
In reference to the SPE soliton shown in figure \ref{SolitonsComparison}, we have chosen the parameter $m$ to be $0.05$. With regards to the NLSE soliton in (\ref{NLSE_BrightSoliton}), figure \ref{SolitonsComparison} depicts the soliton with a unit velocity and an amplitude of $0.2$ unit. 

The term soliton was introduced by Zabusky and Kruskal in 1965 while working on the interaction among KdV solitary waves numerically. These solitary waves were named as solitons because of the particle-like behavior when they collide. The exact solution of the KdV equation was found by Gardner in 1967 through a method called the inverse scattering transform. Other notable nonlinear equations possessing soliton solutions, such as the NLSE and the sine-Gordon (sG) equation, may also be solved analytically by the inverse scattering method.

KdV solitons are different than the optical solitons. They describe the solitary wave of a wave. On the other hand, an optical soliton describes the solitary wave of an envelope in a nonlinear cubic medium. Optical solitons are electromagnetic waves that are self-localized or self-trapped. This means they move at a constant speed in a medium with no change in their shape because of a delicate balance between nonlinearity and dispersion. The first technological application of optical solitons was done in 1973 for pulse propagation in optical fibers. Since then, NLSE solitons have been used in many practical situations. Using ultra-short solitons in data transfer and communication may expand the spectrum of technology \cite{hasegawa-kodama:1995,karasawa-nakamura-etal:2001}. Due to the possibility of a wide-range applicability, we will further investigate ultra-short solitons (the SPE solitons) and show their solitonic properties numerically.


\section{The SPE Solitons}

An analytical solution of the SPE is given by (\ref{spe_soln}) and (\ref{spe_approx_soln}). The latter is an approximation of the exact solution whenever the parameter $m$ takes small values. The parameter $m$ determines the shape of the pulse. 
\begin{figure}[htp]
\begin{center}
\scalebox{0.5}{\includegraphics{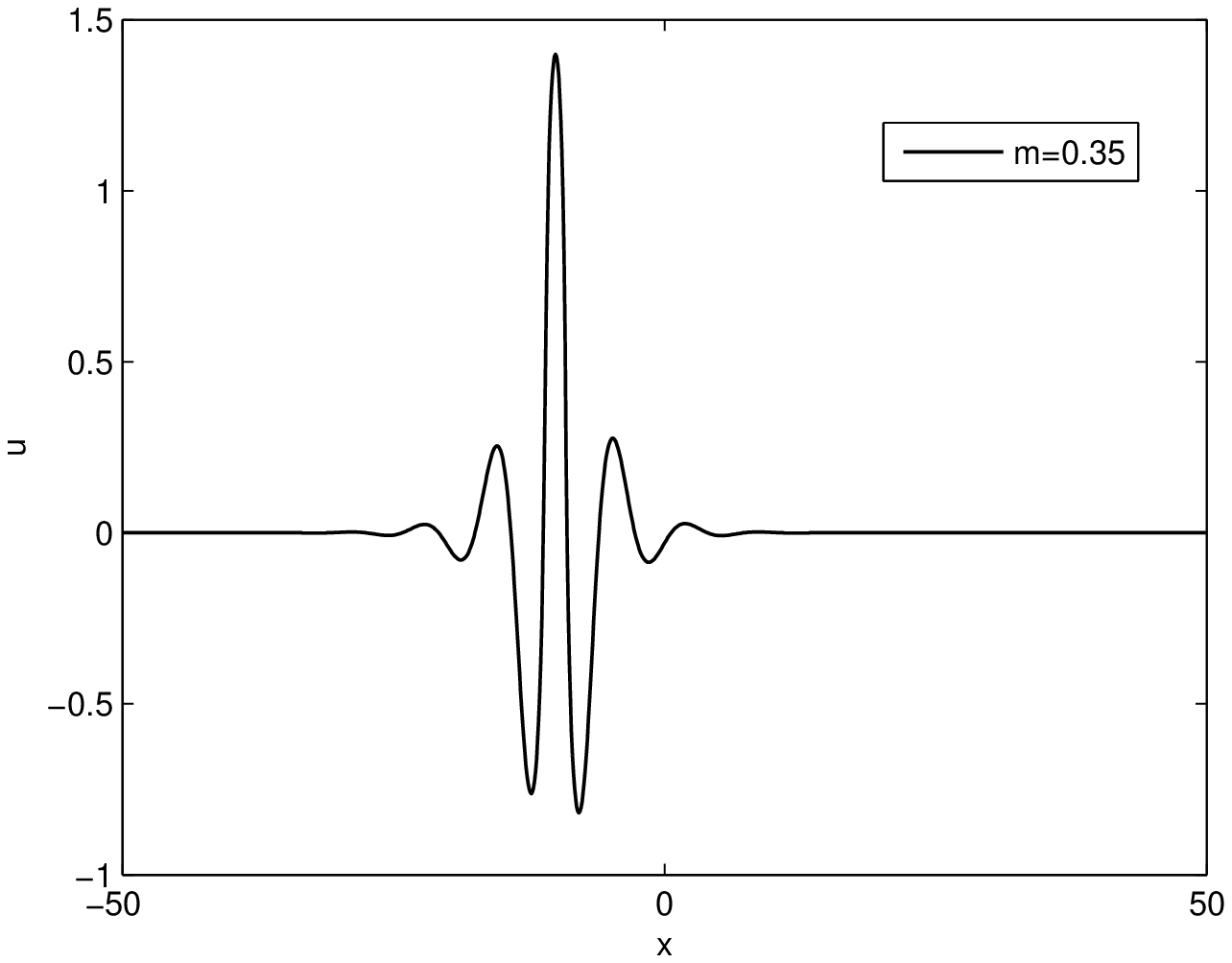}}
\scalebox{0.5}{\includegraphics{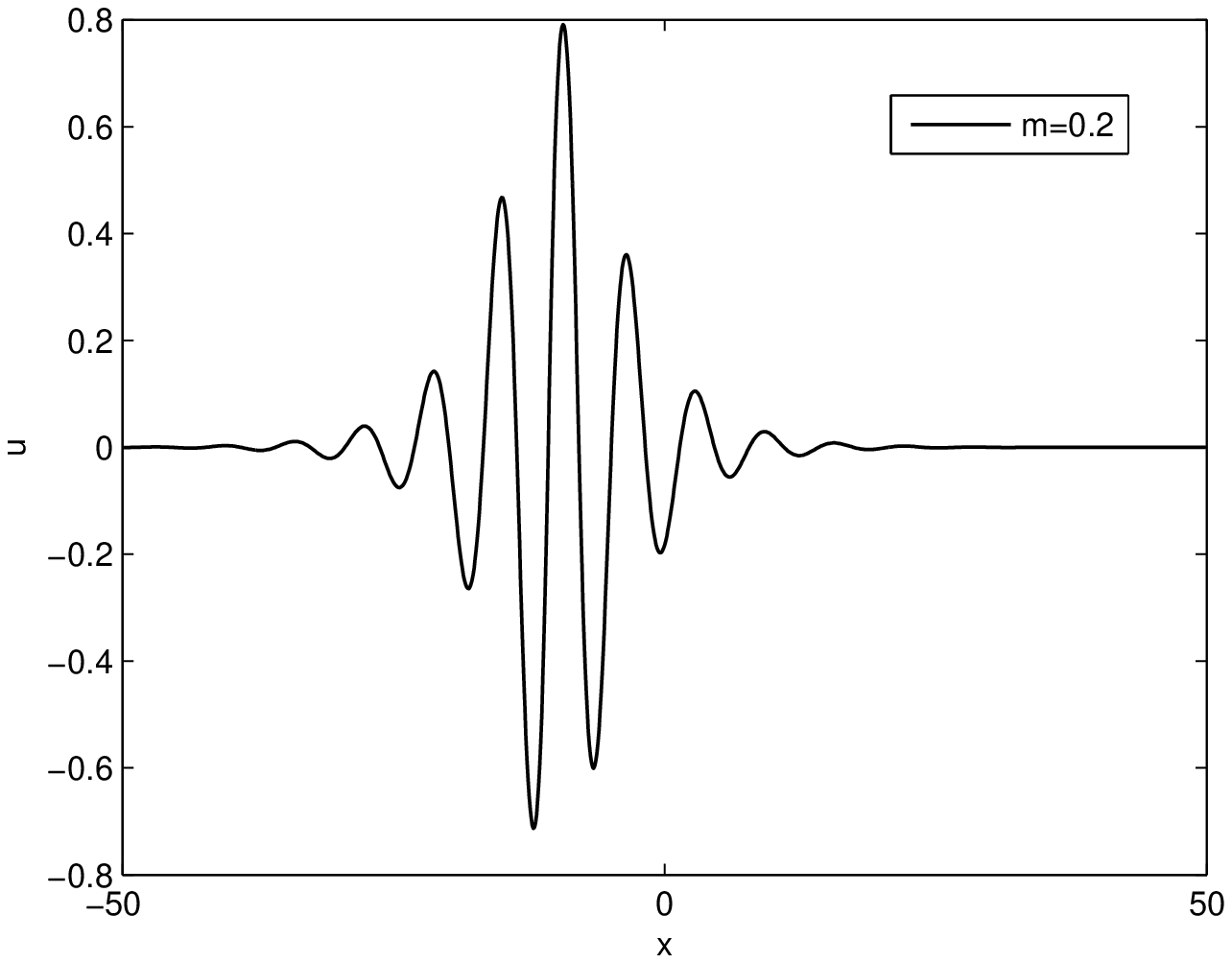}}
\scalebox{0.5}{\includegraphics{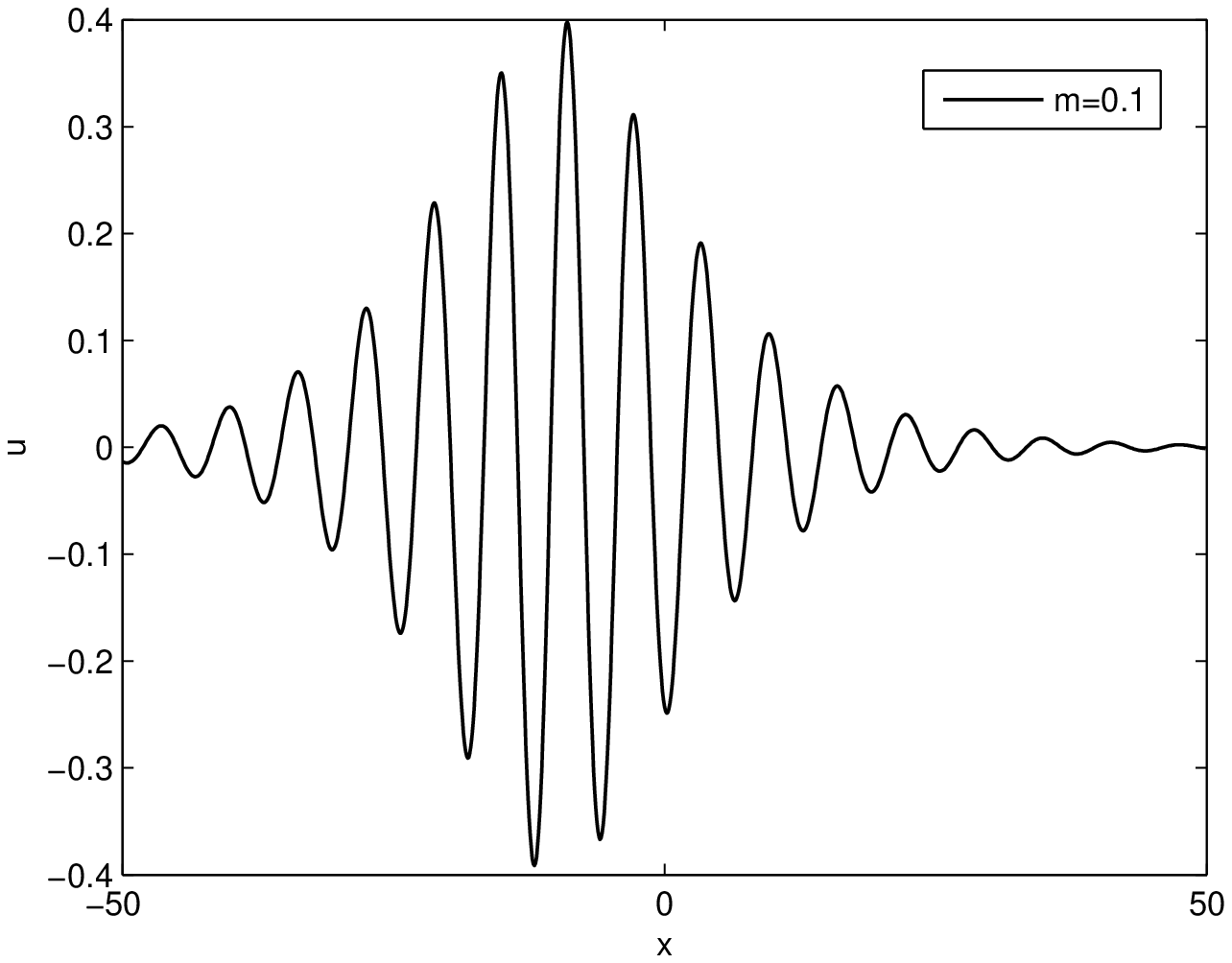}}
\caption{The parameter $m$ in the exact one-soliton solution determines the shape and width of the SPE solitons. Three exact SPE solitons for $m=0.1$, $m=0.2$, $m=0.35$ are shown at $t=10$ units of propagation distance.}
\label{Soliton3mValues}
\end{center}
\end{figure} 
Figure \ref{Soliton3mValues} shows the exact solution (\ref{spe_soln}) for different values of $m$ at $t=10$. As we increase $m$ from $0$ to $0.35$ (see (\ref{mCritical})), the pulse width narrows and the amplitude grows. In the case by which $m$ takes on smaller values, the width widens and decreases in amplitude. For higher values of $m$, we observe pulses as short as three cycles of its central frequency. When we choose a small value for $m$, one can use the approximate solution (\ref{spe_approx_soln}) instead of the exact solution (\ref{spe_soln}).
\begin{figure}[htp]
\begin{center}
\scalebox{0.6}{\includegraphics{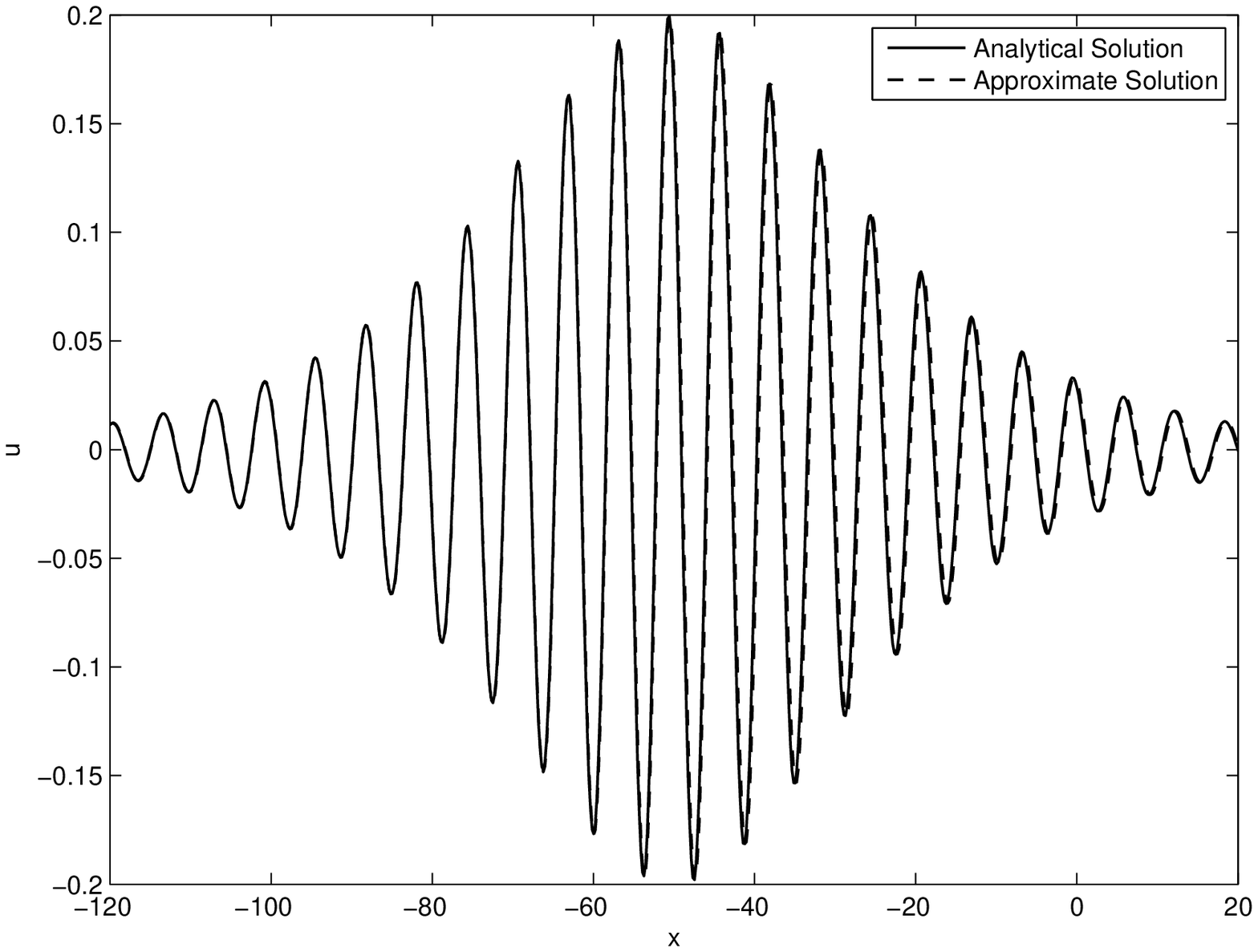}}
\scalebox{0.6}{\includegraphics{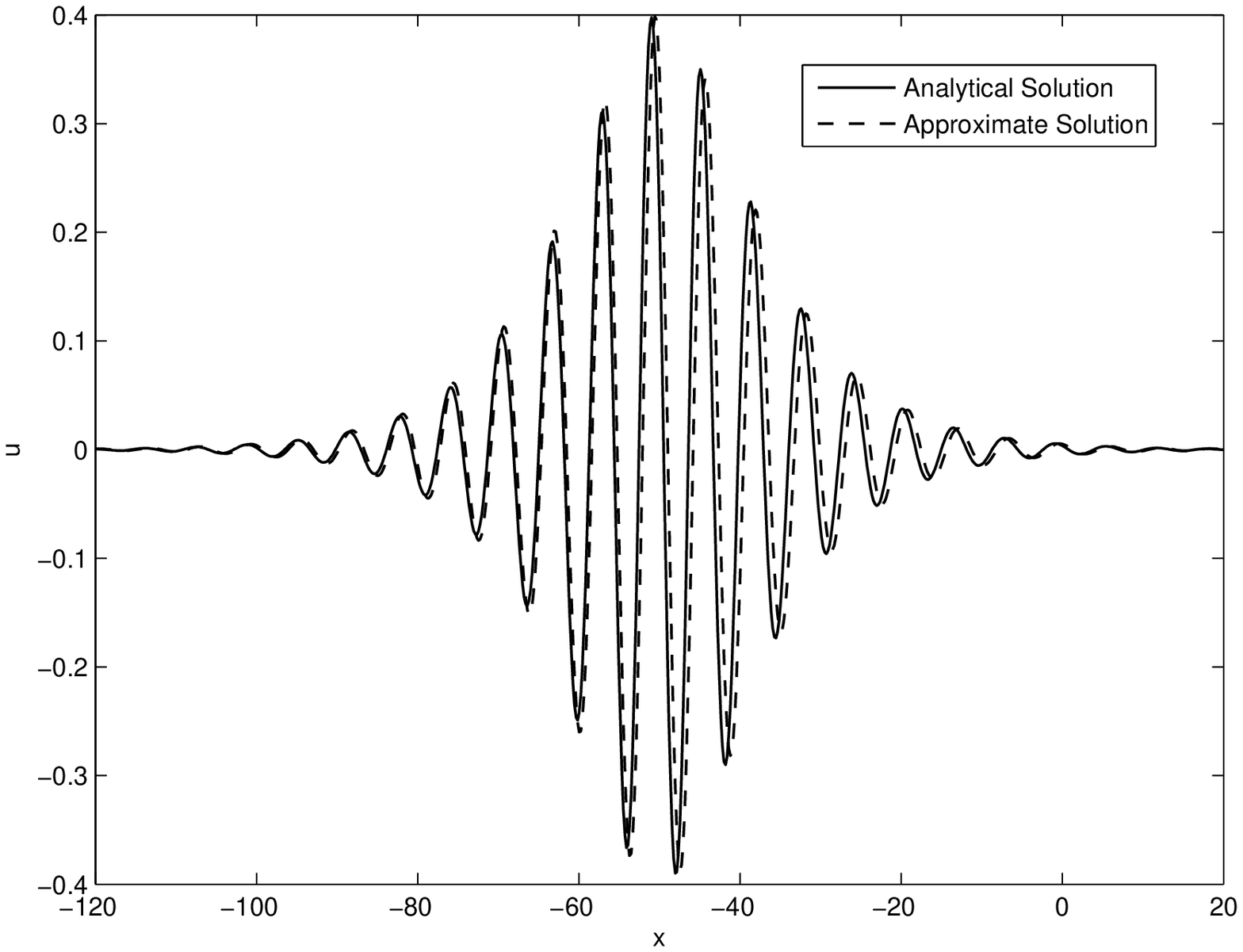}}
\caption{Comparison of the exact analytical and approximate solutions of the SPE for $m=0.05$ (upper graph) and $m=0.1$ (bottom graph) at $t=50$ distance units.}
\label{ComparisonM01M005}
\end{center}
\end{figure} 
The numerical work displayed in figure \ref{ComparisonM01M005} demonstrates that this is a very good approximation for $m=0.05$. However, if $m=0.1$ is chosen, the approximation starts to fail as can be seen in figure \ref{ComparisonM01M005}. For this reason, we suggest that an approximate solution should be used for $m$ values that are equal to or less than $0.05$ if necessary. 

As it was emphasized before, the solitary wave solution of the short pulse equation lends itself to a stable propagation due to a fine balance between dispersion and nonlinearity. Dispersion, on one hand, drives solitons to spread out as they propagate. On the other hand, nonlinearity gives rise to a centralizing effect on the solitons tending to draw them together. The linear broadening of a soliton is canceled out by the nonlinearity of the medium whose origin is the intensity dependence of the refractive index \cite{mollenauer-gordon:2006}. If one of these balancing effects is lost, the result is an unstable soliton which cannot exist over an extended period of time. Our numerical simulations validate such effects for the SPE solitons. 

Let's now switch off the nonlinearity in the SPE. Without the nonlinear term, the SPE takes the form
\begin{equation}\label{spe_NoNL}
U_{XT}=U \,.
\end{equation} 
\begin{figure}[htp]
\begin{center}
\scalebox{0.7}{\includegraphics{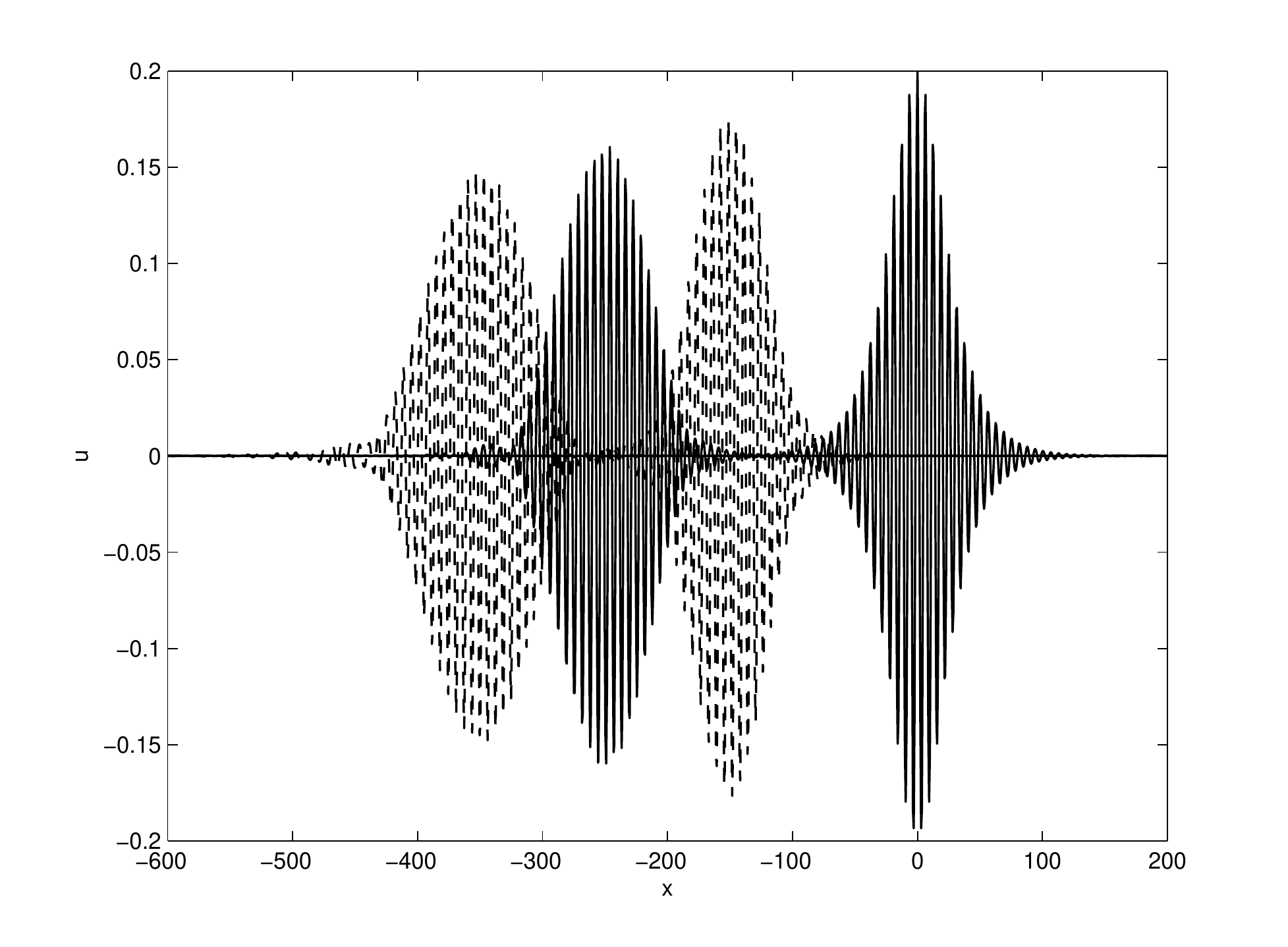}}
\caption{The propagation of the SPE solitary wave in the absence of nonlinearity.}
\label{SolitonNoNonlinearity}
\end{center}
\end{figure} 
Figure \ref{SolitonNoNonlinearity} displays the broadening of the solitary pulse as it propagates along the medium without the nonlinear term. As anticipated, dispersion dissipates broadens the pulse linearly in the absence of nonlinearity.

Leaving on nonlinearity, and now switching off dispersion, the SPE takes the form
\begin{equation}\label{spe_NoDispersion}
U_{XT}=\frac{1}{6}(U^3)_{XX} \,.
\end{equation}
Once we let the pulse propagate in the presence of nonlinearity alone, the nonlinear term forces the pulse to be more concentrated at the center as it tends to travel. This centralizing effect is not balanced by the linear broadening effect due to dispersion and the pulse eventually blows up. 
\begin{figure}[htp]
\begin{center}
\scalebox{0.7}{\includegraphics{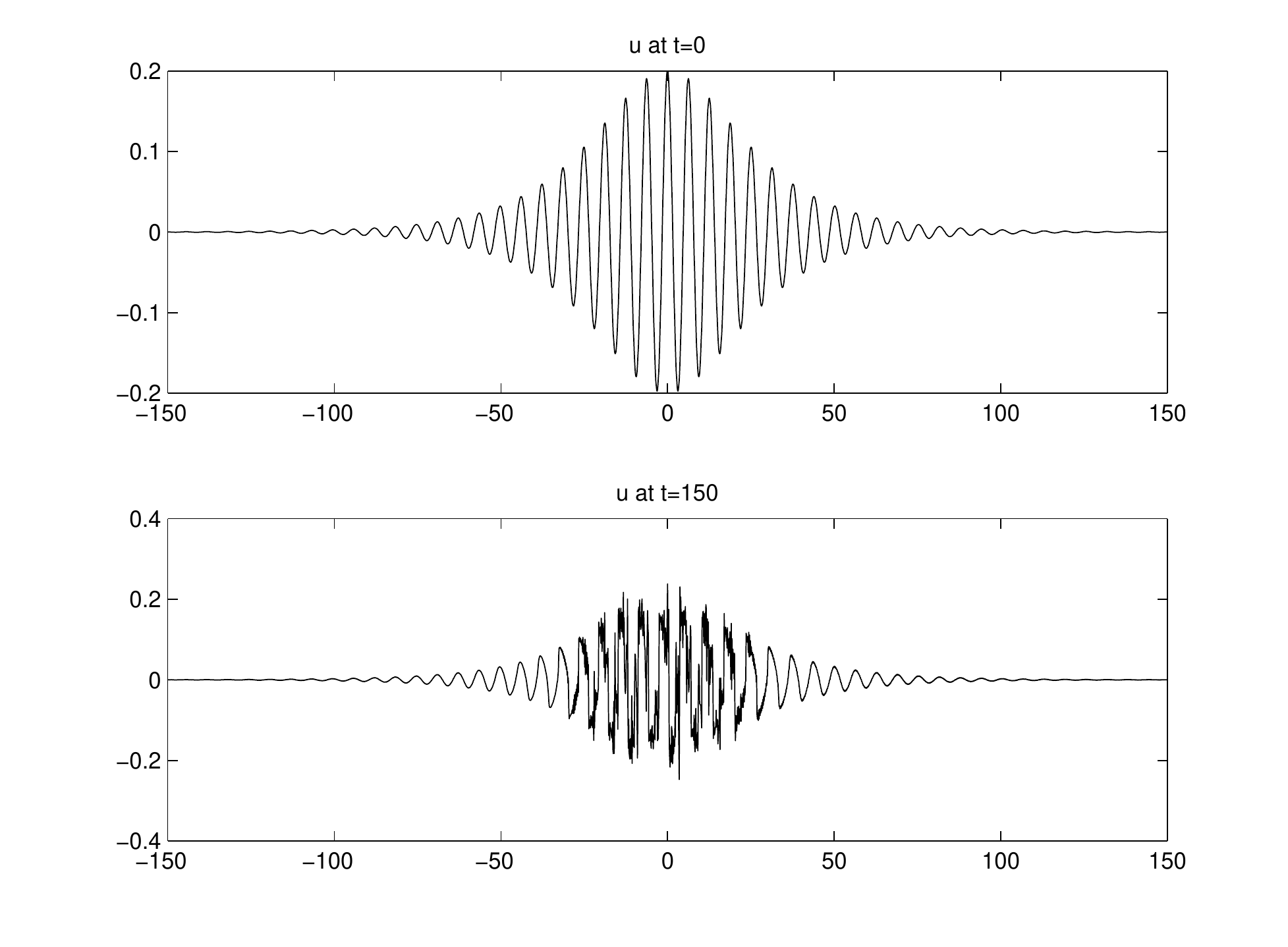}}
\caption{The propagation of the SPE solitary wave in the absence of dispersion.}
\label{solitonNoDispersion}
\end{center}
\end{figure} 
Figure \ref{solitonNoDispersion} shows the centralizing effect of the nonlinear term at the propagation distance $t=150$. The pulse does not move, and it becomes more concentrated in the absence of dispersion. If we allow the solitary wave to propagate further, it would blow up.

We have shown numerically that SPE solitons shows a stable propagation due to the key balance between dispersion and nonlinearity. This is a unique property of solitons. In the next section, we will simulate the particle-like behaviour of the SPE solitons as they collide.


\section{Colliding the SPE Solitons}

Solitons are localized excitations propagating in a system with a constant velocity. They behave like particles \cite{mollenauer-gordon:2006}. When there is a large seperation distance between two solitons, they essentially do not interact. Once we allow for the two solitons to move in opposite directions towards each other, each moves with a constant shape and velocity. As the solitons approach one another, their shapes begin to deform. The waves merge into one another. In the process, a wave packet is formed and as such, it cannot be represented as a linear combination of two solitons. This wave packet, however, soon splits into two solitons each with the same shape and velocity as before. Thereafter, the solitons move along in their respective directions as if nothing had happened. 

We have numerically validated this property of the SPE solitons by colliding the exact solitary waves of the SPE using an exponential time differencing (ETD) method-based algorithm. The numerical experiment validating the soliton interaction employs the two-soliton solution given by equations (\ref{spe_soln_Matsuno_MultiSoln}) and (\ref{spe_soln_Matsuno_MultiSoln_With}). We choose $N=4$ to derive the two-soliton solution of the SPE. The parameters determining the speed of each soliton are chosen such that $a_1=0.1$, $b_1=0.5$, $a_2=0.16$, and $b_2=0.8$. Notice that the condition of having a singular two-soliton solution (\ref{MultiSolitonCondition}) is satisfied. With these particular values of $a$ and $b$ parameters, the speed of the first soliton is $c_1=1.50$ and the speed of the other is $c_2=3.85$ units. 
\begin{figure}[htp]
\begin{center}
\scalebox{0.6}{\includegraphics{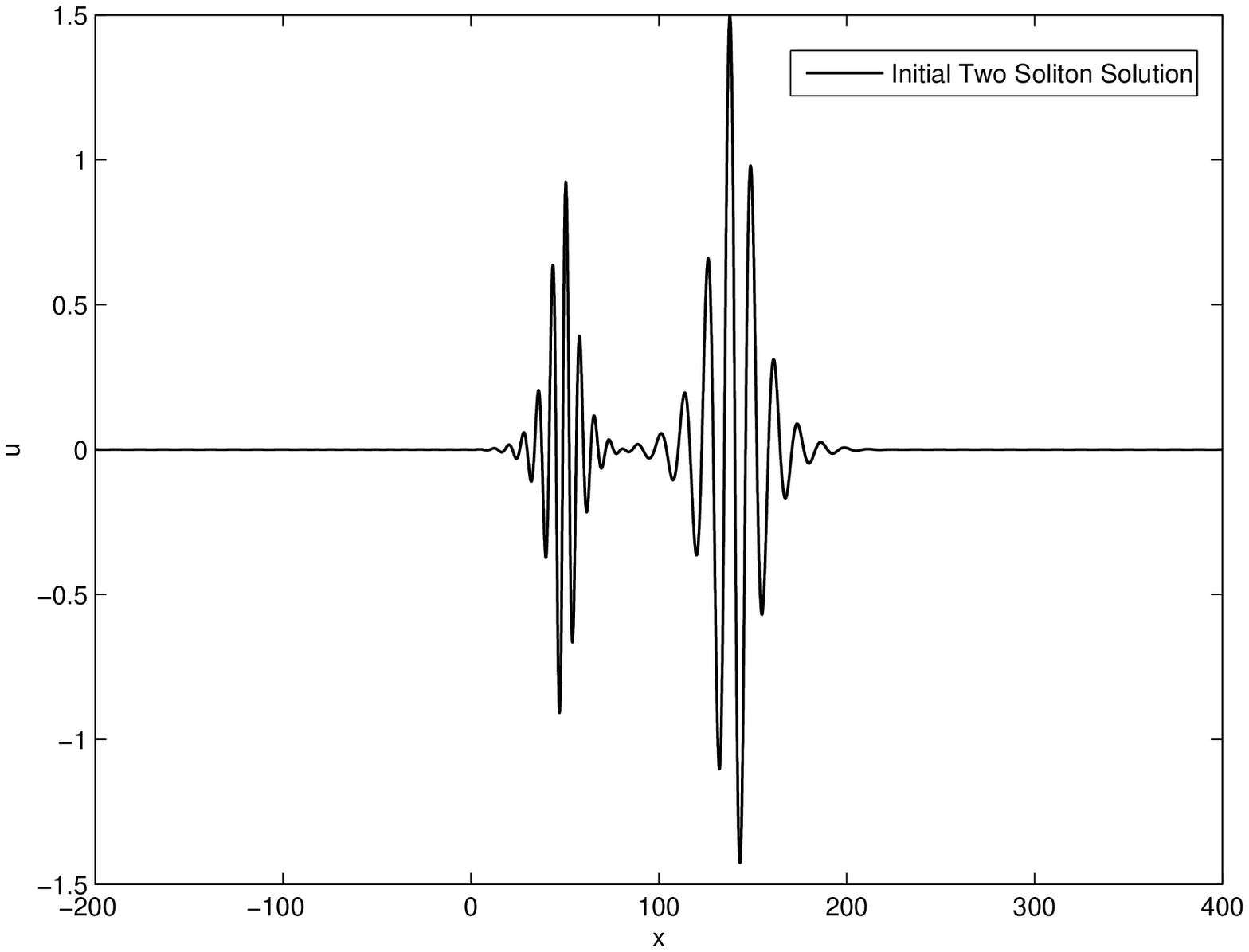}}
\scalebox{0.6}{\includegraphics{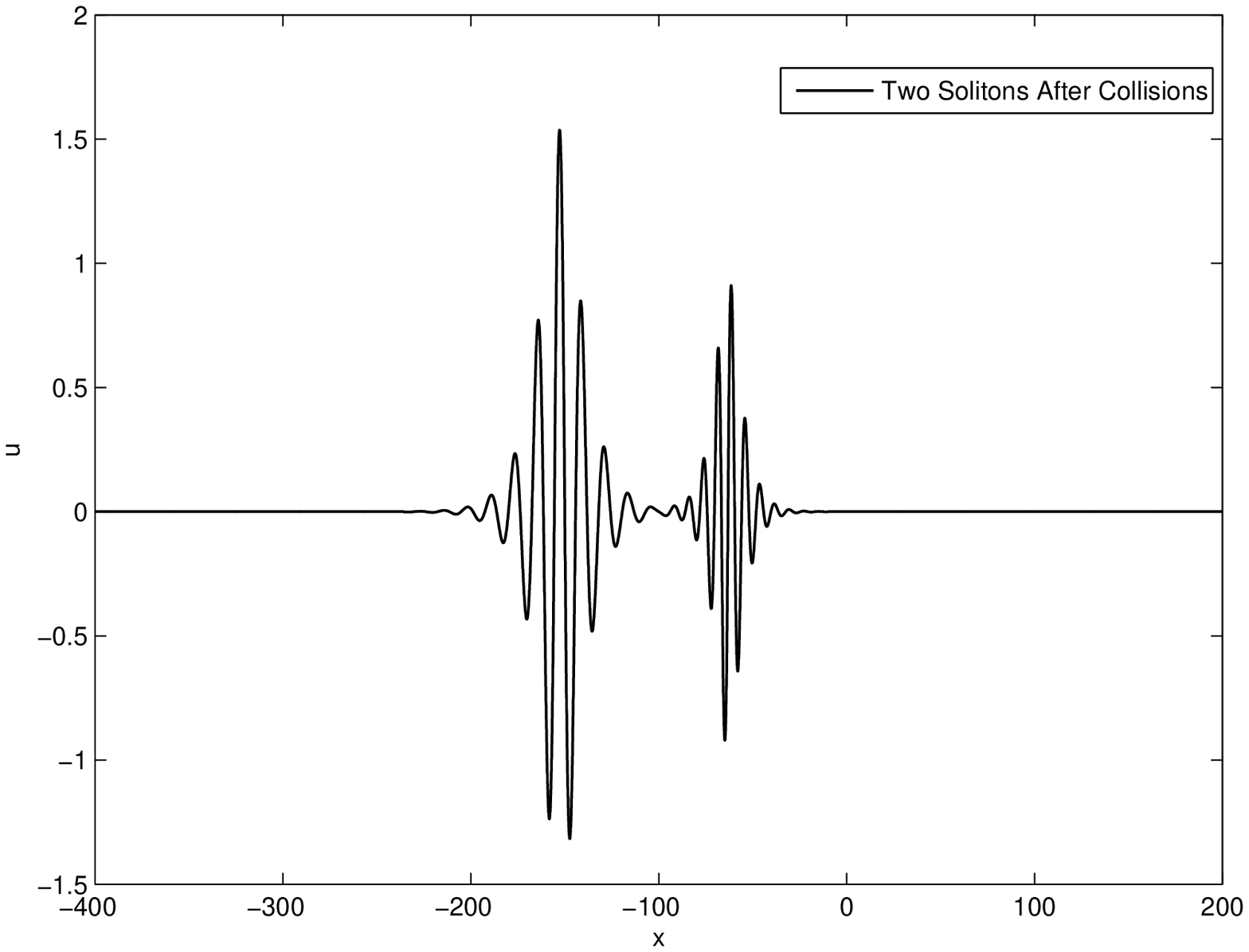}}
\caption{ Two SPE solitons are apart before the collision takes place (upper graph) and two SPE solitons after the collision occurs (bottom graph).}
\label{2solitonICAfter}
\end{center}
\end{figure} 
We show the initial two-soliton solution in figure \ref{2solitonICAfter}. The larger pulse is the one with the higher speed. As they propagate along the line, the faster soliton collides with the slower one ( the smaller of the two) and they merge into each other. They soon split apart into two pulses so that the one traveling with the higher speed is followed by the other as shown in figure \ref{2solitonICAfter}.

The next issue to concern ourselves with is to check whether or not the solitons remain the same after they collide. We then compare the solitons with the analytical result. 
\begin{figure}[htp]
\begin{center}
\scalebox{0.6}{\includegraphics{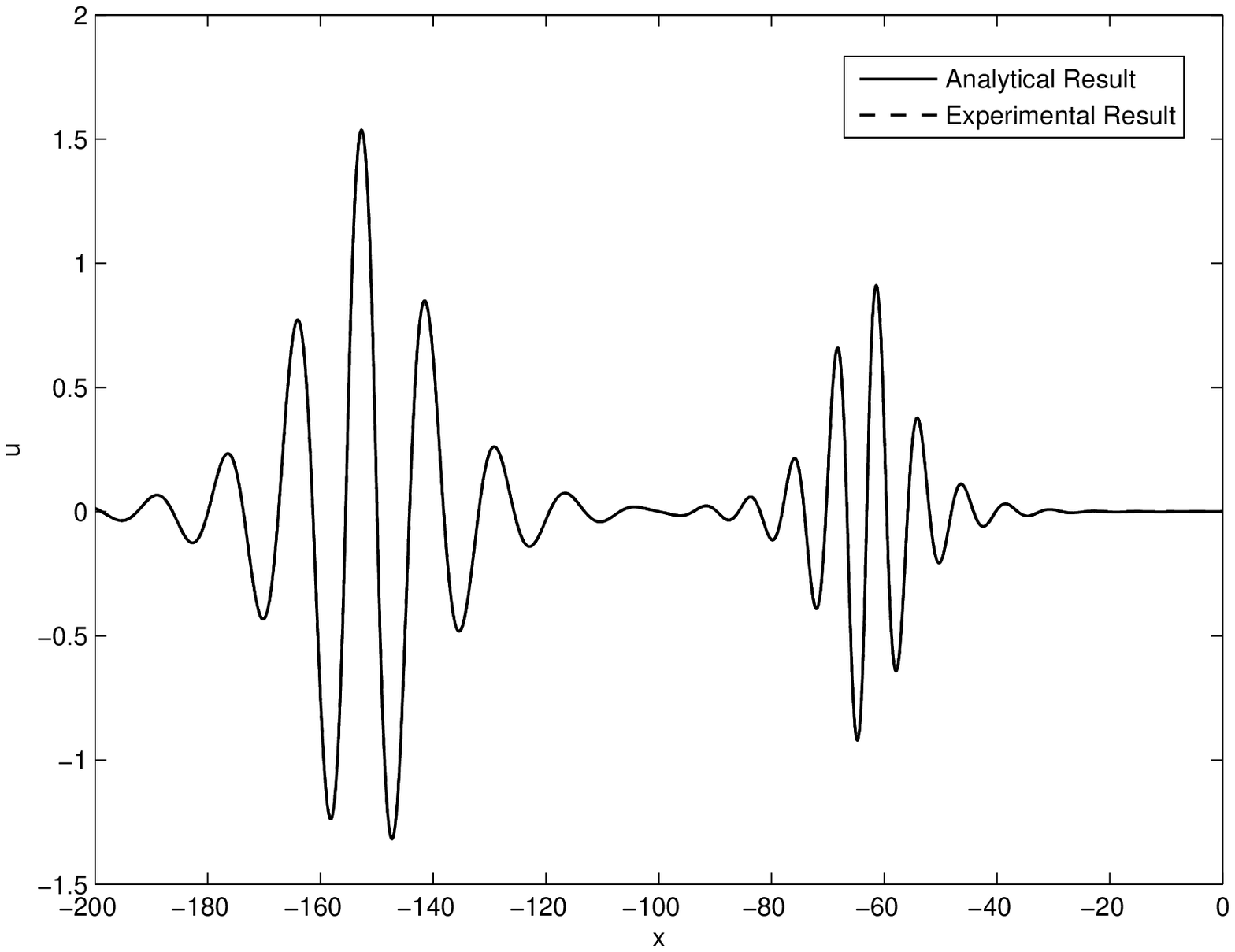}}
\scalebox{0.6}{\includegraphics{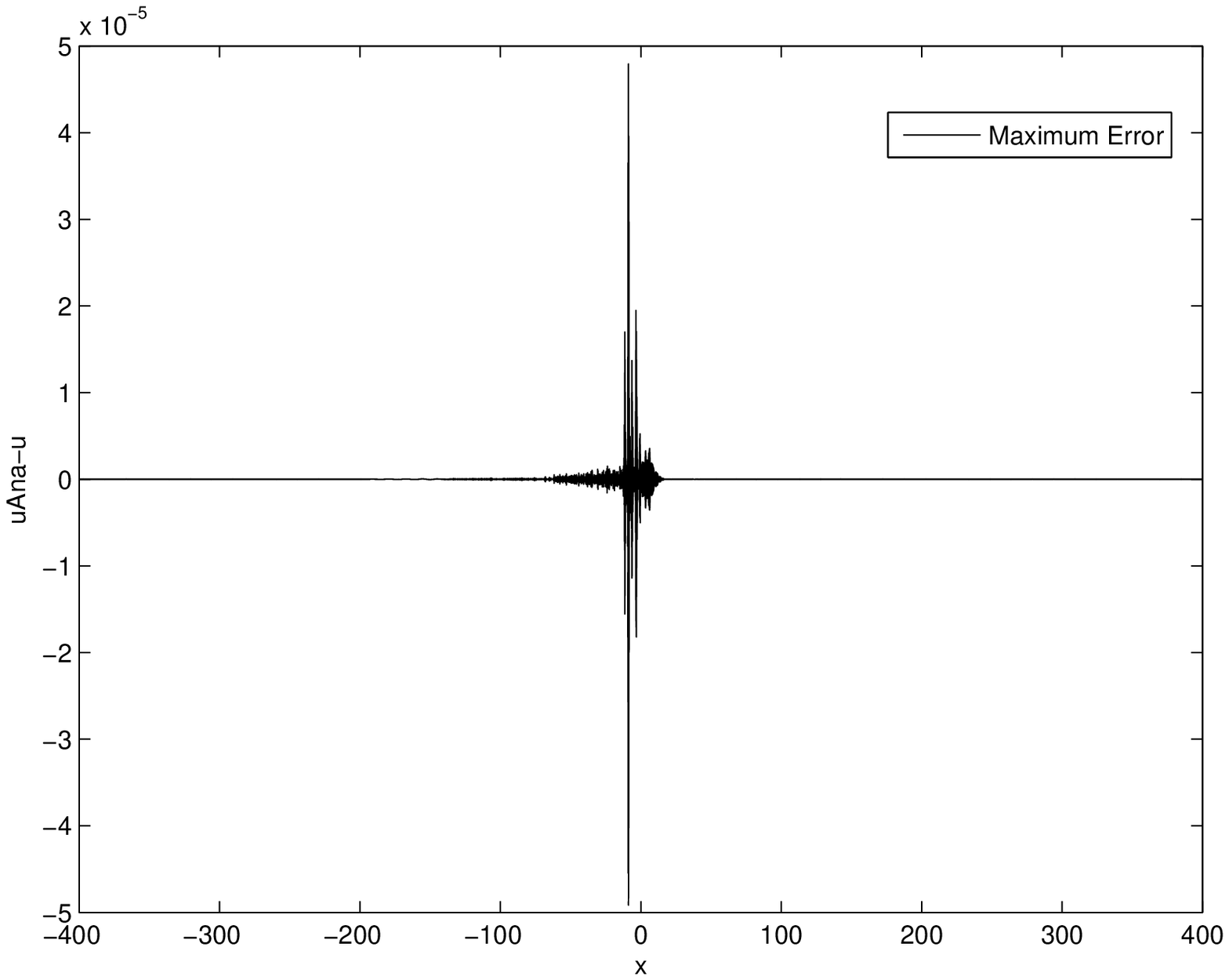}}
\caption{ Comparison of the exact solution of the SPE to the numerical solution after the collision occurs (upper graph) and maximum error between these two solutions (bottom graph)}
\label{2solitonComparison}
\end{center}
\end{figure} 

As we see in figure \ref{2solitonComparison}, we do not observe any discernable difference between the exact solution and the experimental result. We also show the maximum error between these two results in figure \ref{2solitonComparison} since it is almost impossible to distinguish one from the other. 

It may also be noteworthy to mention that one can validate the interaction picture using the nonlinear wave equation (\ref{maxwell_1d}). In that case, one can use the two one-soliton solutions of the form (\ref{spe_soln}) as an initial condition. If the sign in the propagation variable of the one-soliton solution is made minus, the soliton will be traveling in the opposite direction. Therefore, these two one solitary wave solutions can be used as the initial condition of the numerical schemes like two solitons moving towards each other. We have carried out the numerical experiment in colliding two one-soliton solutions by employing the Ablowitz-Ladik algorithm (see chapter seven), however, the results obtained are not included herein as they only reproduce the the particle-like property of the SPE solitons as previously shown.  

%% file: Chap6_StochasticSPEinc.tex
We will derive a stochastic version of the short pulse equation in this chapter. The impact of stochasticity on ultra-short pulse propagation and the comparison of the stochastic SPE with the stochastic nonlinear wave equation will be discussed via the numerical experiments.


\section{White Noise and Discrete Noise}

The physical system subject to random fluctuations are modelled by stochastic differential equations (SDEs). These stochastic equations may be either in the ordinary differential or partial differential form. In general, the random fluctuations in the STDs are repsented by white noise \cite{gardiner:1985}, which is a random process. There are, of course, other types of random processes such as jump processes. 

Let us consider a simple SDE
\begin{equation}\label{SimpleSTD}
\frac{dx}{dt}=a(x,t)+b(x,t)\xi(t) \,,
\end{equation} 
where $\xi(t)$ is a function of time representing the random fluctuations in the system, and $a(x,t)$ and $b(x,t)$ are some given functions depending on the variables $x$ and $t$. This equation is sometimes called the Langevin equation. The function $\xi(t)$ representing the random fluctuations is called the white noise. The derivative of the well-known Brownian motion or Wiener process $W(t)$ may be associated to the white noise such that
\begin{equation}\label{WienerProcessDer}
\frac{dW(t)}{dt}=\xi(t) \quad \mathrm{or} \quad dW(t)=\xi(t)dt \,.
\end{equation}
By employing the relation between the white noise and the Wiener process (\ref{WienerProcessDer}), we can write the SDE (\ref{SimpleSTD}) in the form
\begin{equation}\label{STD_wiener}
dx=a(x,t)dt+b(x,t)dW(t) \,.
\end{equation}
The treatment of the noise in a physical system can also be analyzed from the perspective of the Fokker-Planck equation, which is an equation of motion for the probability distrubution function \cite{risken:1989}. Therefore, the physical systems in which there is fluctuating random noise can be studied through stochastic differential equations or the corresponding Fokker-Planck equations. An ideal mathematical formulation of the white noise is a stochastic process with zero mean such that
\begin{equation}\label{WhiteNoise}
\begin{aligned}
\langle\xi(t)\rangle&=0 \\
\langle\xi(t^{'})\xi(t)\rangle&=\delta(t^{'}-t) \,,
\end{aligned}
\end{equation}
where the Dirac delta function is defined as 
\begin{equation}\label{DiracDelta}
\delta(t^{'}-t)= %
\begin{cases}
0 & \text{if } t^{'} \neq t \\
\infty & \text{if } t^{'}=t 
\end{cases}
\end{equation} 
White noise does not exist in the physical world. The noise in our system has actually a finite correlation. The delta function correlation of the white noise is just the idealization of realistic noise, and it is indeed a good representation of the physical noise in a mathematical sense \cite{gardiner:1985}. 

It may also be remarkable to emphazise that the Fourier transform of the white noise is a diffferent noise, but the noise distribution in the Fourier domain appears to be random noise as well \cite{owen:2007}. 

The origin of the noise in our physical system will be discussed in the next section. We will now show how the white noise as a mathematical representation of the physical noise can be discretized. This is particularly important when stochastic equations are studied numerically. We assume there exists a discrete approximation $\chi$ of the continuous white noise $\xi$ given by (\ref{WhiteNoise}). If the time interval is divided into $N$ intervals, we have a discrete random number for each interval. The collection of $N$ random numbers $\chi_1,\chi_2,...,\chi_N$ drawn at times $t_1=\Delta t,t_2=2\Delta t,...,t_N=t$  has a correlation such that
\begin{equation}\label{DiscreteWhiteNoise}
\langle \chi_i \chi_j \rangle = \sigma^2\delta_{ij}, \quad i=j=1,2,...,N \,,
\end{equation}
where $\sigma^2$ is the variance of the discrete noise and $\delta_{ij}$ is the Kronecker delta function such that
\begin{equation}\label{KroneckerDelta}
\delta_{ij}= %
\begin{cases}
0 & \text{if } i \neq j \\
1 & \text{if } i=j 
\end{cases}
\end{equation}
The Kronecker delta function (\ref{KroneckerDelta}) is just the discrete analog of the Dirac delta function (\ref{DiracDelta}). The variable $W(t)$ in equation (\ref{WienerProcessDer}) is the continuous Gaussian random variable and is the integration of the continous white noise
\begin{equation}\label{WienerProcess}
W(t)=\int_0^t \xi(t^{'})dt^{'}\,.
\end{equation}
The corresponding Gaussian random variable can be defined as
\begin{equation}
Y=\sum_{j=1}^N\chi_j\Delta t \,.
\end{equation}
Since the discrete noise is the approximation of the continuous noise, the statistical properties of the continuous process $W(t)$ and discrete process $Y$ must therefore match. The continuous process has a zero mean and a variance of $t$ as given by equations (\ref{WhiteNoise}) and (\ref{WienerProcess}). The mean of the sum of all random variables will be assumed to be zero. The variance of the discrete process may not be so evident. Hence,
\begin{equation}
Var(y)=(\Delta t)^2\sum_{j=1}^N\sigma^2=N\sigma^2(\Delta t)^2 \,, 
\end{equation}    
where we have used the relations (\ref{DiscreteWhiteNoise}) and (\ref{KroneckerDelta}). This is the variance of the discrete process and must be the same as the variance of the continuous process. Therefore,
\begin{equation}
N\sigma^2(\Delta t)^2=1\,.
\end{equation}
Because $N\Delta t= t$, the discrete variance
\begin{equation}
\sigma^2=\frac{1}{\Delta t}\,.
\end{equation}
If we choose the variance $1/\Delta t$ in the discrete case, we will match the statistical properties of the continuous process with a variance of $t$.
 
We have to emphasize that the strength of the noise is assumed to be one in this discussion (see the relation (\ref{WhiteNoise})). In the cases where the strength of the noise is not one, the variance of the discrete noise would be $\sigma^2=\nu/\Delta t$, where $\nu$ measures the strength of the noise.


\section{Derivation of the Stochastic SPE}

The stochastic pictures of the nonlinear models are a fundamental issue in nonlinear science. Random fluctuations are widely present in nature and the source of the stochasticity may not be spotted easily. Although the deterministic models may work well in many situations, the stochastic perturbations may have non-negligible effects in some cases. The propagation of ultra-short pulses in a nonlinear medium in which stochasticity is taken into account is a more realistic situation. This is the main motivation behind the attempt to find a stochastic model for ultra-short pulse dynamics. 

Optical soliton propagation in fibers in the presence of a stochastic perturbation has been studied in the context of the NLSE model. The nonlinear Schr\"odinger equation with a linear multiplicative stochastic term is a well known model for pulse propagation in nonlinear media that exhibit a stochastic nature. The sources of randomness in optical fibers vary. Stochasticity may cause the phase of the wave fluctuate. The small fluctuations in the pulse size or intensity can grow with propagation and may eventually lead to a pulse collapse. The inhomogeneities in a fiber's core, or the fluctuations in the linear refractive index of the core, are the major source of medium-related stochastic phase fluctuations. The possible sources of the phase fuctuations are stimulated Brillouin scattering, stimulated Raman scattering and medium inhomogeneities \cite{hart-judy-roy-beletic:1998,sulem-sulem:1999}.
The stochasticity may come from the nonlinearity of the medium as well. The dynamical effect of the noise added by the stochastic nature of the nonlinearity is nowhere comparable to the noise due to the inhomogeneities \cite{abdullaev-hensen-bischoff-sorensen:1998}. Nevertheless, there may be other sources of randomness playing a role, and they may originate from the other parts of the system such as the inherent power fluctuation in lasers used as input pumps.
Apart from these, quantum phase fuctuations are also sources of phase noise in optical fibers although they are practically negligible \cite{agrawal:2007}. 
In many aplications, Langevin noise (white noise) is used to study the fluctuations in the sytem \cite{boyd:1992}.

The deterministic equation (\ref{maxwell_1d}) leads to a model that describes ultra-short pulse dynamics in a deterministic way (equation (\ref{SPE_sak})). To derive a stochastic modeling equation for our system, one must first consider whether there exists any fluctuations in the system. As discussed previously, the main source of the noise appears in linear polarization of the medium, and fluctuations in nonlinearity are quite small if one considers the pulse propagation in the context of NLSE. Recall that the nonlinear part of the polarization is treated as a perturbation to the total polarization. The fluctuations in the perturbed term will be ignored here for the reason that already a small noise in a small nonlinear term does not play a significant role in the pulse dynamics. We claim the randomness in nonlinearity is much smaller than the noise in the dispersion term as in the case of NLSE and is insignificant. We can now argue that the linear polarization of the material in response to an applied electric field is not the same everywhere in a nonlinear medium, but fluctuating. Small fluctuations then appear in the dispersion term in equation (\ref{maxwell_1d}) allowing us to cast the deterministic Maxwell equation into a stochastic form. More rigorously, one can add a fluctuating term to the approximate value of the linear polarization in the Fourier domain. Let us introduce the noise in the dispersion term such that the fluctuating linear susceptibility in equation (\ref{linearSuscepApprox}) appears to be
\begin{equation}\label{linearSuscepApproxWithNoise}
\hat{\chi}^{(1)}(w)\approx\hat{\chi}^{(1)}(\lambda)=\hat{\chi}_0^{(1)} - (\hat{\chi}_2^{(1)}+\nu^{'} \hat{\chi}_{rand}) \lambda^2 \,,
\end{equation}
where $\hat{\chi}_{rand}$ represents the small noise in the Fourier domain and $\nu^{'}$ is the strength of the noise. Once we substitute the fluctuating linear susceptibility into equation (\ref{ln_wave_eqn_2}) and follow the same procedure we used to obtain the deterministic Maxwell equation (\ref{maxwell_1d}), the rescaled stochastic Maxwell equation can be written as 
\begin{equation}\label{maxwell_1d_stochastic}
u_{xx}=u_{tt}+(a+\nu \xi(x))u+b(u^3)_{tt} \,,
\end{equation}
where $\nu$ is the rescaled strength of the noise. This is the stochastic version of the nonlinear wave equation. The noise in the system is modelled as a white noise whose statistical properties are defined as 
\begin{equation}
\begin{aligned}
\langle\xi(x)\rangle &= 0 \\ 
\langle\xi(x) \xi(x^\prime)\rangle &= \delta(x-x^\prime)
\end{aligned}
\end{equation}
Since the nature of noise in many physical systems exhibits a normal distribution with zero mean, white noise offers a convenient mathematical implementation as understood by (\ref{maxwell_1d_stochastic_E}). Furthermore, the fact that fluctuations in the linear polarization are small and the average over these fluctuations is zero, averaging the stochastic wave equation (\ref{maxwell_1d_stochastic}) removes fluctuations from the system leading to the deterministic Maxwell equation (\ref{maxwell_1d}).    

We have now reached a point at which we can derive a stochastic short pulse equation. To make a distinction between deterministic and stochastic equations, we will make a notational change and replace the amplitude of the applied field $u(x,t)$ with $E(x,t)$. The stochastic wave equation, in terms of our new notation, is now expressed as  
\begin{equation}\label{maxwell_1d_stochastic_E}
E_{xx}=E_{tt}+(a+\nu \xi(x))E+ b(E^3)_{tt} \,.
\end{equation}
One can often obtain a more useful mathematical expression for a given system by introducing a fast scale as well as a slow scale. We now proceed to derive a stochastic SPE in a similar manner to the way in which we derived the deterministic SPE in chapter four. A multi-scale expansion of the form
\begin{equation}\label{E_expansion}
E(x,t)=\epsilon M_0(\phi,x_0,x_1,x_2,...)+\epsilon^2 M_1(\phi,x_0,x_1,x_2,...)+...
\end{equation}
with new scales
\begin{equation}\label{xt_expansion_Stochastic}
\phi=\frac{t-x}{\epsilon}, \qquad x_n=\epsilon^n x
\end{equation}
is used in the derivation of the stochastic SPE. Notice that $M_0(\phi,x_0,x_1,x_2,...)$ has a dependence on $x_0$, whereas the $A_0(\phi,x_1,x_2,...)$ term in expansion (\ref{u_expansion}) for the deterministic case has no dependence on the scale $x_0$. If $x_0$ had been incorporated in the $A_0$ function ($A_0=A_0(\phi,x_0,x_1,x_2,...)$), the same deterministic equation (the SPE) (\ref{SPE_original}) would have been derived. The reader's curiosity may demand to know what occurs if $x_0$ is removed from the expansion (\ref{E_expansion}) as in the deterministic case (see \ref{u_expansion}). We will soon see that this is not permissible in the stochastic case. To understand exactly why this is, we will proceed to fully carry out the derivation of the stochastic SPE. The first derivatives in the new scales take the form 
\begin{equation}\label{spe_NewScale_1stDer_Stochastic}
\begin{aligned}
\frac{\partial}{\partial x} &= \frac{\partial \phi}{\partial x} \frac{\partial }{\partial \phi} + \frac{\partial x_0}{\partial x} \frac{\partial }{\partial x_0} 
+ \frac{\partial x_1}{\partial x} \frac{\partial }{\partial x_1} \\
 &= - \frac{1}{\epsilon} \frac{\partial }{\partial \phi} + \frac{\partial }{\partial x_0} + \epsilon \frac{\partial }{\partial x_1} \\
\frac{\partial}{\partial t} &= \frac{\partial \phi}{\partial t} \frac{\partial }{\partial \phi} + \frac{\partial x_0}{\partial t} \frac{\partial }{\partial x_0} +  \frac{\partial x_1}{\partial t} \frac{\partial }{\partial x_1} \\
 &= \frac{1}{\epsilon} \frac{\partial }{\partial \phi} \,,
\end{aligned}
\end{equation}
where $\partial \phi/\partial x = - 1/\epsilon$, $\partial x_0/\partial x=1$, $\partial x_1/\partial x=\epsilon$, $\partial x_0/\partial t=\partial x_1/\partial t=0$ and $\partial \phi/\partial t=1/\epsilon$. Note that we keep terms in the expansion up to the order of $\epsilon$. The second derivatives can now easily be written as
\begin{equation}\label{spe_NewScale_2ndDer_Stochastic}
\begin{aligned}
\frac{\partial^2}{\partial x^2} &=\frac{\partial^2}{\partial x_0^2}+2\epsilon \frac{\partial^2}{\partial x_1 \partial x_0}-\frac{2}{\epsilon}\frac{\partial^2}{\partial x_0 \partial \phi} + \epsilon ^2 \frac{\partial^2}{\partial x_1^2} - 2\frac{\partial^2}{\partial x_1 \partial \phi} +\frac{1}{\epsilon^2}\frac{\partial^2}{\partial \phi^2} \\
\frac{\partial^2}{\partial t^2} &= \frac{1}{\epsilon^2} \frac{\partial^2}{\partial \phi^2}
\end{aligned} 
\end{equation}
If the expansion (\ref{E_expansion}) and the second derivatives (\ref{spe_NewScale_2ndDer_Stochastic}) 
are inserted into the stochastic Maxwell equation (\ref{maxwell_1d_stochastic_E}), we obtain 
\begin{equation}
\begin{aligned}  	
(\frac{\partial^2}{\partial x_0^2}+2\epsilon \frac{\partial^2}{\partial x_1 \partial x_0}-\frac{2}{\epsilon}\frac{\partial^2}{\partial x_0 \partial \phi} + \epsilon ^2 \frac{\partial^2}{\partial x_1^2} - 2\frac{\partial^2}{\partial x_1 \partial \phi} + \\
\frac{1}{\epsilon^2}\frac{\partial^2}{\partial \phi^2}) 
( \epsilon M_0+\epsilon^2 M_1+...) = 
\frac{1}{\epsilon^2}\frac{\partial^2}{\partial \phi} \left( \epsilon M_0+\epsilon^2 M_1+...\right) + \\
\left( a+\nu \xi(x) \right)\left( \epsilon M_0+\epsilon^2 M_1+...\right) + b\frac{1}{\epsilon^2}\frac{\partial ^2}{\partial \phi ^2}( \epsilon M_0+\epsilon^2 M_1+...)^3
\end{aligned}
\end{equation}
The terms up to $O(1/\epsilon)$ vanish. The terms of $O(1)$ yield
\begin{equation}
-2\frac{\partial^2 M_0}{\partial x_0 \partial \phi}=-2\frac{\partial}{\partial x_0}\left(\frac{\partial M_0}{\partial \phi}\right)=0\,.
\end{equation}
This implies $M_0$ is independent of $x_0$, i.e., $M_0=M_0(\phi,x_1,x_2,...)$ and the derivative of $M_0$ with respect to $x_0$ goes to zero whenever it appears in the higher order terms. By taking this into consideration, we are left with the equation in the order of $\epsilon$ 
\begin{equation}\label{stochasticSPE_solvabilityCond}
- 2\frac{\partial^2 M_1}{\partial x_0 \partial \phi} = 2\frac{\partial^2 M_0}{\partial x_1 \partial \phi} + \left( a+ \nu \xi(x) \right) M_0 + b \frac{\partial ^2 (M_0 ^3)}{\partial \phi^2} \,.
\end{equation}
This is the first non-trivial order. The solution of this equation requires knowledge of $M_0$. If we eliminate the noise from this equation by setting $\nu=0$, the solvability condition of this equation leads to the deterministic short pulse equation (\ref{SPE_original}), which we have derived in chapter four. 

The question then naturally arises as to how to obtain $M_0$ in the presence of white noise. We first divide the evolution variable $x_0$ into $n$ equal periods such that the $n$ periods are $0$ to $1$, $1$ to $2$, ..., $x_{n-1}$ to $x_n=x_0$. Let's now integrate equation (\ref{stochasticSPE_solvabilityCond}) with respect to $x_0$ from zero to one;
\begin{equation}
\int_0^1 - 2\frac{\partial^2 M_1}{\partial x_0 \partial \phi}\,dx_0 
= \int_0^1 \left[2\frac{\partial^2 M_0}{\partial x_1 \partial \phi} + ( a+ \nu \xi(x)) M_0 + b \frac{\partial ^2 (M_0 ^3)}{\partial \phi^2} \right]\,dx_0.
\end{equation}
It is easy to differentiate the right-hand side because $M_0$ is independent of $x_0$. Carrying out the integration in both sides yields 
\begin{equation}  
\begin{aligned}
-2[\frac{\partial M_1(\phi,x_0=1,x_1,...)}{\partial \phi} - \frac{\partial M_1(\phi,x_0=0,x_1,...)}{\partial \phi}] = \\
[2\frac{\partial^2 M_0}{\partial x_1 \partial \phi}  + ( a+ \nu (\int_0^1\xi(x)\,dx_0)) M_0 + b \frac{\partial ^2 (M_0 ^3)}{\partial \phi^2}] \int_0^1 dx_0 \,.
\end{aligned}
\end{equation}
Note that changing the limits of integration from zero to another $x_0$ (other than one) does not make any difference in our discussion. We require the left-hand side to be zero because $\partial M_1/\partial \phi$ grows unbounded with time. In other words, the term in the left-hand side of equation (\ref{stochasticSPE_solvabilityCond}) is a secular term, and as such it must be removed. As a result, we pin down $\partial M_1/\partial \phi$ to zero at the end points of each period in order to avoid any growth of the function $M_1$. Notice that we can just set the left-hand side of equation (\ref{stochasticSPE_solvabilityCond}) to zero without first integrating the equation. It must be emphasized that the integration of the noise term is to be handled carefully. When we integrate equation (\ref{stochasticSPE_solvabilityCond}) from zero to one (over the first period), we draw a random number $\zeta_1 = \int_0^1\xi(x)\,dx_0 $. In a similar manner, we integrate equation (\ref{stochasticSPE_solvabilityCond}) from one to two (over the second period), kill the growth of $M_1$ and draw another random number  $\zeta_2 = \int_1^2\xi(x)\,dx_0 $. If this process is repeated $n$ times, we obtain $n$ random numbers such that $(\zeta_n)=(\zeta_1,\zeta_2,\zeta_3,...,\zeta_n)$. The collection of these random numbers produce a normal distribution as is the case with white noise. However, white noise is a continuous distribution, whereas the collection of these numbers is a discrete one. If we follow up on the discussion of the discrete approximation of continuous noise mentioned in the previous section, we obtain \cite{kurt-schaefer:2010}        
\begin{equation}\label{StochasticSPEoriginal}
- 2(M_0)_{x_1 \phi} = (a+ \nu \Xi(x_1))M_0+ b(M_0 ^3)_{\phi \phi} \,.
\end{equation}
This is the stochastic version of the short pulse equation corresponding to the deterministic form (\ref{SPE_original}). If the transformation (\ref{spe_trans}) ($A_0$ is replaced with $M_0$) is applied to equation (\ref{StochasticSPEoriginal}), the stochastic equation corresponding to the deterministic equation of the form (\ref{SPE_sak}) can be obtained as 
\begin{equation}\label{StochasticSPE_sak}
U_{XT}=(1+ \frac{\nu}{2} \Xi(T))U+\frac{1}{6}(U^3)_{XX} \,,
\end{equation} 
where we assume $\Xi(T)=\Xi(-T)$. We address this form of the stochastic short pulse equation as the stochastic SPE. It is the governing equation for soliton propagation in the stochastic environment.

It is noteworthy to mention that the strength of the slow-scale noise $\Xi(x_1)$ is governed by the expansion parameter $\epsilon$ as well as $\nu$, whereas the strength of the fast noise $\xi(x)$ is controlled by $\nu$. Since the leading order amplitude $M_0$ is independent of $x_0$ and the randomness is only dispersive, the strength of the slow noise rather takes a simple form. In cases where the leading order may be more complicated, the strength of coarse-graining noise (slow scale noise) requires a more careful treatment \cite{schaefer-moore:2007}.


\section{Numerical Analysis of the Stochastic SPE}

It is not a rhetorical question to ask how the ultra-short solitons of the deterministic short pulse equation propagate in a world that is not perfect.
We have already derived an equation to model the ultra-short pulse propagation in this imperfect world. The origin of the noise and randomness has already been mentioned in the previous section. However, we have not tested how these pulses evolve in a stochastic environment. In this section, we experiment with the pulse propagation in such an environment, and show some of our numerical results that analyze the evolution of the SPE solitons via the stochastic SPE. By doing so, we use the exact solution of the deterministic SPE as an initial condition and let it propagate in the stochastic short pulse equation.
\begin{figure}[htp]
\begin{center}
\scalebox{0.7}{\includegraphics{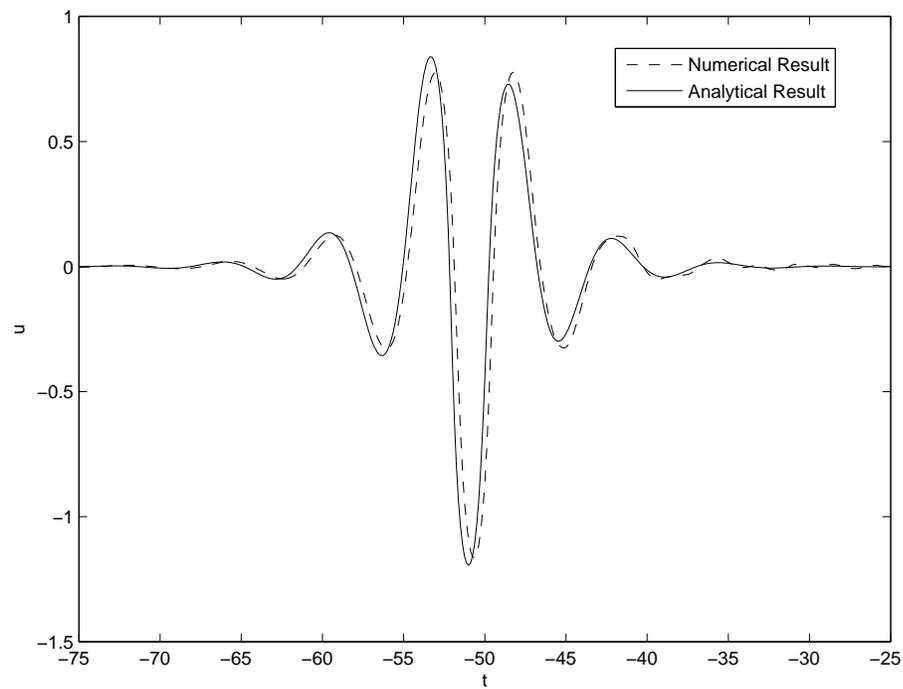}}
\caption{ Evolution of the SPE soliton in the stochastic SPE.}
\label{StochasticSPEpropagation51}
\end{center}
\end{figure} 
Figure \ref{StochasticSPEpropagation51} shows the evolution of the SPE soliton at $t=51.2$ units. The dashed line is the solution of the stochastic SPE at $t=51.2$, and the solid line is the analytical result at the same distance. The noise strength in this experiment is chosen as $\nu=0.05$. We set the soliton parameter $m=0.3$ and the expansion coefficient $\epsilon=0.2$ in a scheme that employs the semi-implicit method (to be discussed in greater detail in the succeding chapter). The noise is generated through the Matlab's random generator $rand$ in a normalized way.
Although there is a small change in the shape of the soliton, it propagates stably in the stochastic SPE. The comparison of the solution of the stochastic equation and the exact result at $t=51.2$ indicates that the effect of random dispersion on the ultra-short pulses is not strong . However, it is known that when the NLSE is used as a modeling equation for very short pulses in optical fibers, the random dispersion is quite strong \cite{wolf:1994}, and in that case the modified NLSE must be used. On the other hand, the SPE solitons persist in the stochastic short pulse equation in the presence of noise. As long as the noise in the physical system does not embody large scale fluctuations, the SPE solitons undergo a stable propagation in a stochastic environment as confirmed by our numerical results.   
\begin{figure}[htp]
\begin{center}
\scalebox{0.7}{\includegraphics{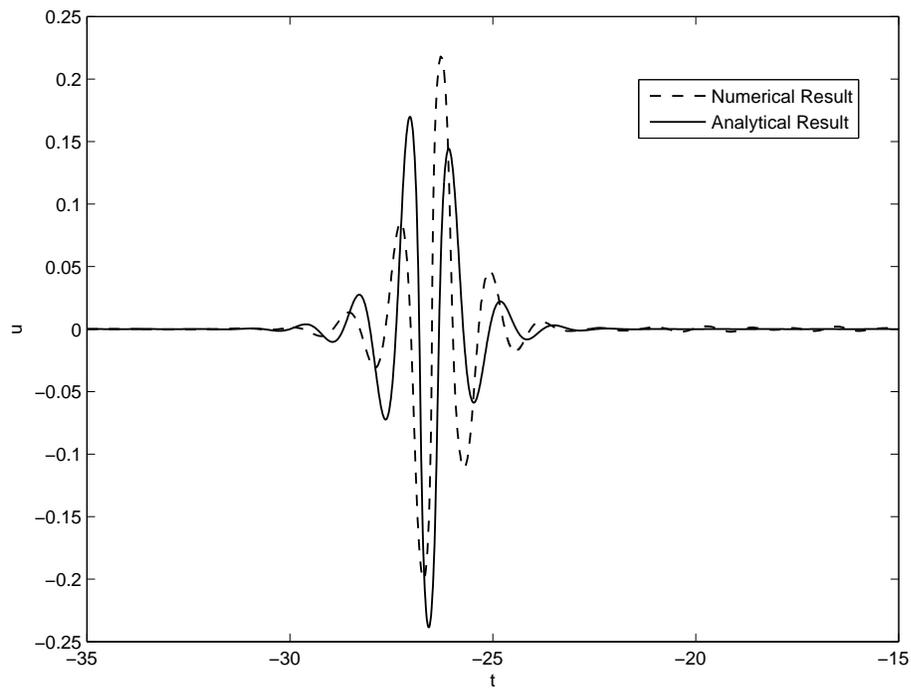}}
\caption{ Evolution of the SPE soliton in the stochastic Maxwell equation.}
\label{StochasticMaxwellpropagation51}
\end{center}
\end{figure} 

In analogy with the deterministic case, we are interested in observing how the SPE solitons evolve in the stochastic nonlinear wave equation if they are used as initial conditions. Figure \ref{StochasticMaxwellpropagation51} displays the numerical result and its comparison to the exact result at the propagation distance $t=25.6$ units. The dashed line is the numerical solution of the stochastic Maxwell equation, and the solid line is the exact solution without noise at $t=25.6$. We use the Ablowitz-Ladik scheme in this experiment. The random numbers incorporated in our experiments are generated similar to the way in which we generate them for the stochastic SPE equation (refer to chapter seven). The noise strength $\nu$, the soliton parameter $m$ and the expansion parameter $\epsilon$ are set to $0.05$, $0.3$ and $0.2$ respectively \cite{kurt-schaefer:2010}. Note again that the initial condition and the exact result are modified, respectively, as  $u_{initial} = \epsilon \,u(x/\epsilon,0)$ and  $\mathrm{u_{analytical}} = \epsilon \, \mathrm{u}\left(\right[(x-25.6)/\epsilon\left],-\epsilon \, 25.6\right)$ as imposed by the multiple scale expansion in the stochastic case like had been shown in the deterministic case. With the appropriate choice of initial condition, we observe in figure \ref{StochasticMaxwellpropagation51} a stable propagation of the solitons  in the stochastic environment through the nonlinear wave equation as well. The comparison of the exact and numerical results clearly show that the impact of the noise somehow affects the soliton as it propagates, but these solitons persist in the stochastic nonlinear wave equation despite the presence of the of noise. On the other hand, it seems as though the impact of the noise on the solitons is more significant in the case of numerical solutions to the stochastic Maxwell equation than those for the  stochastic SPE, which is used to model soliton propagation in the stochastic environment. In the next section, we will dicuss such a significance numerically and qualitatively.


\section{Comparison of the Stochastic SPE and Stochastic Maxwell Equation}

The stochastic SPE (\ref{StochasticSPE_sak}) (or alternatively (\ref{StochasticSPEoriginal})) is expected to be a good approximation of the stochastic Maxwell equation (\ref{maxwell_1d_stochastic_E}) up to at least $O(1/\epsilon)$. We have already shown that the SPE solitary waves propagate in the deterministic nonlinear wave equation to a distance of $O(1/\epsilon^2)\approx 26$ (much larger than the leading order distance) with very good agreement (see figure \ref{MaxwellSPEcomparison}). The error accumulation versus propagation distance was also shown in figure \ref{LNorms}. We now ask how much error is accumulated if we allow the SPE solitons to propagate in the stochastic Maxwell equation.    
\begin{figure}[htp]
\begin{center}
\scalebox{0.65}{\includegraphics{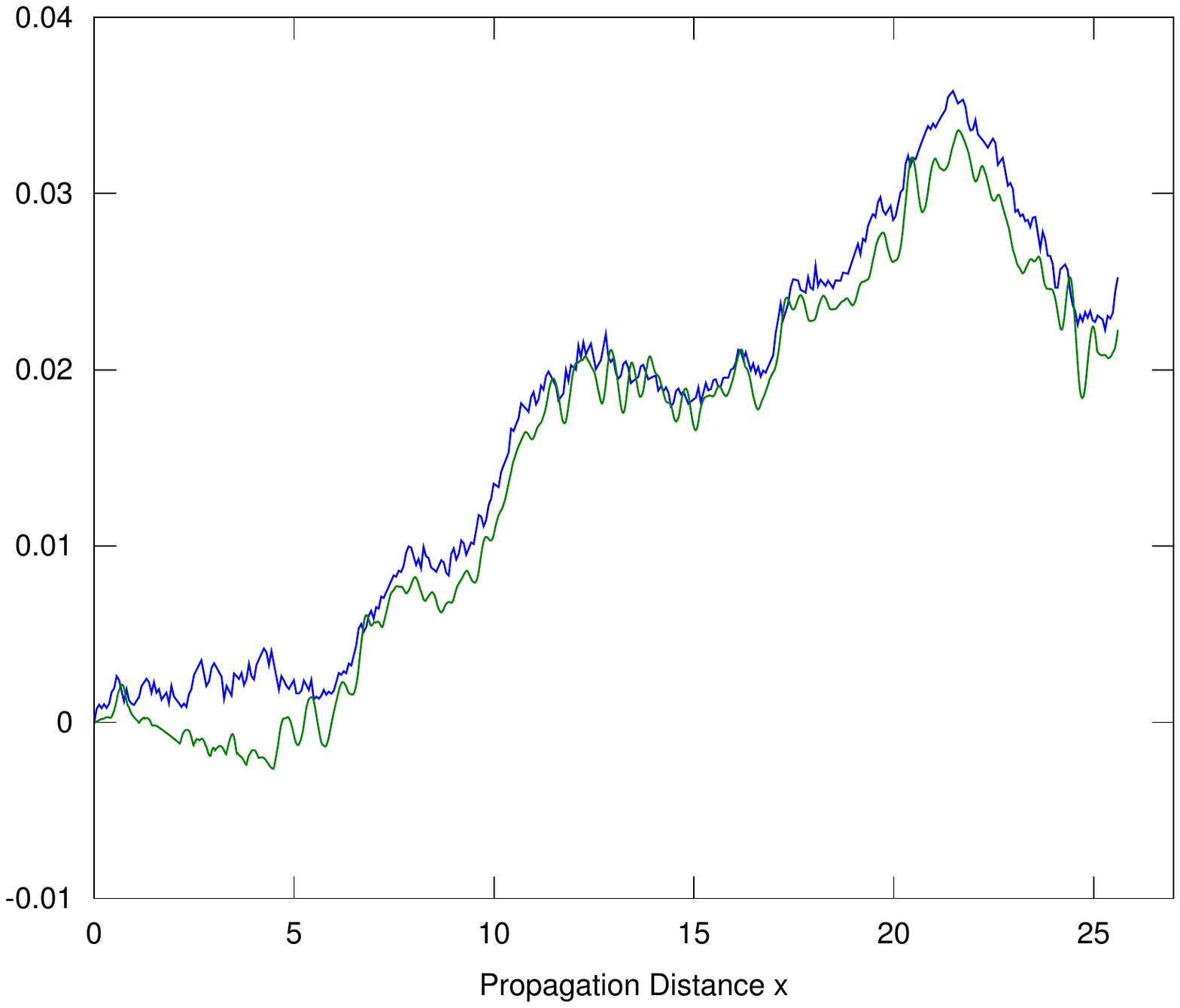}}
\caption{The growth of the deviations for the evolution of the SPE soliton in the stochastic SPE and stochastic Maxwell equation as defined by $L^{\infty}$ norms.} 
\label{figNormDifferenceStoch}
\end{center}
\end{figure} 
Figure \ref{figNormDifferenceStoch} demonstrates the error accumulations between the exact result (deterministic SPE solitons) and the results of the stochastic SPE and the Maxwell equation for one realization of the noise. For the simulation of the Maxwell equation, the error in the deterministic case is subtracted from the error in the stochastic case. The blue line shows the $L^{\infty}$-norm generated for the difference of the exact SPE solitons and the solution of the stochastic Maxwell equation with the deterministic error ($L^{\infty}$-norm shown in figure \ref{LNorms}) removed, whereas the green line is the $L^{\infty}$-norm for the difference of the exact SPE solitons and the solution of the stochastic SPE. We observe an excellent agreement between the fast scale noise and the slow scale noise for this particular realization.  

Since we obtained figure \ref{StochasticSPEpropagation51} and figure \ref{StochasticMaxwellpropagation51} for one realization, repeating the experiments may result in a change in the shape of the pulses for other realizations due to the randomly distributed nature of the white noise in the system. Therefore, the error accumulation in a stochastic environment for one realization may be meaningless. To compare the statistical properties of the fast noise and the slow noise, we have to draw many realizations and see whether or not the statistical properties are in good agreement.

The path-wise correspondence in figure \ref{figNormDifferenceStoch} indicates that the related probability distributions must be in agreement with one another as well. 
\begin{figure}[htp]
\begin{center}
\scalebox{0.6}{\includegraphics{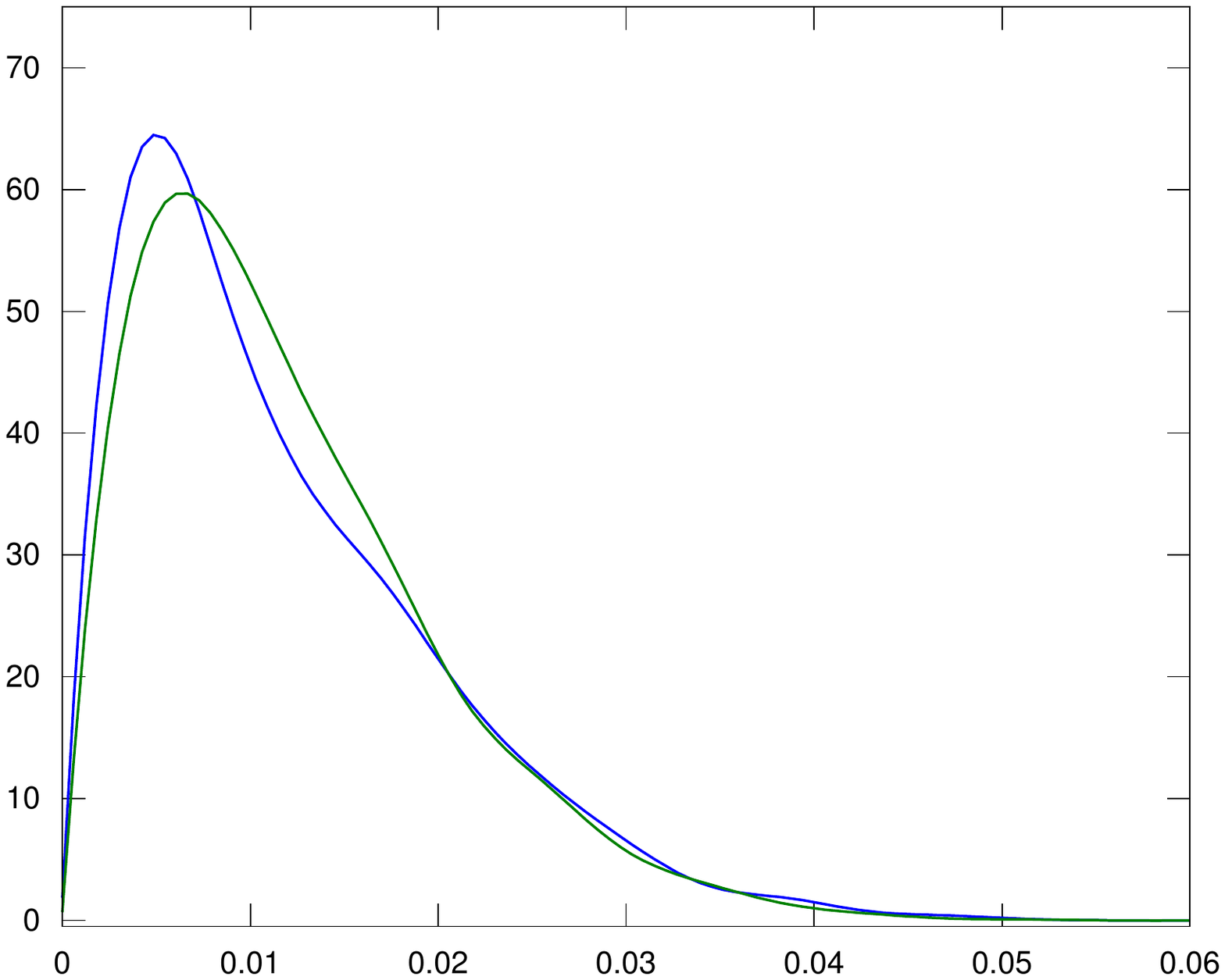}}
\caption{Probability distribution of the deviations of the stochastic results from the deterministic results versus the $L^{\infty}$-norm of the difference between the stochastic PDEs and the exact SPE solitons at the propagation distance $x=25.6$ units}
\label{figStatistics}
\end{center}
\end{figure} 
In figure \ref{figStatistics}, we plot the probability distribution versus the $L^{\infty}$-norm of the difference between the stochastic PDEs and the exact SPE solitons at the propagation distance $x=25.6$ units. The blue line is the probability distribution of the Maxwell equation and the green line is the probability distribution of the stochastic SPE. The plots are obtained for $10,000$ realizations of the stochastic SPE and the stochastic Maxwell equation by joining the midpoints of the histograms of the deviations of the solutions to the stochastic PDEs from the deterministic evolution of the SPE solitons. The probability distributions match to very good accuracy as seen in figure \ref{figStatistics}, and this is indicative of a correlation between the noise over slow scales as well as fast scales in accordance with the discrete-noise approximation, coarse-graining noise and multiple scaling expansion (see chapter seven for more details). The tail of each distribution curve reflects the fact that any deviation between the numerical results and the exact result at $x=25.6$, which happens to be greater than $0.02$, becomes less probable \cite{kurt-schaefer:2010}.        

%% file: Chap7_NumericalMethodsinc.tex
It is not rare to come across a problem whose analytical solution may either not exist or  be too tedious to obtain by hand in applied sciences and mathematics. Numerical analysis is the branch of mathematics and applied sciences that is used to find approximations to such difficult problems. These problems may include finding the roots of non-linear equations, solving differential equations, complex integration, numerical differentiation, Fourier analysis, finite differences and so on \cite{mathews-fink:1998,hunt-lipsman-rosenberg:2001}. One can, on the other hand, use numerics as a tool to validate the analytical answers of mathematical objects and equations. In this sense, numerical methods are the experiments, and computers are the laboratories of mathematics. We are fortunate enough that there are a variety of programing languages available to those which deal with such experiments in applied fields \cite{pao:1999}. The software we have used extensively in the preparation of this thesis is Matlab. Matlab is a high-level technical computing language and interactive environment for algorithm development, data visualization, data analysis, and numeric computation \cite{mathworks:online}. One can either use Matlab's built-in-commands to perform tasks or write their own codes to execute \cite{hunt-lipsman-rosenberg:2001}. Although compiled languages such as C, C++, and Fortran are the most efficient options in terms of execution speed, the most important advantage of Matlab is that it is a very simple, yet powerful, programming language. 

A number of different numerical schemes is widely available for solving numerical problems. In the effort of understanding the short pulse dynamics, we have used the schemes employing Euler's Method, Runge-Kutta Method, the Semi-Implicit Method, Ablowiz-Ladik Scheme and the exponential time differencing method . Each scheme has its advantages and disadvantages in terms of numerical stability and computational errors. We shall discuss, in this chapter, the numerical tools we have used in our codes and the application of each numerical method to the Maxwell and short pulse equations in both the deterministic and stochastic cases.


\section{Fourier Transform and FFT Algorithm}

The Fourier transform is a powerful technique that applies to a wide variety of problems in mathematics and applied sciences, and therefore, is widely used tool in numerical calculations as well. We discuss the details of the Fourier analysis, and how it is implemented in our numerical schemes in this section. 

The French mathematician, Fourier, found that any periodic waveform can be expressed as a series of harmonically related sinusoids, i.e., sine and cosine waves, whose frequencies are multiples of its fundamental frequency or the first harmonic. The basic idea behind this technique is to look at the problem from a different perspective. The periodic function is first transformed to a new space, called the Fourier space, in which it is represented as the sum of the sines and cosines. The manipulation of the function may be carried out with the new look of the function in the Fourier space,  and a solution is sought. The Fourier transform is reversible and one can always transform the function back to the original space via the inverse Fourier transform once the mathematical operation(s) in the Fourier domain is done. 

Mathematically speaking, a periodic function $f(x)$ can be expressed as a series of the sines and cosines \cite{arfken-weber:1995,mathews-walker:1971} as
\begin{equation}
f(x)=\frac{1}{2}a_0+\sum\limits_{i=0}^{n=\infty} a_n cos(nx) + \displaystyle\sum\limits_{i=0}^{n=\infty} b_n sin(nx) \,,
\end{equation}
where
\begin{eqnarray}
a_0 &=& \frac{1}{\pi}\int_{-\pi}^{\pi} f(x) dx \nonumber \\
a_n &=& \frac{1}{\pi}\int_{-\pi}^{\pi} f(x) cos(nx) dx \nonumber \\
b_n &=& \frac{1}{\pi}\int_{-\pi}^{\pi} f(x) sin(nx) dx \nonumber 
\end{eqnarray}
and $n=0,1,2,3,...$. The first term $a_0/2$ in the series  is a constant and represents the average component of the function $f(x)$. The terms with the coefficients $a_1$ and $b_1$ in the series represent the fundamental frequency component w. Likewise, the terms with the coefficients $a_2$ and $b_2$ represent the second harmonic component 2w, and so on. If the periodic function $f(x)$ has the even symmetry, or in other words, if it is an even function, i.e., $f(-x)=f(x)$, the series consists only of the cosine terms with zero or nonzero $a_0$. If it has the odd symmetry, that is, if it is an odd function ($f(-x)=-f(x)$), the series includes only the sine terms. 

The Fourier series can be generalized to the complex numbers, and further generalized to derive the Fourier transform. We will not show the derivation of the Fourier transform here, but we will only give the definitions. The Fourier transform and the inverse Fourier transform are defined respectively as 
\begin{equation}
\begin{aligned}
F(w)&=\int_{-\infty}^{\infty} e^{iwx}f(x)dx \\
f(x)&=\int_{-\infty}^{\infty} e^{-iwx}F(k)dk \,,
\end{aligned}
\end{equation}
where $w=2\pi k$ is the angular frequency, $k$ is the Fourier frequency and $i=\sqrt{-1}$\,. The Fourier transform maps a time series into the series of frequencies of the amplitudes and phases. The inverse Fourier transform maps the series of frequencies (their amplitudes and phases) back into the time series. The derivatives of the periodic functions can be also be written in the Fourier space. Let us show the Fourier transform of the first derivative as an example
\begin{equation}\label{Fourier_1stDer}
\begin{aligned}
F^\prime(w)&=\widehat{f^\prime(x)}= \int_{-\infty}^{\infty} e^{iwx}\widehat{f^\prime(x)}dx \\
&=f(x)e^{iwx} \Big |_{-\infty}^{\infty} -iw\int_{-\infty}^{\infty} e^{iwx}f(x)dx \,.
\end{aligned}
\end{equation}
Since $f(x) \rightarrow 0$ as $x \rightarrow \pm \infty $, equation (\ref{Fourier_1stDer}) becomes
\begin{equation}
F^\prime(w)=-iwF(w)\,.
\end{equation} 
We can generalize to the nth derivative;
\begin{equation}\label{Fourier_nthDer}
F^{n}(w)=(-iw)^nF(w)
\end{equation} 
Note also that the condition for the Fourier transform to be applied is that $f(x)\rightarrow$0 as $x\rightarrow \pm \infty$, and $f(x)$ must be a periodic function. 

There is also the discrete counterpart of the Fourier transform \cite{moler:2004,arfken-weber:1995} and is called the discrete Fourier transform (DFT). The DFT can be turned into the numerical language easily and efficiently. Suppose we truncate the Fourier series at the $Nth$ term and use the $N$ number of harmonics;
\begin{equation}\label{DFT}
F_n=\displaystyle\sum\limits_{k=0}^{N-1} f_k e^{2\pi ink/N}, \qquad n=0,1,...,N-1 
\end{equation}
This is the forward discrete fourier transform equation. The complex numbers
$f_0,f_1,...,f_N$ are transformed into the complex numbers $F_0,F_1,..., F_n$. The backward formula, which is called inverse discrete fourier transform (IDFT), can be written as
\begin{equation}\label{IDFT}
f_k=\frac{1}{N}\displaystyle\sum\limits_{n=0}^{N-1} F_n e^{-2\pi ikn/N}, \qquad n=0,1,...,N-1 
\end{equation}
The complex numbers $F_0,F_1,...,F_n$ are transformed into complex numbers
$f_0,f_1,...,f_N$ by IDFT. These two formulae (\ref{DFT}) and (\ref{IDFT}) are the basis of computer algorithms for the fourier analysis. 

The DFTs and IDFTs are computed using the so-called fast fourier transform (FFT) algorithms in modern numerical applications \cite{yang-cao-chung-morris:2005}. The FFT algorithms have been discovered independently by several researchers, and many people have since contributed to the development of the FFT schemes. The starting point for the modern usage of the FFT dates back to the seminal paper published by John Tukey of Princeton University and John Cooley of IBM Research in 1965. An FFT algorithm re-expresses the DFT of a size $N = N_1N_2$ in terms of the smaller DFTs of sizes $N_1$ and $N_2$ recursively to reduce the computation time. The FFT is a very efficient technique and replaces the DFT mainly because of two reasons such that the FFT generates very accurate results and is a quite fast algorithm. The computational time for the DFT algorithms is proportional to $N^2$, where N being the number of discretized points. On the other hand, the FFT of a size $N$ does $Nlog_2(N)$ number of operations to carry out the Fourier transform \cite{moler:2004}. For example, if the number of data points is $1000$, i.e., $N=1000$, then the algorithm requires $1,000,000$ operations to take the Fourier transform in case of using the DTF. However, an FFT based-algorithm would do approximately $10,000$ operations. This means that a hundred times less operations done by the FFT algorithm. This is an enormous computational cost saving. 

The derivation of the fast algorithm FFT starts with the definition of the discrete Fourier transform such that 
\begin{eqnarray}
F_n &=& \displaystyle\sum\limits_{k=0}^{N-1} f_k e^{2\pi ink/N} \nonumber \\
    &=& \displaystyle\sum\limits_{k=0}^{N/2-1} f_{2k} e^{2\pi in(2k)/N} + \displaystyle\sum\limits_{k=0}^{N/2-1} f_{2k+1} e^{2\pi in(2k+1)/N} \nonumber \\
    &=& \displaystyle\sum\limits_{k=0}^{N/2-1} f_{2k} e^{2\pi ink/(N/2)} + e^{2\pi in/N} \displaystyle\sum\limits_{k=0}^{N/2-1} f_{2k+1} e^{2\pi ink/(N/2)}\label{FFT_algorithm} 
\end{eqnarray}
This is the FFT algorithm. Notice that $n$ runs from $0$ to $N$, not just to $N/2$ in the last line of (\ref{FFT_algorithm}). The discrete Fourier transform of $N$ length is reduced to the sum of two the Fourier transforms of lengths $N/2$. The reduction of each Fourier transform to a Fourier transform of a smaller size can be done recursively. Although there are different adaptation of the FFT, the case where N equals to a power of 2 is especially attractive. If $N$ is an integer number of power $2$, the FFT of length $N$ is expressed in terms of two FFTs of lengths $N/2$, then four FFTs of lengths $N/4$, then eight FFTs of lengths $N/8$ and so on until we obtain $N$ numbers of FFTs of length one. An FFT of length one is just the number itself. If $N$ is not an integer power of $2$, it is still possible to express the FFT of length $N$ in terms of the several shorter lenghts of FFTs. Breaking a big size of discrete fourier transform into a number of smaller sizes of DFTs has an enormous impact on computational time. For each value of $n$ in (\ref{DFT}), computation of $F_n$  requires $N$ complex multiplications and $N-1$ complex additions. Therefore, computation of length $N$ requires approximately $N^2$ complex operations for big values of $N$ whenever a DFT algorithm is utilized. However, if $N=2^p$ was chosen in (\ref{FFT_algorithm}), the number of steps in the recursion would be $p$. There are also $N$ number of complex operations in the final stage of the FFT making the total number of computation $Np=N(log_{2}N)$ for the FFT algorithm. 

Let us mention how we implement FFTs and IFFTs in our numerical schemes before moving into the next section. We have extensively used Matlab's fast Fourier transform algorithm by means of fft and ifft built-in commands in our research endeavour. The built-in fft and ifft functions are based on the FFTW, the fastest Fourier transform in the West, developed at MIT by Matteo Frigo and Steven G. Johnson. The periodic $u$ function of the SPE and Maxwell equation with vanishing boundary conditions is manipulated by forward and backward Fourier transforms via fft and ifft commands. Each Fourier transformed function is multiplied by the angular frequency. Since the short pulse equation (\ref{SPE_sak}) and Maxwell equation (\ref{maxwell_1d}) include the first and the second derivatives of time, we also apply the derivative of the Fourier transform to these equations according to (\ref{Fourier_1stDer}) and (\ref{Fourier_nthDer}) as well. The numerical schemes employing the numerical methods (except the ETD method) that will be discussed in the next sections implement Matlab's built-in commands fft and ifft to solve the short pulse equation and Maxwell equation.


\section{Solving Differential Equations: Euler's Method and Midpoint Method}

The differential equations are commonly used for mathematical modelling of scientific inquiries \cite{salsa:2008},  and play a prominent role in many fields such as physics, engineering, and economics \cite{farlow:1993}. Whenever there is no analytical solution available, the numerical approximations are required. We will, in this section, discuss the Euler's numerical schemes, which we used them in the numerical analysis of Maxwell and the short pulse equations extensively. 

Let us mention Euler's methods briefly first. Consider the first order differential equation 
\begin{equation}\label{Diff_eqn}
\frac{dy}{dt}=f(y,t)\,.
\end{equation}
Let $[a,b]$ be the interval over which we want to find the solution to the above initial value problem (IVP) with a given initial value $y(a)=y(0)$. Since we want to find a numerical solution, our aim is not to seek  a differential function that satisfies the IVP. Instead, we will generate a set of points $[(y_n, t_n)]$ which will be used to satify the differential equation. Using the formal definition of the derivation and the Taylor expansion \cite{mathews-fink:1998}, one can approximate the left hand side of (\ref{Diff_eqn}) for the small increment of time $t$, i.e., $\Delta t$ as
\begin{equation}
\left.\frac{dy}{dt} \right|_{t=t_n} \approx \frac{y_{n+1} - y_n }{\Delta t} \,,
\end{equation}
where we use the notation $y_n$ for $y\left(t_n\right)$ and $y_{n+1}$ for $y\left(t_n+\Delta t\right)$. The differential equation (\ref{Diff_eqn}) evaluated at time $t=t_n$ is then be
\begin{equation}
\frac{y_{n+1} - y_n}{\Delta t} = f(y_n, t_n), 
\end{equation}
which can be rearranged to obtain Euler's approximation
\begin{equation}\label{Euler's_Method}
y_{n+1} = y_n + {\Delta t} f(y_n, t_n). 
\end{equation}
At $t=0$, we have $y_1 = y_0 + \Delta t f(y_0, t_0) $. Since the initial value $y(0)$ is given at $t=0$, one can calculate $f(y_0,t_0)$ and, therefore, the first iterated value $y_1$. This iteration process is repeated until the iterated values approximate the solution curve $y = y(t)$. This method is called the finite difference method, and equation (\ref{Euler's_Method}) is the difference equation. For the reason that the difference equations approximate derivatives, the finite-difference methods approximate the solutions of the differential equations. 

The numerical techniques approximating the solutions of the differential equations may result in different results. Hence, error analysis is very important to see how good the numerical results are. Apart from having the round-off errors when using any finite difference methods to approximate the solution of the differential equations, there is also a discretization error and truncation error. The discretization error can be defined as the difference between the analytical solution and the numerical result obtained by the difference method. The truncation error arises from truncating the Taylor series at the first derivative in case of Euler's approximation (\ref{Euler's_Method}). Both the discretization and truncation errors depend on the step size $\Delta t$ or $dt$. The Euler's method approximating the first derivative of a function accumulates truncation error in the order of the step size or discretized interval, i.e., $O(\Delta t)$. The smaller the discretized interval $\Delta t$ is, the smaller the truncation and discretization errors are. The first order Euler scheme (\ref{Euler's_Method}) may therefore give relatively big errors as the process proceeds, and therefore, it has limited usages. The error accumulation of the Euler's scheme can also be qualified if it is applied to the SPE. Forexample, when the initial ultra-short pulse obtained from equation (\ref{Sak Solution}) with $m=0.3$ is allowed to propagate in the first order Euler scheme with a step size $dt=0.001$, the result does not agree with the analytical solution with a good accuracy. The error accumulation is much bigger than the one obtained by Euler's midpoint formula even at the shorter propagation distances
\begin{figure}[htp]
\begin{center}
\scalebox{0.75}{\includegraphics{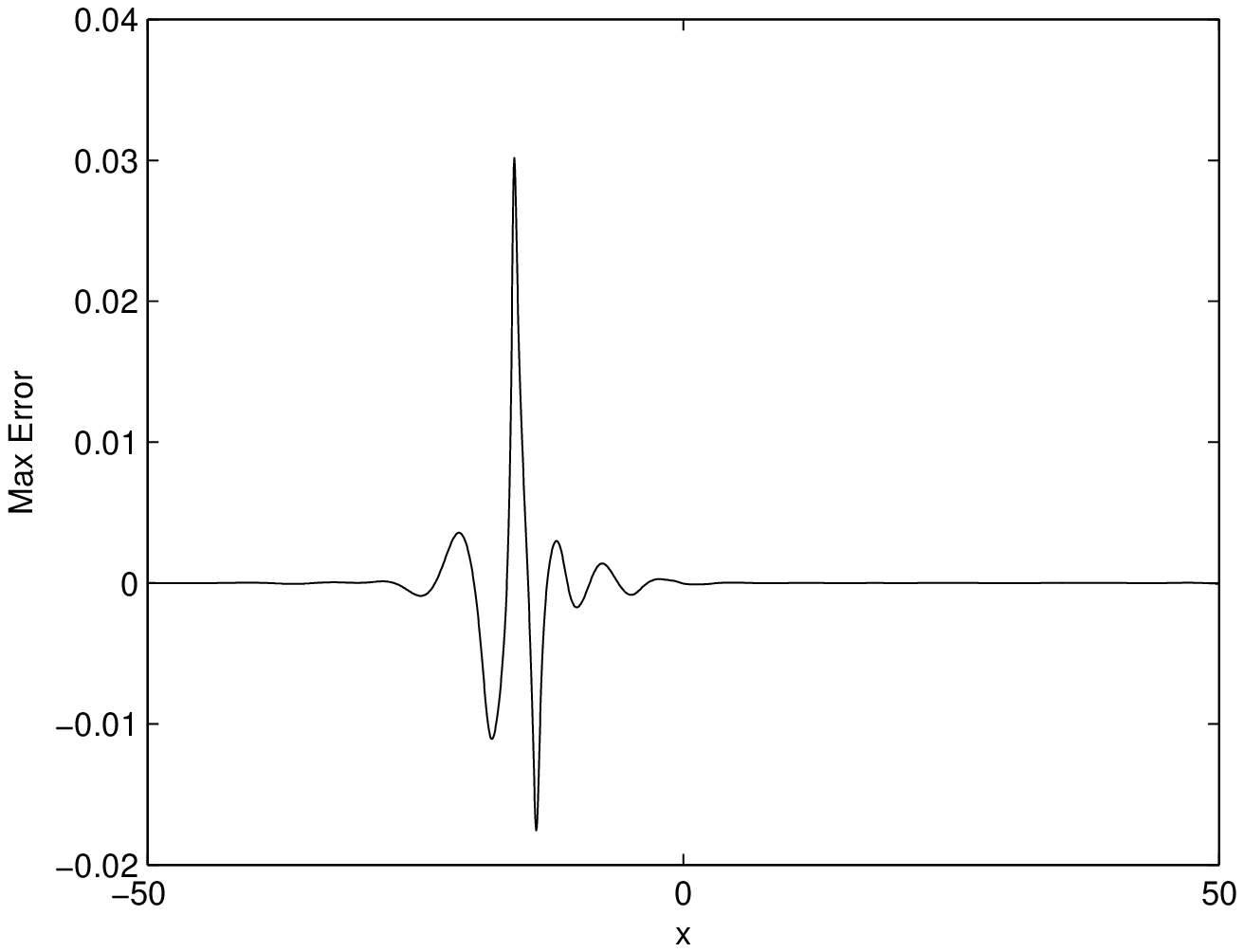}}
\scalebox{0.75}{\includegraphics{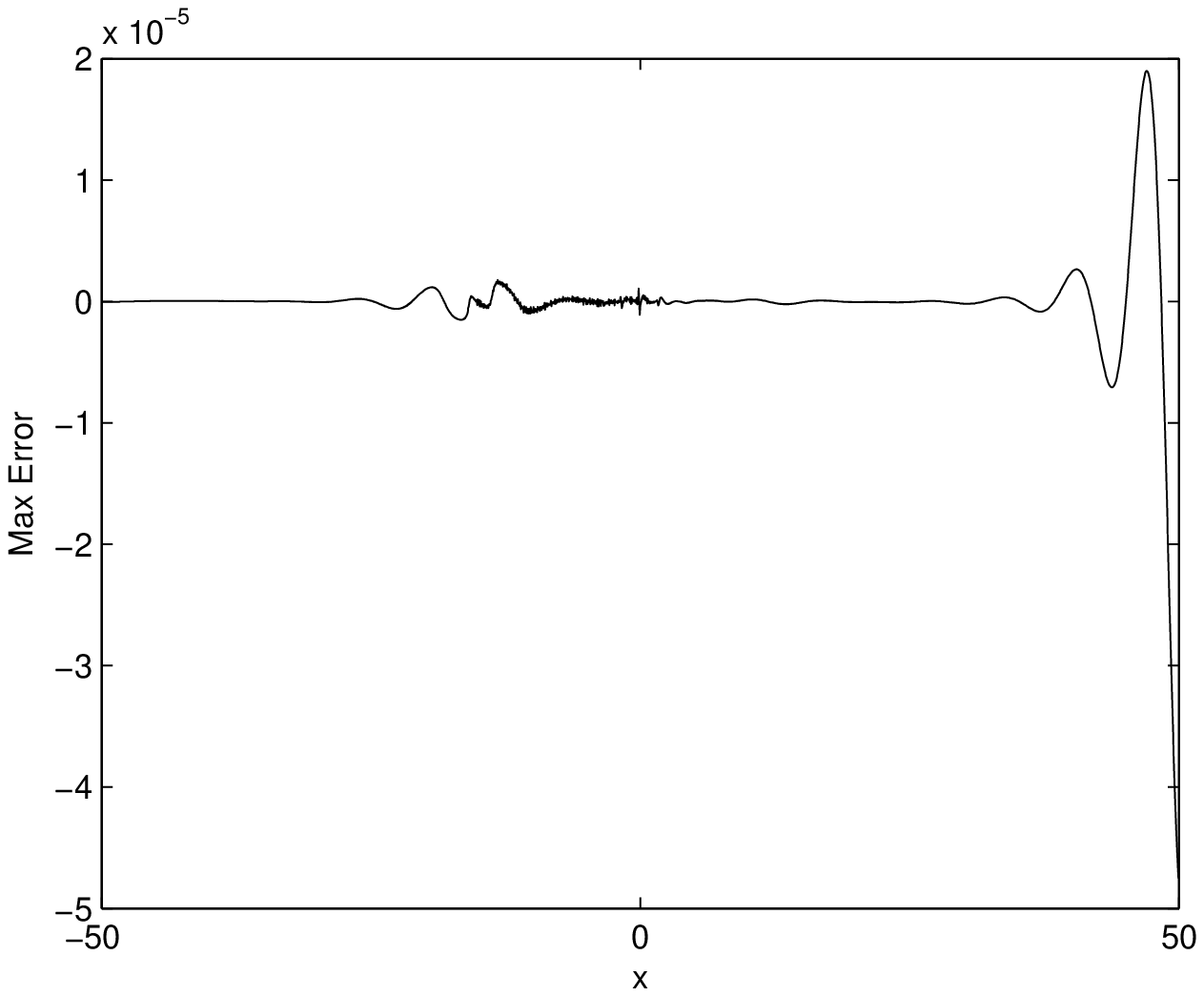}}
\caption{The maximum error between the numerical solution of the Euler's first order scheme and the analytical result (upper graph), and between the solution of the midpoint formula based scheme and the analytical result (bottom graph) at the propagation distance $t=15$ units.}
\label{Eulers1stError}
\end{center}
\end{figure}
as shown in figure \ref{Eulers1stError}. The graph displays the difference between the numerical and the analytical value of $u$ (maximum error) at the propagation distance $t=15$ ($x$ is the temporal variable in the SPE (\ref{SPE_sak})). 

We use Euler's method in one step only to generate an initial value. Euler's method (\ref{Euler's_Method}) is implemented in the SPE numerics together with the midpoint method. The midpoint formula for numerical integration of (\ref{Diff_eqn}) is
\begin{equation}\label{MidPoint_Formula}
y_{n+1} = y_{n-1} + 2{\Delta t} f(y_n, t_n).
\end{equation}
The midpoint formula is the first central difference approximation and is a second order formula. It is just the improvement of the first order Euler formula (\ref{Euler's_Method}) and can be derived in a similar manner (see the next section for the derivation). The midpoint formula may also be classified as the second order Runge-Kutta formula \cite{kiusalaas:2005} and is sometimes refered as the leapfrog method. The leapfrog method is widely used because of its good stability when solving the partial differential equations with the oscillatory solutions. The error at each step of the midpoint method is of the $O(\Delta t^2)$. As shown in figure \ref{Eulers1stError}, midpoint formula generates much less error than Euler's method does, and it may generate very good results at the expense of some more computational effort. 

The application of the Euler's method and Midpoint method is straightforward in our numerical codes for the SPE. The Fourier transform of the SPE (\ref{SPE_sak}) with respect to $x$ variable is first taken via fft command of Matlab, which then yields                                                   $\hat{u}_t=\hat{u}/iw+[(iw)/6] \hat{u^3}$. The zero frequency mode must be exluded in the above equation because the first term in the right hand side otherwise becomes infinity in the fourier domain. Once we finish with the Fourier transform, we can take the inverse Fourier transform via ifft and be back in the spatial domain with only one $t$ (evolution variable) derivative left, i.e., $u_t=\alpha u+ \beta u^3$. The $\alpha$ and $\beta$ are the new coefficents after the Fourier transforms being done. The iteration schemes can now be applied. The time and space domains can be chosen and discretized in a desired way in the codes. These choices have to be made by considering numerical stability and error accumulations. If $n=0$ is chosen in (\ref{Euler's_Method}), we have the first iteration equation
\begin{equation}\label{1stIteEuler1st}
u(y_1,t_1)=u(y_0,t_0)+\Delta t (\alpha u(y_0,t_0) + \beta u(y_0,t_0)^3).
\end{equation} 
This is an initial value problem and one can use the analytical solution (\ref{spe_soln}) at $t=0$ for $u(y_0,t_0)$. Once the initial value is substituted into (\ref{1stIteEuler1st}), the first iterated value is generated. This is the entire usage of the Euler's first order formula in our numerical scheme. Following the application of the Euler's method, we can now apply the midpoint formula. If we choose $n=1$ in equation (\ref{MidPoint_Formula}), we get 
\begin{equation}\label{midpointN=2}
u(y_2,t_2)=u(y_0,t_0)+2\Delta t (\alpha u(y_1,t_1) + \beta u(y_1,t_1)^3).
\end{equation} 
Using the initial value and Euler's result for $u(y_0,t_0)$ and $u(y_1,t_1)$ respectively, we obtain the second value $u(y_2,t_2)$. We get the third from the first and the second,the fourth from the second and the third, and so on by the successive application of equation (\ref{MidPoint_Formula}). Note that Fourier tranform and inverse fourier transform is applied at each step. This iteration repeated until the process stops. Overall process of this scheme is a stable numerical propagation of the SPE solitions in the short pulse equation. Figures produced by midpoint formula are shown in chapter four already. 

At this point, one may argue the initial value obtained from the analytical solution. Note that the spatial variable $t$ and the temporal variable $x$ are the free variables of the partial differential equation (\ref{SPE_sak}). The initial condition (\ref{spe_soln}) does not depend on the free variable $x$ directly. Instead, it depends on another variable $y$ from which we can get initial contion $u(y,t)$. We cannot chose $y$ values in an arbitrary way because it is not a free variable. It depends on $x$, $t$ and itself, i.e., $y(x,y,t)$ ( See equation (\ref{spe_parameters}) ). The first approach coming to mind might be writing $y$ values in terms of $x$ and $t$. A blind look at the equation (\ref{spe_soln}) shows that it may not be possible to take the inverse of the equation (\ref{spe_soln}). The $y$ values are approximated from the free variables $x$ and $t$ numerically. To do this, we first set $t=0$ and obtain $y$ from equation (\ref{spe_soln}) as
\begin{equation}\label{y_values}
y = x-\frac{2mn ((msin(2ny)-nsinh(2my))}{(m^2sin(ny)^2+n^2cosh(my)^2))} \,,
\end{equation}
where $m$ and $n$ are the equation parameters and may take different values according to (\ref{spe_parameters}). We choose an interval for free $x$ values and set $y=x$ at first. Having $x$ and $y$ values in hand by this choice will let us to apply equation (\ref{y_values}) to get a new interval for $y$ values. This new set of $y$ values along with the $x$ values can be used to generate another new set of $y$ values. This process has been repeated until we generate a precise set of $y$ values. How good the final values of $y$ is the next question to ask. A simple way to check is to insert numerically produced $y$ values in equation (\ref{spe_soln}) and re-generate free $x$ values from these $y$ values.  
\begin{figure}[htp]
\begin{center}
\scalebox{0.8}{\includegraphics{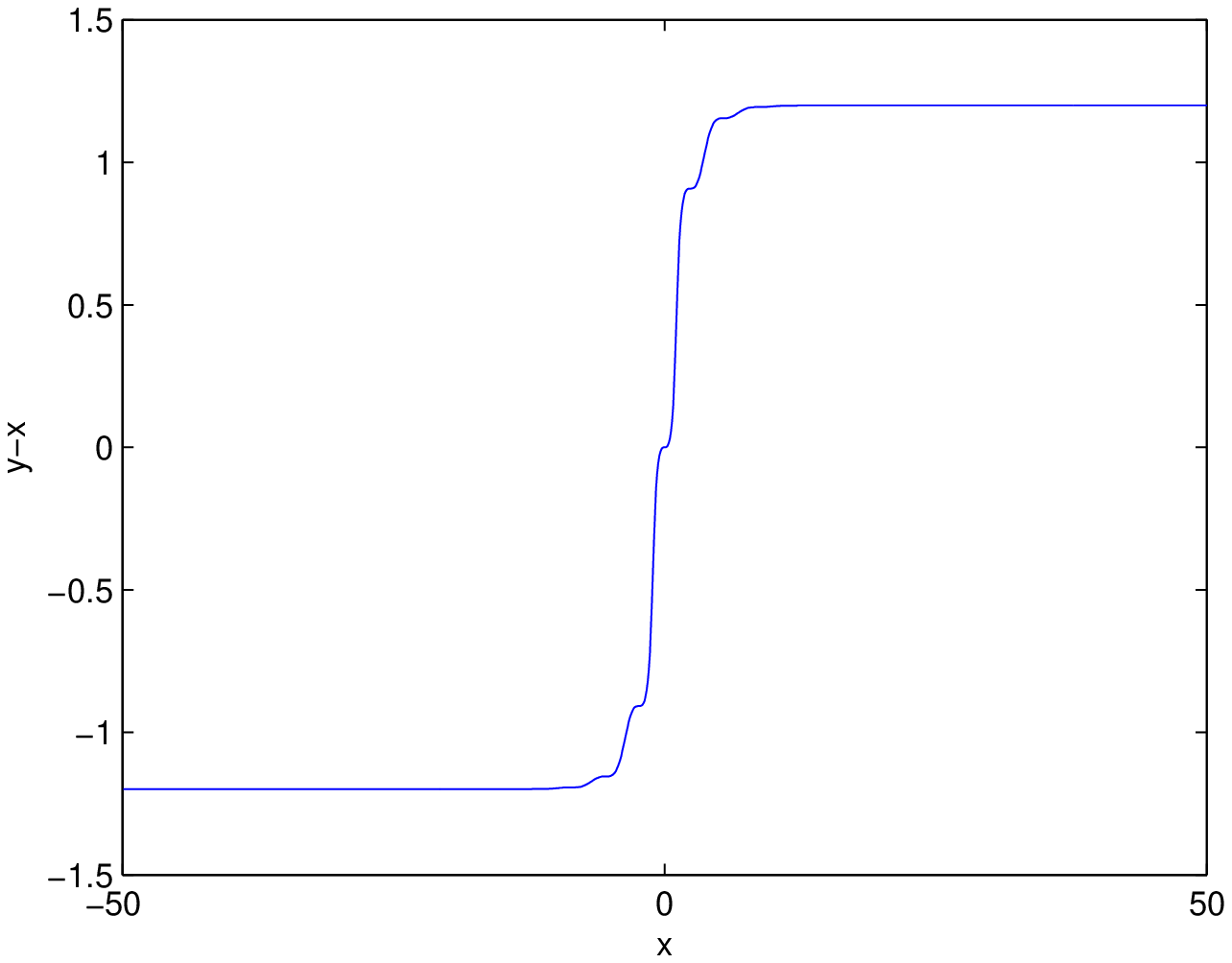}}
\scalebox{0.8}{\includegraphics{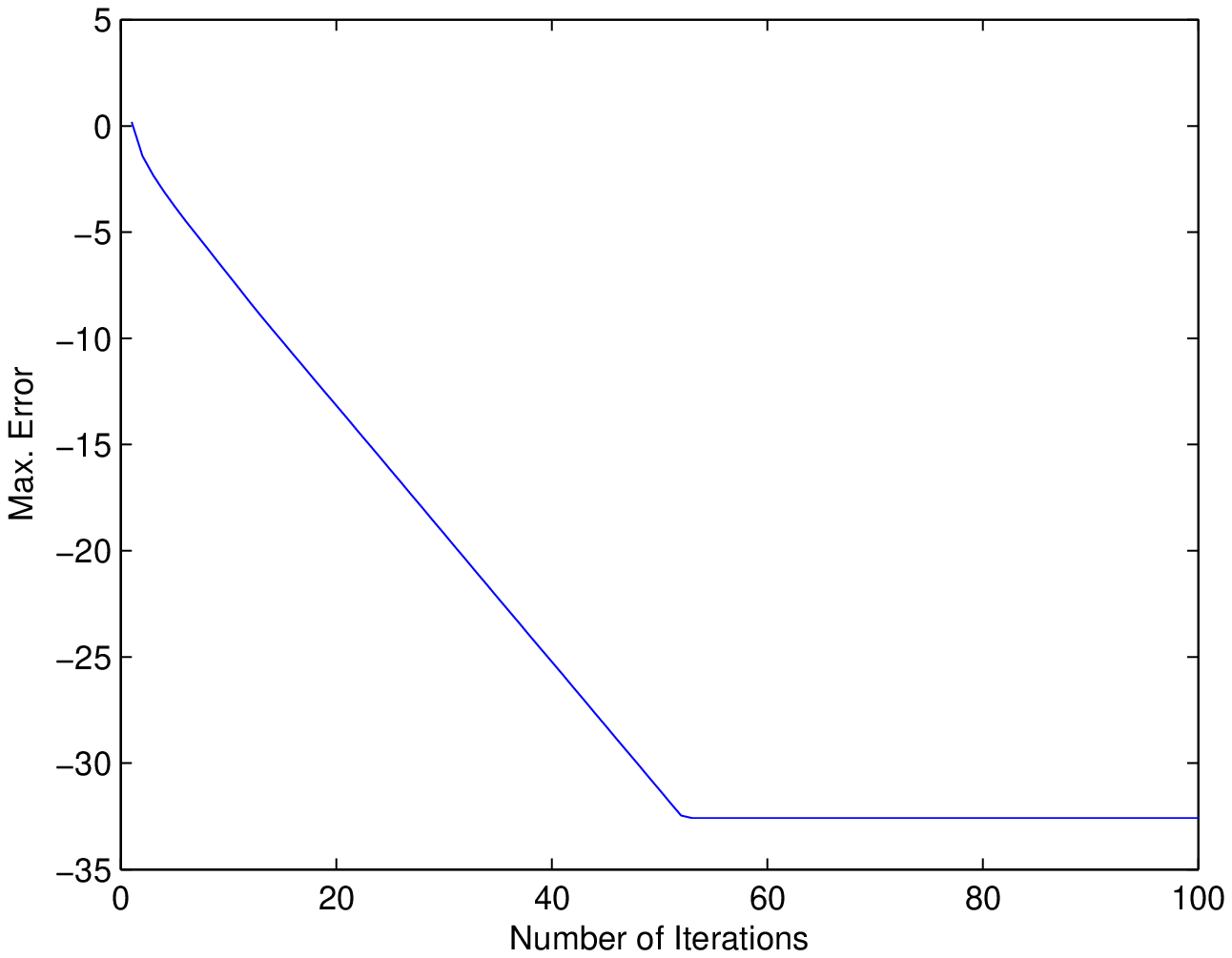}}
\caption{The plot of the difference between the numerical computation of the variable $y$ and the free variable $x$ versus $x$ (upper graph), and the plot of the logarithmic difference versus the number of iterations (bottom graph).}
\label{yvalues}
\end{center}
\end{figure} 
The plot $y-x$ versus $x$ in figure \ref{yvalues} demonstrates that $y$ and $x$ values are not the same and, therefore, the necessity of the $y$ values for the numerical experiments is not questionable. The logarithmic plot of the maximum error versus the number of iterations in figure \ref{yvalues} displays that one has to repeat the iterations about sixty times to get precise values of $y$. Once the minimum number of iterations are executed, the numerical $y$ values may be safely used in the experiments of the short pulse dynamics. In the case of approximate solution (\ref{spe_approx_soln}), such a problem does not arise because $y \approx x$. Before closing this section, we may make our final remark that our numerical scheme based on Euler's formula and midpoint formula can be used for further research applications.


\section{Semi-Implicit Method and Euler Central Formula}

The first numerical scheme adopted for the Maxwell equation (\ref{nonlinear_maxwell}) is the Euler's central method. We have already discussed the finite difference approximation for the first derivative in the previous section. In dealing with the SPE, we only need a difference formula based on the first derivative of the function $u$. However, the numerical technique needed for the nonlinear wave equation requires the second derivative of the function $u(x,t)$. 

The derivation of the finite difference approximations for the derivatives of a function $f(x)$ are based on forward and backward Taylor series expansions of $f(x)$ about $x$ \cite{mathews-fink:1998,karris:2007,otto-denier:2005}. We will only illustrate the derivation of the difference formula for $f^{''}(x)$ here. Let us start with Taylor expansion of $f(x+h)$
\begin{equation}\label{taylor_frwd}
f(x+h)=f(x)+h f^{'}(x)+\frac{h^2}{2!} f^{''}(x)+\frac{h^3}{3!} f^{'''}(x)+... \,.
\end{equation}
Similarly,
\begin{equation}\label{taylor_bkwrd}
f(x-h)=f(x)-h f^{'}(x)+\frac{h^2}{2!} f^{''}(x)-\frac{h^3}{3!} f^{'''}(x)+... \,.
\end{equation}
Here, $h$ can be taken as the difference between the successive step in space or time domain, i.e., the step size. Adding up equations (\ref{taylor_frwd}) and (\ref{taylor_bkwrd}) eliminates the odd derivatives $f^{'}(x)$,$f^{'''}(x)$,.... If we truncate the series at the fourth derivative, we obtain the relation for the second derivative $f^{''}(x)$
\begin{equation}\label{Euler_Central}
f^{''}(x)=\frac{f(x+h)-2f(x)+f(x-h)}{h^2}-\frac{2h^2f^{(4)}(x)}{4!} \,.
\end{equation}
The first term in (\ref{Euler_Central}) is the desired formula and called Euler's central formula. The second term is the error term $E(f,h)=\left(h^2/12\right)f^{(4)}(x)$ due to the truncation. If we subtract equation (\ref{taylor_bkwrd}) from (\ref{taylor_frwd}) and truncate the series at the third derivative, we will obtain the midpoint formula (\ref{MidPoint_Formula}) with the error term $E(f,h)=\left(h^2/6\right)f^{'''}(x)$. Notice that the first order non-central forward finite difference (see equation (\ref{Euler's_Method})) can be obtained from equation (\ref{taylor_frwd}) with the truncation error $E(f,h)=\left(h/2\right)f{''}(x)$. The power of $h$ in each error term shows the order of the truncation error for these finite difference methods. As it is clearly seen in the formula (\ref{Euler_Central}), the formula is truncated at the fourth derivative and the error due to that is in the $O(h^2)$. The truncation error is not the only error source in the numerical work. We also encounter various kinds of errors when using an algorithm for the computations such as the discretization and round-off errors. The truncation error in computation arises from having a finite number of terms instead of infinitely many terms we have in theory. On the other hand, the round-off error is caused by storing numeric data in finite bits, and the discretization error results from the fact that a function of a continuous variable is represented in the computer by a finite number of evaluations. The discretization and truncation errors can usually be reduced by using a smaller step size. Such a decrease in the step size increases the computational cost in return. Note that 
the sum of the coefficients of $f$ functions ($f(x)$ and $f(x\pm h)$) is zero in all finite difference expressions. If $h$ is very small, the values of $f(x)$ and $f(x \pm h)$ will be approximately equal and the effect on the roundoff error can be profound. On the other hand, we cannot make $h$ too big because the truncation error would be intolerable. This means as we decrease the step size, we decrease the truncation and discretization error and increase the round-off error. Therefore, we must find an optimum value of the step size which balances out these errors. One partial solution to this problem is to use a formula of higher order so that a larger value of $h$ will generate a better accuracy. There may also be error due to the noise in the system and due to the instability of the algorithm. Although we cannot get rid of these kinds of inevitable errors completely, we should minimize the error by choosing the optimum value of step size and by having a stable algorithm. The step size $dt=0.001$ is in general used in the schemes employing the semi-implicit method and the Euler's central formula. This choice of $dt$ is small enough and computationaly affordable. 

Let us now look at the details of the scheme employing the Euler's central formula (\ref{Euler_Central}) for the nonlinear wave equation (\ref{nonlinear_maxwell}). To apply the second order finite approximation, we take the Fourier transform of equation (\ref{maxwell_1d}) and apply the Euler central formula (\ref{Euler_Central}), which then leads to a difference equation in the Fourier domain as
\begin{equation}
\hat u_{n+1}=2 \hat u_n -\hat u_{n-1}+ \Delta t^2 \left[ (-w^2+b)\hat u_{n-1}+(-c*w^2)(\widehat {u^3})_{n-1}\right] \,,
\end{equation}
where $w$ is the angular frequency, $n=1,2,3...N$ and $\hat u_{n+1} = \hat u(w_n,t_{n+1})$. The choice $n=1$ is the first iteration equation 
\begin{equation}\label{1stIte}
\hat u_2=2 \hat u_1 -\hat u_0+ \Delta t^2 \left[ (-w^2+b)\hat u_0+(-c*w^2)(\widehat{ u^3})_0\right] \,,
\end{equation}
where $\hat u_2 = \hat u(w,2 \Delta t)$, $\hat u_1 = \hat u(w,\Delta t)$ and $\hat u_0 = \hat u(w,0)$. One must have the knowledge of $\hat u_1$ and $\hat u_0$ to get the first iterated value $\hat u_2$. We use the Fourier transfom of the SPE initial condition as the initial condition of the Maxwell equation in (\ref{1stIte}). This is done for two reasons; the analytical solution of the Maxwell equation is unknown as of yet, and the SPE is derived from the Maxwell equation. It may be worty to mention that the transformation rule (\ref{spe_trans}) must be applied to the initial condition because this is not the exact solution of the nonlinear wave equation. If the SPE initial condition is not rescaled according to the transformation rule and is used as the initial condition of the Maxwell equation in the numerical experiments, the instablity is inevitable. The reader may ask the following question; how do we get $\hat u_1$? This is the other initial condition which includes the first derivative of $u$ function. The easiest way of getting this second initial condition is to compute it numerically.  The simple shift of $\hat u(w,0)$ by $\Delta t$ in the numerical code generates $\hat u(w,\Delta t)$. We also wanted to get $\hat u_1$ analytically. This analytical implementation introduces a first derivative of $u$ function (\ref{spe_soln}). Since analytical solution (\ref{spe_soln}) of the SPE does not depend on free variables $x$ and $t$ directly, a simple application of chain rule includes the time derivative of $y$ function (\ref{y_values}). After dealing with partial derivative of $y$, the first time derivative of $u$ function can be written as
\begin{equation}
\frac{du}{dt} = u_t + u_y(-f_t/(1+f_y))
\end{equation}    
where $u_t$ and $u_y$ are the partial derivative of $u$ with respect to $t$ and $y$ respectively, and $f_t$ and $f_y$ are the partial derivatives of the second term in equation (\ref{spe_soln}) ($x=y+f(y,t)$) with respect to $t$ and $y$ respectively. Once we get the first derivative, $u_1$ can be obtained from the first order Euler's method $u^{'}(t) = [u(t+h) -u(t)]/h$. Note that $n=0$ choice instead of $n=1$ as the starting value of $n$ does not mean that we are stuck. In that case, backward Euler's formula,  $u^{'}(t) = u(t) -u(t-h)/h$, should be utilized to start the iteration scheme. The first iteration (\ref{1stIte}) with Fourier transform of $u_1$ and $u_0$ gives $\hat u_2$, which then together with $\hat u_1$ generates $\hat u_3$. This iteration process is repeated many times until we hit the most precise value of the solution. Finally, we take the inverse Fourier transform of the latest iterated value and we have the numerical solution of $u$. As for the results produced by Euler central formula, first of all when the coefficients of nonlinearity in nonlinear wave equation (\ref{maxwell_1d}) is taken $0.005$, it seems like the numerical scheme is doing a good job. Initial SPE solitary pulse propagates stably and is compared with the analytical solution very well at even long propagation distances such as $40$ or $50$ units distances. This way of comparing the numerical work with the analytical value can be misleading and deceptive. To be able to observe the effect of nonlinearity with these coefficients of nonlinearity, one may need to propagate the initial pulse a much longer distances (forexample, $400$ or $500$ unit distances). Propagation to such a big distance will increase the computational cost and computational time. Besides, numerical instability may interfere the numerical propagation for such a big propagation journey. Therefore, the coefficient of nonlinearity should be big enough in order to observe the effect of dispersion and nonlinearity within few units of propagation distance. Remember a soliton propagates stably because of the delicate balance between dispersion and nonlinearity. At the same time, we cannot choose these coefficients arbitrarily. For instance, if we just made both coefficients of dispersion and nonlinearity one, we would hurt such delicate balance between dispersion and nonlinearity, which then leads to an unstable propagation of solitary wave. For that reason, we rescale the coefficients of the nonlinear wave equation (\ref{maxwell_1d}) such that the coefficient of the dispersion becomes two, i.e., $a=2$ and the coefficient of the nonlinearity becomes one third, i.e., $b=1/3$. With these coefficients, we can now observe the effects of nonlinearity when we let the soliton progates a few units. As a matter of fact, this way of checking the propagation produces results within few units of evolution about the reliability of the code. The instability in the propagation of the initial pulse with these choice of coefficients starts even at two units of propagation slightly and builds up much more leading to the loss of the pulse as the pulse keeps moving more and more. It is not easy to spot what causes this instability right away. Because there are two different ways of getting $\hat u(w,dt)$, and the usage of numerical shift $\hat u(w,dt)$ in the code (instead of the value $\hat u(w,dt)$ obtained by the analytical first derivative $du/dt$ ) does not remove the instability in the code either, we cannot suspect wrong calculation of the analytical derivation of $u$ function (\ref{spe_soln}) with respect to $t$. Implementation of filtering was not a remedy to further application of our Euler's central formula based numerical scheme. For this reason, another scheme is employed in the numerical analysis of the nonlinear wave equation. 

In the next section, we will discuss this new scheme, called Ablowitz-Ladik scheme. On the other hand, Euler's central formula (midpoint formula) for the first derivative applies to the short pulse dynamics as discussed in the previous section and this implementation produces reliable results for spe. Before going into the next section, let's also mention how it is implemented in a different way than the one mentioned in section $8.1$. The central formula for the first derivative follows the basic manipulation of Taylor forward and backward manipulation. If the equation (\ref{taylor_bkwrd}) is subtracted from the equation (\ref{taylor_frwd}) and if the series is truncated at the third derivative, we get
\begin{equation}\label{euler_central1st}
f^{'}(x)=\frac{f(x+h)-f(x-h)}{2h}-\frac{h^2}{6}f^{(3)}(x).
\end{equation}
The $h^2$ dependence of truncation error means that truncation error here is in order of $O(h^2)$. The iteration equation follows the application of equation (\ref{euler_central1st}) to spe (\ref{spe_sak}). A simple treatment of a short pulse numerics using Euler's central formula requires the fourier transform of the spe first. We can take care all integration in the fourier domain and transform back to the time domain after the integration is complete. For integration in Fourier space, we need initial contions $\hat u(y_0,w_0)$ and $\hat u(y_1,w_1)=\hat u_1$. Fourier transform of analytical solution at $t=t_0$ gives $\hat u(y_0,w_0)=\hat u_0$ and application of (\ref{Euler's_Method}) to Fourier tranform of the spe (\ref{SPE_sak}) gives $\hat u(y_1,w_1)$. Once we have the initial conditions, iteration scheme
\begin{equation}\label{EulerCentfourier}
\hat u_{n+1}=\hat u_{n-1}+2h[\frac{\hat u_n}{iw} + (iw) \hat u^3_n ]
\end{equation}
is ready to use. This formula is very similar to the equation (\ref{MidPoint_Formula}). It may be called midpoint method in Fourier space alternatively. Here, $n=1,2,...,N$ and $w$ is the Fourier frequency as mentioned before. Note that we can start $n$ values from zero. In that case, we need to apply Euler's backward formula to get $u_{-1}$. When $n=1$, we get an equation in Fourier domain similar to equation (\ref{midpointN=2}). Successive application of (\ref{EulerCentfourier}) results in $\hat u(y_n,w_n)$. Inverse Fourier transform of $\hat u(y_n,w_n)$ would be the final step in this scheme. The advantage of using this scheme is that it is much faster because Matlab does not apply Fourier and inverse Fourier transform $n$ times, but only once. As for the results generated by these two schemes, they come out almost same. The ignorable difference between two results may stem from making coefficent of the fourier transform of spe (\ref{SPE_sak}) with respect to $x$ zero whenever $w=0$ at each fourier step and/or some numerical uncertainty that we cannot state exactly. The other numerical technique that has extensively been used in the numerical analysis of the short pulse dynamics is semi-implicit method. It is the second order nonlinear method that is solving nonlinear partial differential equations as well as linear partial differential equations, and is simply an improved version of the Euler's formula (\ref{Euler's_Method}). There are also many other improved versions of Euler's formula exist in the literature such as Runga-Kutta methods and Heun's method \cite{otto-denier:2005,mathews-fink:1998}. This modified version is given as
\begin{equation}\label{semi-imp_meth}
\frac{y_{n+1}-y_n}{\Delta t}=f(\frac{y_{n+1}+y_n}{2},t_n)
\end{equation}
where $n=0,1,2,...,N$ and $\Delta t=h$ is the step size. The difference between semi-implicit method and Euler's method can be seen by comparing the equations (\ref{semi-imp_meth}) and (\ref{Euler's_Method}). When $y$ and $\Delta t$ are replaced with the short pulse variables $u$ and $h$ respectively, and equation (\ref{semi-imp_meth}) is manipulated, the iteration scheme becomes
\begin{eqnarray}\label{semi-imp_scheme}
u_{n+1} &=& u_n+hf(\frac{u_{n+1}+u_n}{2},t_n) \nonumber \\
        &=& u_n+h[\alpha(\frac{u_{n+1}+u_n}{2})+\beta (\frac{u_{n+1}+u_n}{2})^3].
\end{eqnarray}
where $\alpha$ and $\beta$ are the new coefficients as they appear in $u_t=\alpha u + \beta u^3 $. Note that $t$ dependence in the second line in the above relation is embedded in the $u$ function. The modification in the function $f(u,t)$ (the first derivative of the $u$ function with respect to $t$ in our case) makes a scheme that works much better than the first order Euler's scheme.
A comparison between a partial modification and a full modification in the nonlinear short pulse equation demonstrates the degree of improvement in the numerical result. We obtain the scheme 
\begin{equation}\label{semi-imp_scheme_Lin}
u_{n+1}=u_n+h[\alpha(\frac{u_{n+1}+u_n}{2})+\beta u_n^3].
\end{equation}
after we modify the linear part and leave the nonlinear part unchanged.   
\begin{figure}[htp]
\begin{center}
\scalebox{0.8}{\includegraphics{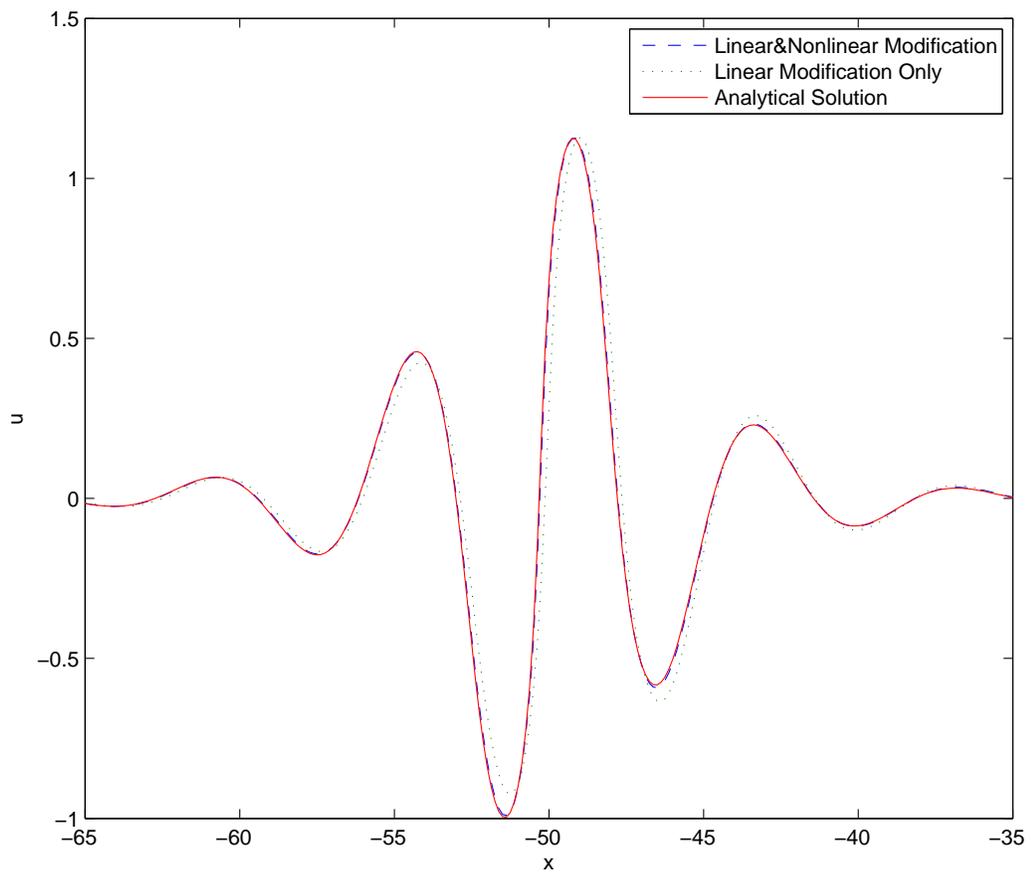}}
\caption{ Comparison of the solutions of the semi-implicit formula to the exact result at $t=50$ units of propagation distance. The solid line is the exact solution, and the dashed line displays the solution of the semi-implicit formula. The dotted line represents the solution of the numerical experiment with the semi-implicit method being only applied to the linear term of the short pulse equation.}
\label{SemiImpComparison}
\end{center}
\end{figure} 
Numerical schemes using iteration equations (\ref{semi-imp_scheme}) and (\ref{semi-imp_scheme_Lin}) solve the SPE (\ref{SPE_sak}) and the results are shown in figure \ref{SemiImpComparison}. An initial solitary wave of the short pulse equation propagates $50$ units in these schemes and is compared with the analytical solution at $t=50$ units. Dashed line shows the result obtained by the scheme (\ref{semi-imp_scheme}), whereas the dash-dotted line represents the result of (\ref{semi-imp_scheme_Lin}) at $t=50$. The solid line in figure \ref{SemiImpComparison} is the analytical solution at the same propagation distance. Our numerical experiment indicates that the modification in both the linear and nonlinear parts approximates the exact solution almost perfectly. 

It may be useful to mention more about the details of the adopted semi-implicit scheme (\ref{semi-imp_scheme}) in our numerics. The analytical SPE solution is used as the initial condition and the Fourier transform is applied to it before the semi-implicit iteration takes place. The choice of $n=0$ in (\ref{semi-imp_scheme}) leads to an
\begin{equation}\label{semi-imp_Ite1st}
u_1=u_0+h[\alpha(\frac{u_1+u_0}{2})+\beta (\frac{u_1+u_0}{2})^3].
\end{equation} 
where $u_0$ is the initial condition and $u_1$ is the first iterated value. Unlike Euler's first order method (\ref{Euler's_Method}) and Euler's central method (\ref{EulerCentfourier}), semi-implicit iteration equation (\ref{semi-imp_Ite1st}) requires $u_1$ as well as $u_0$ as initial conditions to get the first iterated value. To resolve the problem of not having $u_1$ as the initial value,  we choose $u_1=u_0$ initially and apply semi-implicit method.  The loop is executed three times to improve the accuracy of the value $u_1$. After obtaining $u_1$, the iteration is repeated for $u_2$ in the similar way. The process goes on until we obtain $u_N$. We choose the step size $dt=0.01$ in our code. The execution time is small. Although the step size is relatively large, the semi-implicit numerical implementation is fast and the highly accurate algorithm. 

Let us now compare the solutions of the semi-implicit formula with Euler's central scheme (\ref{EulerCentfourier}). 
\begin{figure}[htp]
\begin{center}
\scalebox{1}{\includegraphics{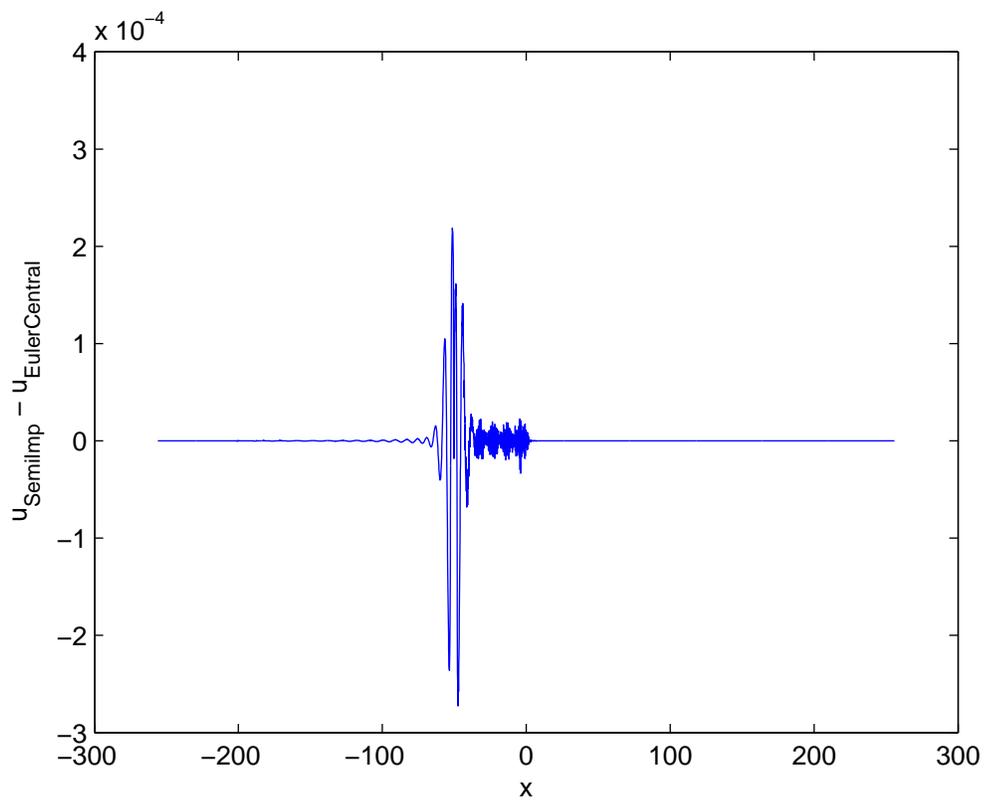}}
\caption{Maximum difference error between the numerical solutions obtained by the semi-implicit formula and the Euler's central formula at $t=50$ units of propagation distance.}
\label{SemiImpCentralComparison}
\end{center}
\end{figure} 
Figure \ref{SemiImpCentralComparison} displays the difference  of the numerical results obtained by these two schemes at $t=50$. $u_{SemiImp}$ is the numerical solution given by semi-implicit method and $u_{EulerCentral}$ is the solution  obtained by Euler's central method. The numerical difference at $t=50$ is almost ignorable as seen in figure \ref{SemiImpCentralComparison}. It should be noted that both numerical results match with the analytical solution at $t=50$ almost perfectly (see figure \ref{SemiImpComparison}). Furthermore, the execution time of semi-implicit scheme to propagate the initial soliton fifty units is about four times faster than the execution time of Euler's central scheme ( 38 seconds versus 135 seconds) to propagate it the same distance.


\section{Ablowitz-Ladik Method}

The Ablowitz-Ladix (AL) scheme is a fast and efficient numerical method for nonlinear evolution equations such as the nonlinear Schr\"odinger, Korteweg-deVries and modified Korteweg-deVries equations. For that reason we apply this scheme for the nonlinear Maxwell equation. 

Let us start our discussion by first introducing the AL scheme. Consider the nonlinear maxwell equation (\ref{maxwell_1d}) in the form
\begin{equation}\label{nonlinear_maxwell}
u_{tt}-u_{xx}=F(u(x,t)) \,,
\end{equation}
where $F(u)=bu+c(u^3)_{tt}$ , and $b$ and $c$ are constant coefficients. To solve this equation numerically, we approximate all the derivatives by finite differences. The domains in space and time are partitioned uniformly so that we have a mesh $x_0,x_1,...,x_J$ for space and a mesh $t_0,t_1,....,t_N$ for time. The difference between two consecutive space points is $dx$ and between two consecutive time points is $dt$.
The solution $u(x,t)$ at space point $x_j$ and at time $t_n$ along with space and time discretization can be written as
\begin{equation}
\begin{aligned}
x_j&=x_0+dx, \qquad &j=(0,1,2...J)  \\
t_n&=t_0+dt, \qquad &n=(0,1,2...N)  \\
u_j^n&=u(x_j,t_n)
\end{aligned} 
\end{equation}
We write finite difference second order derivatives (central differece formulas) for the space at position $x_j$ and for the time at time $t_n$ respectively as
\begin{eqnarray}
\frac{\partial^2 u }{\partial x^2}=\frac{u_{j+1}^n - 2u_j^n + u_{j-1}^n}{dx^2}\label{x_2ndDer} \\
\frac{\partial^2 u }{\partial t^2}=\frac{u_{j}^{n+1} - 2u_j^n + u_{j}^{n-1}}{dt^2}\label{t_2ndDer}
\end{eqnarray} 
Once we substitute equations (\ref{x_2ndDer}) and (\ref{t_2ndDer}) into equation (\ref{nonlinear_maxwell}) and rearrange terms, we obtain the following equation for $u_j^{n+1}$
\begin{equation}
u_j^{n+1}=-u_j^{n-1}+r^2(u_{j+1}^n+u_{j-1}^n)+2(1-r^2)u_j^n+dt^2F(u_j^n).
\end{equation} 
This equation can be stabilized by setting $r=1$ and using an average of the spatial coordinates for the function $F(u)$  as shown by Ablowitz, Kruskal and Ladik \cite{ablowitz-kruskal-ladik:1979} and such a choice leads to the final form of the equation
\begin{equation}\label{AL_scheme}
u_j^{n+1}=-u_j^{n-1}+u_{j+1}^n+u_{j-1}^n+dt^2F(\frac{u_{j+1}^n+u_{j-1}^n}{2}).
\end{equation}
This is the Ablowitz-Ladik (AL) scheme.  This scheme has been tested on traveling wave and periodic breather problems over long time intervals and has given good results in terms of computational accuracy and computational costs \cite{duncan:1997,taha-ablowitz:1988}. The comparison between the Ablowitz-Ladik scheme and the finite difference schemes for the nonlinear Schr\"odinger equation indicates that the Ablowitz-Ladik scheme is faster than the latter \cite{taha-ablowitz:1988}. We adapt the AL scheme for the Maxwell equation. The implementation of the AL scheme requires an initial condition. The SPE soliton solution is used as the initial contion. As discussed earlier, the SPE soliton stably propagates in the nonlinear wave equation, and therefore, this choice is not random. Nonetheless, we have to modify this initial condition acording to the multiple-scale parameters. The parameter $\epsilon$, which arises in the derivation of the SPE from the wave equation (see (\ref{u_expansion})), is chosen $0.2$. As it is forced by multiple-scale expansion technique, $\epsilon$ must be a small number and much less than one. There is no specific reason why we choose $\epsilon=0.2$ other than $\epsilon$ being just a small number. On the otherhand, one may choose $\epsilon=0.1$ and should generate the similar results we got with $\epsilon=0.2$. We also have to modify the spatial and time domains according to (\ref{xt_expansion}). The numerical scheme without proper scaling makes initial spe solitary wave condition spreaded around as it propagates. With the right way of choosing initial conditions, the Ablowitz-Ladik numerical scheme is stable. As a stability requirement, we choose $dx=dt$ and equate them to $0.0125$ in our code. The step size in this scheme is almost ten times bigger than the choice of the step size done in the Euler schemes. Such a change in the step size would effect the computational time incredibly if the Euler's schemes are used. However, changing the step size from $0.0125$ to $0.0031$ (approximately four times smaller now) in the AL code does not really alter the result, but increase the computational time only about twenty times more. Most importantly, the initial solitary wave propagates stably in the AL scheme unlike in Euler's central scheme. The AL schemes' results  compare with the analytical solution at a very good accuracy at any propagation distances. We must note that the filtering in the Fourier domain removes the noise in the numerical scheme. The scheme is not stable without the filtering. Filter in the Fourier domain applies at every step of the iteration. We iterate once, and then we filter out. This is the way we go until the very end. 

The first iteration equation follows the choice of $n=0$ and $j=0$;
\begin{eqnarray}\label{AL_scheme1st}
u_0^1 &=& -u_0^{-1}+u_{1}^0+u_{-1}^0+dt^2F\left(\frac{u_{1}^0+u_{-1}^0}{2}\right) \nonumber \\
      &=& -u_0^{-1}+u_{1}^0+u_{-1}^0+dt^2 \left[\alpha\left(\frac{u_{1}^0+u_{-1}^0}{2}\right)+\beta \left( 
\frac{u_{1}^0+u_{-1}^0}{2}\right)^3 \right] \,,
\end{eqnarray}     
where $\alpha$ and $\beta$ are constant coefficients of the Fourier tranformed SPE. The terms $u_{1}^0$ and $u_{-1}^0$ are the initial conditions obtained from the exact solution such that $u_{1}^0=u(x,t=0)$ and $u_{-1}^0=u(x,-dt)$. As for the term $u_0^{-1}$, it is obtained by shifting the exact solution by one temporal step, i.e., $u_0^{-1}=u(x-dx,t)$. In a similar fashion, the next iteration formulae can be obtained and used to finish the integration.

\section{Exponential Time Differencing Method}

The exponential time differencing (ETD) numerical technique is another powerful method that solves nonlinear partial differential equations. We implement the modified exponential time-differencing fourth-order Runge–Kutta method to solve the short pulse equation. Although the schemes employing the semi-implicit and Euler central methods produce very reliable results for the short pulse dynamics we have discussed so far, they are not stable enough to test the particle properties of the SPE solitons, i.e., collisions of the SPE solitons. We implement the ETD method to allow the SPE solitons collide. The results presented in chapter five regarding the collision of the two -soliton solution of the SPE employ the ETD method. We will only discuss the ETD method itself in the rest of this section.

The short pulse equation can be given in the general form
\begin{equation}\label{spe_ETD_General}
u_t = \mathbf{L}u + \mathbf{N}(u,t) \,,
\end{equation}
where $\mathbf{L}$ and $\mathbf{N}$ are the spatially discretized linear and nonlinear operators respectively. Let us define
\begin{equation}\label{vIntegFactor}
v = e^{-\mathbf{L}t}u \,. 
\end{equation}
The factor $e^{-\mathbf{L}t}$ is called the integrating factor. The time derivative of the integrating factor gives
\begin{equation}\label{vDer}
v_t=e^{-\mathbf{L}t} \mathbf{N}(e^{-\mathbf{L}t}v) \,.
\end{equation}

For the time stepping of the $v$ function, the fourth order Runge-Kutta method is used and given as
\begin{equation}
\begin{aligned}
a&=dtf(v_n,t_n) \\
b&=dtf(v_n+a/2,t_n+dt/2) \\
c&=dtf(v_n+b/2,t_n+dt/2)  \\
d&=dtf(v_n+c,t_n+dt)  \\
v_{n+1}&=v_n+\frac{1}{6}(a+2b+2c+d) \,,
\end{aligned}
\end{equation}
where $dt$ is the time-step (discretization) and $f$ is the right-hand side of equation (\ref{vDer}).
If the $v$ function is integrated over a single time step $dt$, we get
\begin{equation}
u_{n+1}=e^{\mathbf{L}dt}u_n+e^{\mathbf{L}dt}\int_0^{dt}e^{\mathbf{-L}\tau}\mathbf{N}(u(t_n+\tau),t_n+\tau)d\tau \,.
\end{equation}
The integration on the right-hand side of the above equation has to be discretized so that the proposed generating formula \cite{kassam-trefethen:2005} becomes
\begin{equation}
u_{n+1}=e^{\mathbf{L}dt}u_n+dt\sum_{m=0}^{s-1}g_m\sum_{k=0}^m (-1)^k 
\binom{m}{n} \mathbf{N}_{n-k}\, ,
\end{equation}
where $s$ is the order of the scheme. When the integration is carried out, the complex analysis is used to compute the coefficients via contour integrals in the complex plane. We use the fourth order Runge-Kutta method so that $s=4$ in our case. Therefore, the numerical iteration formula becomes
\begin{equation}
\begin{aligned}
a_n&=e^{\mathbf{L}dt/2}u_n+\mathbf{L}^{-1}\left(e^{\mathbf{L}dt/2}-\mathbf{I}\right)\mathbf{N}(u_n,t_n)\\
b_n&=e^{\mathbf{L}dt/2}u_n+\mathbf{L}^{-1}\left(e^{\mathbf{L}dt/2}-\mathbf{I}\right)\mathbf{N}(a_n,t_n+dt/2)\\
c_n&=e^{\mathbf{L}dt/2}a_n+\mathbf{L}^{-1}\left(e^{\mathbf{L}dt/2}-\mathbf{I}\right)
\left(2\mathbf{N}(b_n,t_n+dt/2)-\mathbf{N}(u_n,t_n)\right)\\
u_{n+1}&=e^{\mathbf{L}dt}u_n+(dt)^{-2}\mathbf{L}^{-3} \bigg( \left[ -4 - \mathbf{L} dt + e^{\mathbf{L}dt} \left( 4-3\mathbf{L}dt+ \left(\mathbf{L}dt\right)^2 \right) \right] \mathbf{N}(u_n,t_n) \\
&+2\left[2+\mathbf{L}dt+e^{\mathbf{L}dt}\left(-2+\mathbf{L}dt\right)\right]
\big(\mathbf{N}(a_n,t_n+dt/2)+\mathbf{N}(b_n+t_n+dt/2) \big) \\
&+\left[-4-3\mathbf{L}dt-\left(\mathbf{L}dt\right)^2+e^{\mathbf{L}dt}\left(4-\mathbf{L}dt\right)\right]
\mathbf{N}(c_n,t_n+dt) \bigg) \,.
\end{aligned}
\end{equation}
This is the iteration formula we use in our numerics. The initial condition is chosen as the two-soliton solution obtained from equations (\ref{spe_soln_Matsuno_MultiSoln}), (\ref{spe_soln_Matsuno_MultiSoln_With}) and (\ref{MultiSolitonCondition}) by setting $N=4$ to produce figures \ref{2solitonICAfter} and \ref{2solitonComparison} in chapter five. The number of points on the contour integral and the time-step were chosen, respectively, as $m=32$ and $dt=0.0125$ in producing these figures. The maximum error ($4.913710^{-5}$) was also shown in chapter five for the choices of the number of discretization $N=2^{15}$ and the time-step $dt=0.0125$. The maximum error, on the other hand, are $0.0020939$ and $0.027373$ units for the choices of $N=2^{14}$ and $dt=0.025$, and $N=2^{13}$ and $dt=0.05$ respectively. The execution time for the choice of $N=2^{15}$ and $dt=0.0125$ is about $1568$ seconds, whereas the  execution time for $N=2^{14}$ and $dt=0.025$ becomes only $330$ seconds. 
\begin{figure}[htp]
\begin{center}
\scalebox{0.6}{\includegraphics{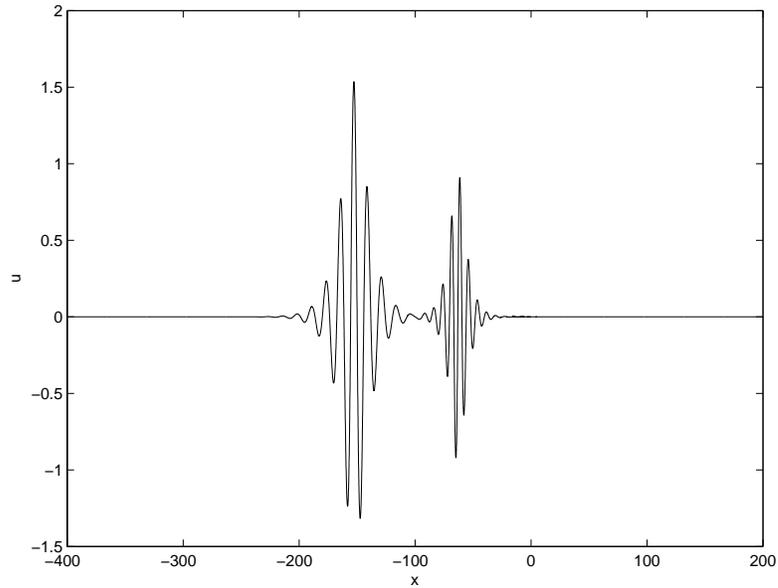}}
\caption{Interaction of the two-soliton solution of the SPE with the time step $dt=0.025$ and the number of discretized temporal interval $N=2^{14}$}
\label{2solitonAfterCollision2ndNdt}
\end{center}
\end{figure} 
The numerical experiment in figure \ref{2solitonAfterCollision2ndNdt} is the repetition of the experiment of figure \ref{2solitonICAfter} with the different discretized temporal domain and the time step, i.e., $N=2^{14}$ and $dt=0.025$. Although the maximum error is bigger in this case, the results seems to be approximating the true values at a good accuracy as seen in figure \ref{2solitonAfterCollision2ndNdt}.    
\begin{figure}[htp]
\begin{center}
\scalebox{0.6}{\includegraphics{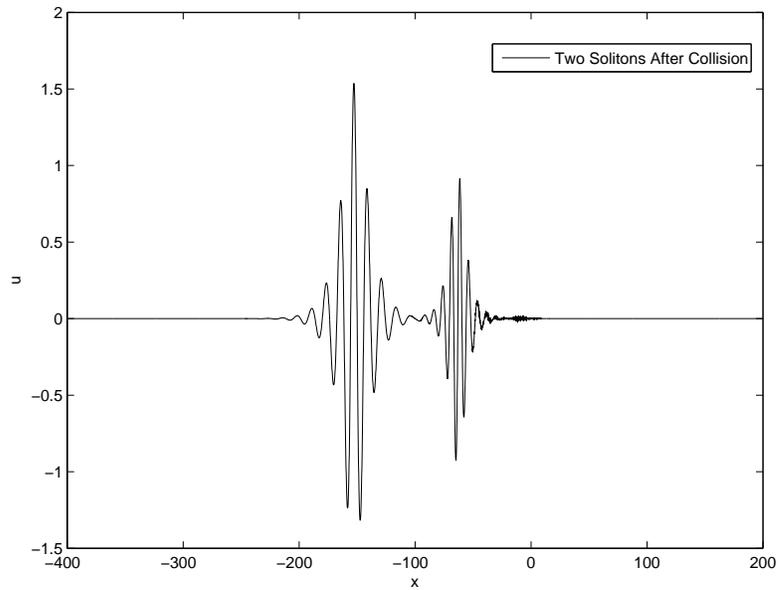}}
\caption{Interaction of the two-soliton solution of the SPE with the time step $dt=0.05$ and the number of discretized temporal interval $N=2^{13}$}
\label{2solitonAfterCollision3rdNdt}
\end{center}
\end{figure} 
If we set $N=2^{13}$ and $dt=0.05$, the integration time decreases to only $82$ seconds. However, figure \ref{2solitonAfterCollision3rdNdt} indicates that the scheme is not stable. We observe some noise in the right tail of the slower soliton after the interaction. The noise like fluctuations shall be associated to the numerical instability.  
.


\section{Random Numbers and Random Number Generators}

The schemes approximating the solutions of the stochastic short pulse equation and stochastic nonlinear wave equation implement random numbers to account for the random variable and randomness in the equations. This section describes random number generators and how randomness is implemented in our numerical methods \cite{otto-denier:2005,moler:2004,numericalrecipes:online}. 

A random variable is a variable whose possible values are numerical outcomes of a random phenomenon. There are two types of random variables; discrete random variable and continuous random variable. Numerical methods utilize the discrete one. A discrete random variable is a random variable that is countable and takes discrete values such as $0,1,2,3,...$. We use simulated random variables to explain statistical pattern recognition and measures. The ability to generate random variables from known probability distributions is the subject of the computational statistics \cite{martinez-martinez:2002}. The statistical analysis of a random process in which there is a random variable that exhibits stochasticity and probabilistic features is called a probability measure. We have already shown probability measures and statistical features of the stochastic short pulse (\ref{StochasticSPEoriginal}) and Maxwell (\ref{maxwell_1d_stochastic_E}) equations in chapter six. When dealing with a random dataset related to the evolution of a random variable, we consider such dataset as one realization of an ensemble that consists of a large number of realizations of a generating process. This is the so-called random process or stochastic process.  A random dataset displays the values of random variables. Such random values may be generated by several different ways. In our numerical methods, we use Matlab's built-in random number generator $randn$ that generates random numbers having a normal (Gaussian) distribution with a mean of zero and a variance of unity. In other words, $randn$ returns a scalar value drawn from a normal distribution with mean zero and standard deviation one. The $randn$ generator should not be mixed up with the $rand$ generator which produce random numbers having uniform distribution. Note that the command $rand$ returns a random number between zero and one. The choice between the usage of these two generators is related to the physics of a problem. As mentioned in chapter six, we model the noise in the system with the white noise whose statistical properties obey normal distribution. The Gaussian white noise is a good approximation of many real-world situations and provides maneuverings for mathematical models. 

The numbers generated by Matlab's generators $rand$ and $randn$ are not truly random numbers. They are, instead, pseudorandom numbers. Pseudorandom numbers generated by the random number generator is a sequence of numbers that approximates the properties of random numbers. This means there is a period of repeating the sequence of random numbers. The sequence of numbers produced by Matlab's pseudorandom generator $randn$ is determined by the internal state of the generator. There are two methods, $state$ and $seed$, that determines the internal state. The method $state$ uses Marsaglia's ziggurat algorithm, which is the default in Matlab's versions $5$ and later. The period for $state$ generator is approximately $2^{64}$. The other method $seed$ uses the polar algorithm, which is the default in Matlab's version $4$ and its period of repeating the pseudorandom numbers is approximately $(2^{31}-1)*(\pi/8)$. One may set the internal state to either one. However, changing states does not improve any statistical properties. Furhermore, $randn$ will generate the same sequence of numbers in each session unless the state is changed since Matlab resets the state at start-up \cite{mathworks:online}. 

When using the $randn$ generator with the either state in the stochastic SPE solvers, one must be very cautious. First of all, the period of the states may seem good enough, but it may not be. The generator $randn$ may draw big random numbers quickly from the normally distributed pseudonumbers. The impact of these relatively big numbers may diverge the pulse propagation since these numbers would be interpreted as a strong noise pertubations. Secondly, the discrete noise approximation necessary to implement the numerical method will modify the variance of the $randn$ generator and, thirdly, the strenght of the noise will be modified as imposed by the multiple scale expansion. Therefore, we have to multiply the $randn$ generator by the coefficient $\epsilon\sqrt{NoiseAmplitude.dt}$\,. The coefficient will surely take different values for the different amplitudes of the noise strength (noise amplitude) and $dt$. One may choose the noise amplitude and $dt$ in a way that the relatively big random numbers drawn by the $randn$ are yet to be rescaled properly.

There are many different way of generating pseudorandom numbers \cite{martinez-martinez:2002}. We also implement $rand$ generator in our numerical schemes. We force the $rand$ generator to produce negative and positive pseudorandom numbers between zero and one. Such an implementation serves as a safety net to block the any unwanted big pseudorandom number in the system's evolution.  
It may be worth to mention that the mean of these pseudorandom numbers may not most likely be zero. We produced the figures in chapter six regarding the stochastic dynamics using the forced $rand$ generator. The results clearly show the impact of the stochasticity on the ultra-short solitons.     

%% file: Chap8_Conclusion.tex
The short pulse equation (SPE) possessing the exact solitary wave solution may be a better modelling equation for the ultra-short pulse propagation in the optical fibers. 
Our numerical work validates the exact derivation of the analytical solution of the SPE. We have also numerically shown that the SPE solitons approximates the solution of the Maxwell equation in one dimension. The higher order SPE is derived. The numerical validation of the higher order SPE and its comparison to the nonlinear wave equation are yet to be done. 

The solitonic properties of the SPE solitons have been tested by allowing them to collide numerically. The particle-like behaviour of the SPE solitons shall be experimented in the presence of stochastic perturbations.

The SPE is a one dimensional model. The short pulse equation in at least two dimensions may be more appealing. Although it seems that the multiple scale expansion ansatz used in the derivation of the one dimensional SPE may not be the multiple scales for a two or three dimensional modelling equation, there may still exist the two or three dimensional SPE.  
 
Stochastic perturbations are present in all physical systems. The randomness has been introduced in the dispersion coefficient of the Maxwell equation based on the arguments made earlier. We have derived a stochastic version of the SPE and shown the results of the numerical experiments to underpin the impact of the stochastic perturbations on the SPE solitons. The future pursuit of the short pulse dynamics in the presence of the randomness may include the introduction of the stochasticity in the nonlinearity. In this case, the study of the statistics of the coarse-graining noise will require a more cautious approach. 

The transformation rule between the SPE and the sine-Gordon (sG) equation opens a channel by which the exact kink and anti-kink solutions of the sG equation are used to obtain the exact solutions of the SPE. The same transformation rule may be applied to the stochastic sG equation to derive a stochastic form of the SPE. Whether the randomness is introduced in the dispersion or in the nonlinearity of the Maxwell equation, the stochastic SPE and its counterparts, if they so happen to exist, may be compared to the other stochastic models such as the stochastic NLSE and the stochastic KdV. The higher order stochastic SPE may be derived and experimented upon so as to improve the numerical results pertaining to the leading order stochastic SPE.